\documentclass[11pt,a4paper]{article}
\usepackage{jheppub}
\usepackage{booktabs}
\usepackage{mathtools}
\usepackage{enumitem}
\usepackage{bm}
\usepackage{array}
\usepackage{multirow}
\usepackage{longtable}
\usepackage{tikz}
\usetikzlibrary{decorations.markings,arrows.meta,calc}
\usepackage{comment}
\usepackage{fontawesome5}
\usepackage[percent]{overpic}

\makeatletter
\newcommand{\github}[1]{%
   \href{#1}{\faGithub}%
}
\makeatother

\newcommand{\Ycal}{\mathcal{Y}}

\newcommand{\Ncal}{\mathcal{N}}

\newcommand{\Dcal}{\mathcal{D}}

\newcommand{\phys}{\mathrm{phys}}

\newcommand{\Del}{\Delta}
\newcommand{\Res}{\mathop{\mathrm{Res}}}
\newcommand{\ds}{\mathrm{dS}}

\newcommand{\Sp}{\mathrm{Sp}}

\newcommand{\Dphys}{\Dcal_{\mathrm{phys}}}
\newcommand{\ksq}[1]{k_{#1}^2}

\newcommand{\intL}{\int_{\ell}}
\newcommand{\bl}{\boldsymbol{\ell}}
\newcommand{\bk}{{\boldsymbol{k}}}
\newcommand{\bp}{{\boldsymbol{p}}}
\newcommand{\bP}{{\boldsymbol{P}}}
\newcommand{\bq}{{\boldsymbol{q}}}
\newcommand{\bx}{{\boldsymbol{x}}}

\newcommand{\der}{\text{d}}

\newcommand{\minor}[2]{[\genfrac{}{}{0pt}{1}{#1}{#2}]}

\newcommand{\boxmark}{\tikz[baseline=-0ex,line width=0.7pt]{\draw (0,0) rectangle (1.0ex,1.0ex);}}
\newcommand{\pentmark}{\tikz[baseline=-0.5ex,line width=0.7pt]{%
  \draw (90:0.62ex)--(162:0.62ex)--(234:0.62ex)--(306:0.62ex)--(18:0.62ex)--cycle;}}

\makeatletter
\g@addto@macro\bfseries{\boldmath}
\makeatother

\title{Higher-Spin Correlators in dS}

\author[\hskip 1pt\boxmark,\hskip 1pt \pentmark]{Shounak De}
\emailAdd{sde25@sas.upenn.edu}

\author[\hskip 1pt\boxmark]{and Hayden Lee}
\emailAdd{haydenhl@sas.upenn.edu}

\affiliation[\boxmark]{Center for Particle Cosmology, Department of Physics and Astronomy, University of Pennsylvania, Philadelphia, PA 19104, USA}

\affiliation[\pentmark]{Institute for Advanced Study, Einstein Drive, Princeton, NJ 08540, USA}

\abstract{
We study momentum-space correlators in minimal higher-spin gravity in $\mathrm{dS}_4$ using its holographic vector-model description. 
Connected $n$-point functions in this model are represented by three-dimensional one-loop integrals and are rational functions of the boundary kinematics. 
We show that the four-point function displays a nontrivial geometric feature: its physical singularity is governed by a Ptolemy relation for the dual momentum quadrilateral. 
At five points, the pentagon integral introduces an apparent leading Landau singularity, which we show is spurious and disappears once the Gram constraint is imposed. 
We derive a representation of the five-point function free of spurious singularities and organized in terms of graph-theoretic building blocks.
We then verify the conformal invariance of these correlators and discuss their behavior in the soft, collapsed, and Ptolemy limits. 
This structure suggests a combinatorial bootstrap for higher-spin correlators at general multiplicity, in which conformal invariance and the physical singularities emerge from the underlying graph combinatorics.}

\preprint{}

\makeatletter
\def\@fpheader{\ }
\makeatother

\begin{document}
\maketitle
\setlength\parskip{4pt}
\newpage
\section{Introduction}

Cosmological correlators are among the sharpest observables of quantum field theory in an expanding universe. These encode inflationary dynamics and are ultimately constrained by observations of the late-time matter distribution. 
Over the last decade, a great deal has been learned about cosmological correlators, leading to a broad bootstrap program for de Sitter (dS) and inflationary observables~\cite{Arkani-Hamed:2015bza,Arkani-Hamed:2018kmz,Baumann:2019oyu,Baumann:2020dch,Sleight:2019hfp,Goodhew:2020hob,Pajer:2020wxk,Melville:2021lst,Baumann:2021fxj,Jazayeri:2021fvk,Sleight:2021plv,Bonifacio:2021azc,Hogervorst:2021uvp,Pimentel:2022fsc,Jazayeri:2022kjy,Baumann:2022jpr}. 
The goal, in analogy with the modern on-shell approach to scattering amplitudes~\cite{Arkani-Hamed:2012zlh,Dixon:2013uaa, Elvang:2013cua,Cheung:2017pzi,Caron-Huot:2020bkp}, is to characterize late-time spatial correlators during inflation directly, rather than reconstructing them from bulk time evolution one diagram at a time. 
This shift in perspective has uncovered hidden structures in cosmological correlators that are nearly invisible in the Lagrangian formalism, revealing a deeper combinatorial, geometric, and differential organization of these observables~\cite{Arkani-Hamed:2017fdk,Arkani-Hamed:2023kig,Arkani-Hamed:2023bsv,Arkani-Hamed:2024jbp,De:2025bmf,Ardila-Mantilla:2026cbo}.

In weakly coupled effective field theories (EFTs) of inflation, these observables admit a conventional perturbative organization in terms of finitely many fields. At tree level, the resulting correlators are now known to display a remarkably rigid pattern of singularities and factorization. This understanding has been especially powerful for computing and classifying signatures of cosmological collider physics, where the precise shape of the correlator encodes the particle spectrum and interactions present during inflation~\cite{Chen:2009zp,Chen:2009we,Baumann:2011nk,Assassi:2012zq,Noumi:2012vr,Chen:2015lza,Lee:2016vti,Chen:2016uwp,Wang:2019gbi,Bodas:2020yho,Qin:2022xrs,Jazayeri:2023xcj,Bodas:2025vpb,You:2026xoq}. More recently, similarly constrained patterns have begun to emerge beyond tree level, where loop correlators exhibit their own distinctive singularities and consistency conditions~\cite{Chowdhury:2023ssc,Benincasa:2024lpy,Arkani-Hamed:2025mce,Pimentel:2026abc,Farren:2026hao,Chowdhury:2026upp,Chowdhury:2026dwm}. 

There is, however, a far more radical possibility: the inflationary scale may sit not far below the string scale, or more generally below a dense tower of higher-spin states. In that regime, an EFT involving only finitely many fields is no longer the natural description. The correlator must instead know about an entire tower of particles at once, with couplings tied together by the consistency of the underlying theory. From the bulk perspective, this would require summing infinitely many exchange diagrams of increasing spin, which may be technically challenging. Any finite truncation risks missing the cancellations and structural relations present in the full tower, and the complete correlator can exhibit qualitatively different behavior from any individual contribution or partial sum.\footnote{For phenomenological studies of cosmological correlators generated by continuous or infinite spectra of intermediate states, see~\cite{Green:2013rd,Kumar:2018jxz,Aoki:2023tjm,Hubisz:2024xnj,Kumar:2025anx,Chakraborty:2025myb,Pimentel:2025rds,Jiang:2025mlm,Aoki:2026yrb}.}

In flat space, this type of resummation is famously realized by the Veneziano amplitude~\cite{Veneziano:1968yb}. There, an infinite tower of higher-spin string states is packaged into a single crossing-symmetric function, whose poles and residues are tied together in a way that is difficult to infer from any finite truncation. More recently, this rigidity has also been approached from a bootstrap perspective, where stringy amplitudes are derived or strongly constrained by supplementing crossing and factorization with assumptions about the spectrum and high-energy behavior~\cite{Caron-Huot:2016icg,Cheung:2022mkw,Cheung:2023adk,Arkani-Hamed:2023jry,Cheung:2023uwn,Cheung:2024uhn,Cheung:2024xuo,Wan:2026pjq}. This naturally raises a similar question in cosmology: can analogous principles single out an intrinsically stringy ``Veneziano correlator'' in dS space, whose analytic structure makes the infinite tower manifest directly on the boundary? Such an object would provide a bottom-up characterization of stringy physics in cosmology, without relying on a complete top-down construction.

Unfortunately, we do not yet know the rules of this game well enough to formulate the problem sharply, but several features are clearly necessary. For instance, the correlator should be conformally invariant and factorize correctly in operator product expansion (OPE) limits. More importantly, if the resummed tower is genuinely stringy, the correlator should inherit the ultraviolet softness of the flat-space amplitude, which in cosmology means softening or eliminating the usual total-energy singularity as $E\to0$~\cite{Arkani-Hamed:2018kmz}.\footnote{In anti-de Sitter space, the equivalent condition is the absence of the bulk-point singularity~\cite{Maldacena:2015iua}.} At present, however, these broad principles offer little guidance on where to begin. To make progress, we thus ask a more modest but concrete question: {\it is there an exactly solvable toy model in dS space, with an infinite higher-spin spectrum and calculable $n$-point functions whose analytic structure can be understood in complete detail?}

Minimal higher-spin gravity in dS~\cite{Vasiliev:1990en,Vasiliev:2003ev} offers a rare setting in which this question can be addressed exactly. This theory may be viewed as a tensionless limit of string theory, where the entire higher-spin tower becomes massless.
The enlarged higher-spin symmetry sets the theory apart from an ordinary local EFT, while imposing much stronger constraints on its observables.\footnote{For a complementary approach to answering this question from the Euclidean
partition function perspective, see~\cite{Giombi:2013fka,Anninos:2020hfj,Anninos:2026hia}.}
In AdS, the theory is conjectured to be holographically dual to vector models on the boundary~\cite{Sundborg:2000wp,Mikhailov:2002bp,Klebanov:2002ja,Sezgin:2002rt,Giombi:2009wh,Giombi:2011kc}, and its dS counterpart~\cite{Anninos:2011ui,Anninos:2017eib} admits exactly computable late-time observables, which in momentum space are represented by one-loop polygon integrals. In~\cite{De:2026shn}, the four-point function was shown to have a strikingly simple organization in a Grassmannian representation~\cite{Arundine:2026fbr,Bala:2026hdm,Bala:2026bdx,Huang:2026tsh,Arundine:2026myr}. 
The goal of this paper is to take a complementary momentum-space point of view and understand how this simplicity is reflected in the analytic structure of higher-point correlators.

The surprise already appears at four points. The correlator is represented by a boundary box integral, and its singularities are not at all those expected from a finite collection of local bulk diagrams. In an ordinary EFT description, cosmological correlators are organized around total- and partial-energy singularities, reflecting the local time evolution of bulk fields. Once the full tower has been summed, however, the correlator is no longer organized by individual bulk diagrams. The singularities are instead governed by the geometry of the dual momentum quadrilateral, distinguishing the resummed higher-spin correlator from those of ordinary perturbative EFTs.

At five points, the geometry of physical kinematics begins to play an essential role. The correlator can be obtained from a pentagon integral through a reduction process, but intermediate expressions contain an apparently spurious pole that is an artifact of the parametrization. We show explicitly that this cancels once the Gram relation appropriate to three-dimensional momentum space is imposed. In the final spurious-free form, the five-point function is organized in terms of certain graph-theoretic structures associated with the five-point dual configuration. Consequently, this reveals a beautiful connection between higher-spin conformal correlators and the algebraic structure of physical kinematics~\cite{DAndrea:2004,CalvoCortes:2025ks}. This gives us a concrete hope that the full physical $n$-point correlator can be bootstrapped, allowing de Sitter higher-spin gravity to be exactly solved at all multiplicities.

The paper is organized as follows. 
In Section~\ref{sec:qformalism}, we review the higher-spin/vector-model setup in de Sitter and recall how scalar boundary correlators are represented by polygon integrals in momentum space.
We discuss the reduction mechanism employed in this paper and show that all polygon integrals are rational functions of the kinematics. 
In Section~\ref{sec:fourpoint}, we analyze the scalar four-point function and discuss its analytic structure. 
In Section~\ref{sec:fivepoint}, we study the five-point function in detail. We show that the Gram constraint is essential for obtaining a physical representation and that the spurious singularities in intermediate expressions cancel only on the physical kinematic locus.
In Section~\ref{sec:symmandsing}, we shed light on the conformal symmetry and the singularity structure of general $n$-point scalar correlators. 
We conclude in Section~\ref{sec:conclusions} with open directions.

The paper also contains three appendices for additional technical details. In Appendix~\ref{app:mel-reductionderivation}, we review the Melrose formalism for the reduction of $n$-gon integrals. 
In Appendix~\ref{app:vv-reduction}, we present an alternative reduction mechanism that yields identical results. 
In Appendix~\ref{app:distance-cwi}, we reformulate the conformal Ward identities in terms of scalar kinematic variables.

\section{Higher-Spin Correlators in dS}
\label{sec:qformalism}

In this section, we set up the computational framework used throughout the paper. 
In~\S\ref{sec:review}, we review the $Q$-model construction of \cite{Anninos:2017eib} and show that the connected $n$-point scalar correlator is given by an $n$-gon integral. 
We then introduce the Melrose reduction~\cite{Melrose:1965} in~\S\ref{sec:setup-cayley}, which expresses an $n$-gon integral in terms of lower-point integrals. 

\subsection{Boundary CFT}
\label{sec:review}

\paragraph{$O(N)$ model in AdS.}
We first recall the AdS version of the story (see~\cite{Giombi:2016ejx,Didenko:2014dwa} for reviews). 
Vasiliev higher-spin gravity in AdS$_4$ is
conjecturally dual to the three-dimensional $O(N)$ vector model~\cite{Klebanov:2002ja, Sezgin:2002rt},\footnote{For discussions regarding bulk non-locality in higher-spin theories, see~\cite{Diaz:2006nm,Sleight:2017pcz,Ponomarev:2017qab,Ponomarev:2019ltz}.} with the action
\begin{equation}
S = \frac12\int \text{d}^3x\,\partial_\mu\phi^i\partial^\mu\phi^i\,,
\label{eq:ON-action}
\end{equation}
where $i=1,\cdots\hskip -1pt,N$. 
The theory may be considered either at the free fixed point or at the Wilson-Fisher fixed point obtained by adding the quartic interaction $(\phi^i\phi^i)^2$.
The two models are dual to the bulk theory with two different boundary conditions on the scalar, corresponding to scaling dimensions $\Delta=1$ and $\Delta=2$.
The single-trace operators are the (normal-ordered) scalar bilinear $:\!\phi^i\phi^i\!\!:$, together with an infinite tower of conserved even-spin currents, schematically of the form $:\!\phi^i\partial^s\phi^i\!:\!-$\,trace. These operators are dual to the bulk scalar and the tower of higher-spin gauge fields, respectively.

\paragraph{$Q$-model in dS.} The de Sitter realization is obtained by analytically continuing this duality. It was first formulated in terms of the $\Sp(N)$ model~\cite{Anninos:2011ui}, which computes the boundary wavefunction. This picture was later sharpened in the $Q$-model formulation~\cite{Anninos:2017eib}, which instead gives a holographic description directly for boundary expectation values and endows the theory with a microscopic Hilbert space. 
We work with the free CFT in a Gaussian Hartle--Hawking state, in which case the boundary correlators are computed exactly by Wick contractions. The scalar operator has    dimension $\Delta=1$, while the dual bulk theory remains an interacting higher-spin system of a scalar, the graviton, and an infinite tower of massless even-spin fields.

The basic variables are $2N$ bosonic fields $Q^\alpha$ on the boundary, with
$\alpha=1,\cdots\hskip -1pt,2N$. 
In momentum space, the boundary scalar bilinear operator $\varphi$ of dimension $\Del =1$ is given by\footnote{The scalar operator used here differs from the operator $B_0$ in~\cite{Anninos:2017eib} by the normalization
$\varphi = B_0/(2\sqrt{2})$.} 
\begin{equation}
 \varphi_\bk = \frac{1}{N}\int\!\frac{\text{d}^3p}{(2\pi)^3} :Q^\alpha_\bp\,Q^\alpha_{\bk-\bp}:.
\label{eq:B0-momentum}
\end{equation}
This scalar operator is related to the $\Del=2$ boundary operator dual to the bulk (conformally coupled) scalar field by a shadow transform. Since $Q$ is Gaussian, all expectation values are built from Wick contractions using the two-point function
\begin{equation}
 \langle Q^\alpha_{\bk} Q^\beta_{\bk'}\rangle  =  \frac{1}{2k^2}\delta^{\alpha\beta}\delta_{\bk+\bk'}\,,
\label{eq:Q-propagator}
\end{equation}
where we use the shorthand notation $\delta_{\bk} \equiv (2\pi)^3\delta^{(3)}(\bk)$ for the momentum-conserving delta function. Momentum conservation at each vertex collapses the $n$ loop variables to a single loop momentum $\bl$, and the connected scalar $n$-point correlator takes the form 
\begin{equation} 
 \langle \varphi_{\bk_1}\cdots \varphi_{\bk_n}\rangle_{\text{conn.}}
 = \frac{1}{4N^{n-1}}\, \underbrace{(I_{12 \cdots n} +  \text{perms.})}_{\equiv\,I_n}\delta_{\bk_1+\cdots+\bk_n}\,, \label{eq:connectedscalarcorr}
\end{equation}
for $n\ge 3$. Here 
$I_n$ denotes the full Bose symmetric sum consisting of $(n-1)!/2$ permutations, while $I_{12\cdots n}$ denotes a single ordered contribution to the full correlator.
It is given by the scalar $n$-gon integral in $D=3$ as
\begin{equation}
 I_{12\cdots n}
 = \intL\;\frac{1} {\ell^2(\bl+\bk_1)^2\cdots(\bl+\bk_1+\cdots+\bk_{n-1})^2} \equiv \intL\; \frac{1}{\bq_1^2 \bq_2^2 \cdots \bq_n^2}\,,
\label{eq:ngon-def}
\end{equation}
where
$\intL \,\equiv\, 8\int \frac{\text{d}^3\ell}{(2\pi)^3}$, with a factor of 8 included for later convenience, and $\bq_i = \bl + \sum_{a=1}^{i-1} \bk_a$ denotes the internal momentum of the $i$-th propagator connecting vertex $i$ to $i+1$.

\begin{figure}[t]
    \centering
    \includegraphics[width=1\linewidth]{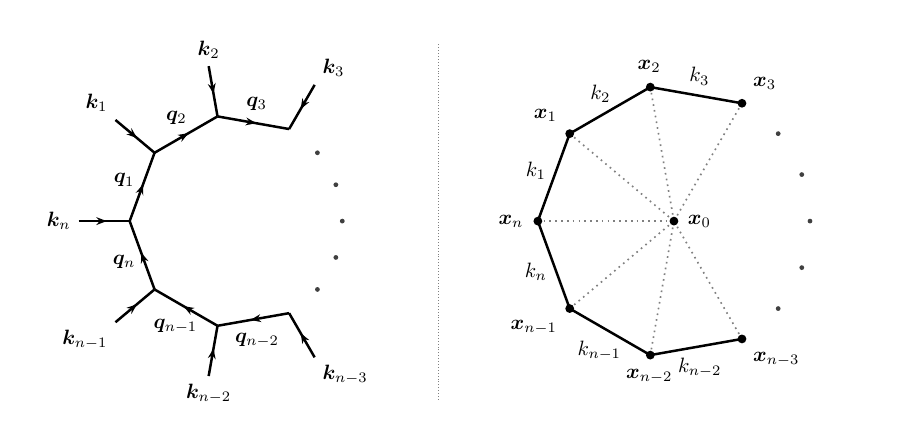}
    \caption{A cyclically ordered $n$-gon (left) and the corresponding dual $n$-gon (right).}
    \label{fig:ngon}
\end{figure}

\paragraph{Boundary kinematics.}
It will be useful to express the above $n$-gon integral in dual coordinates.
Let us introduce $n$ dual points $\bx_i$ defined by the relation 
\begin{align}
    \bx_{i+1} - \bx_i = \bk_i\,,
    \label{eq:dualvertdef}
\end{align}
for $i=1,\cdots\hskip -1pt,n$ with $\bx_{n+1} \equiv \bx_1$, consistent with momentum conservation. 
Introducing an additional dual point $\bx_0$, one can define the loop momentum $\bl = \bx_1 - \bx_0$ such that the $i$-th propagator of~\eqref{eq:ngon-def} becomes $\bq_i \equiv \bx_i -\bx_0$.
In dual coordinates,~\eqref{eq:ngon-def} is rewritten as
\begin{equation}
I_{12\cdots n}=\int_{x_0}\frac{1}{x_{01}^2\,x_{02}^2\cdots x_{0n}^2}~,
\quad x_{ij}\equiv|\bx_i-\bx_j|\,,
\label{eq:ngondual}
\end{equation}
where $\int_{x_0}\equiv 8\int\!\frac{\der^3x_0}{(2\pi)^3}$.
Figure~\ref{fig:ngon} shows both the ordered $n$-gon and its dual counterpart.

\newpage
The dual distance variables $x_{ij}$ specify the kinematics of the ordered correlator $I_{12 \cdots n}$.
They are given by the $\binom{n}{2}$ edges of the complete graph $K_n$ on the dual vertices. 
Writing $k_{i \cdots j} \equiv |\bk_i +\cdots + \bk_{j}|$ for the norm of a sum of consecutive external momenta, the edge set $E_n$ of $K_n$ is given by
\begin{equation}
    E_n=\{x_{ij}\}_{1\leq i<j\leq n} = \{k_1,\cdots\hskip -1pt,k_n\}\, \cup\, \{k_{i\cdots j}\}_{1\leq i<j\leq n-1}\,.
\label{eq:invariantsetEn}
\end{equation}
The $n$ sides are the external energies, $x_{i(i+1)} = k_i$ and $x_{1n}=k_n$ by momentum conservation, and the $\frac{1}{2}n(n-3)$ diagonals $x_{ij} = k_{i \cdots j-1} $ are the partial sums.
In $D=3$ the ordered $n$-gon has $3n-6$ kinematic degrees of freedom, while $E_n$ has $\binom{n}{2}$ elements. 
Consequently, for $n\geq 5$, the invariants in $E_n$ are overcomplete and obey $\frac{1}{2}(n-3)(n-4)$ Gram relations, as we will explicitly see in the case of the pentagon in Section~\ref{sec:fivepoint}.
The complete graph $K_n$ thus controls the boundary kinematics at every multiplicity, and its structure will govern the analytic properties of the scalar correlators throughout this paper. 

\subsection{One-Loop Integral Reduction} 
\label{sec:setup-cayley}

We now discuss the strategy to solve the $n$-gon integrals~\eqref{eq:ngon-def}.
A classic result of Melrose~\cite{Melrose:1965} states that, in $D$~dimensions, any $n$-gon integral can be reduced to a sum of $D$-gon integrals, with coefficients given by ratios of minors of Cayley matrices built out of external kinematics.
The construction is naturally formulated in terms of three nested Cayley matrices. We introduce these matrices below, and then state the linear and quadratic reduction formulas that follow from their rank properties.  
The derivations are reviewed in Appendix~\ref{app:mel-reductionderivation}. 

In three dimensions, the reduction hierarchy is especially simple. The linear reduction expresses each pentagon in terms of boxes, while the quadratic reduction expresses each box in terms of triangles.
The remaining triangle integral is elementary, and we evaluate it explicitly below as a rational function of the Euclidean kinematics.
Thus the full $n$-gon reduction hierarchy terminates in closed form, giving rational expressions for all scalar boundary correlators considered in this paper.\footnote{An alternative reduction scheme based on the van Neerven--Vermaseren basis~\cite{vanNeerven:1984} is presented in Appendix~\ref{app:vv-reduction}. This reduction scheme was used in a similar context in~\cite{Jain:2020rmw}.}

\subsubsection{Cayley Matrices} 
The reduction mechanism is controlled by three nested Cayley matrices, all built from the kinematics of the $n$-gon. 
The first is the \textit{Cayley matrix} $Y_n$, an $n\times n$ symmetric matrix with rows and columns labeled by the propagator indices $i,j\in\{1,\cdots\hskip -1pt,n\}$, whose entries are\footnote{We use a convention in which the Cayley entries are given by minus the squared distances, which absorbs the extra signs that would otherwise appear in the relation to Gram determinants.}
\begin{equation}
(Y_n)_{ij}=\begin{cases}
    0\,, & i=j\,,\\
    -(\bq_i-\bq_j)^2= -(\bk_i+\cdots+\bk_{j-1})^2\,, & i\ne j\,.
\end{cases}
\label{eq:Yn-def}
\end{equation}  
Geometrically, $\bk_i+\cdots+\bk_{j-1}$ is the total external momentum inserted between the two propagators labeled by $\bq_i$ and $\bq_j$ along the cyclic ordering of the polygon. 
Also, since $\bq_i \equiv \bx_i -\bx_0$, the Cayley matrix may equivalently be viewed as minus the Euclidean distance matrix of squared distances between the dual vertices $\bx_1, \cdots,\bx_n$,~i.e.,~$(Y_n)_{ij} = -x_{ij}^2$ with vanishing diagonal entries as above.

Bordering the Cayley matrix $Y_n$ with one extra row and column of unit off-diagonal entries defines the \emph{modified Cayley matrix} $\Ycal_n$,
\begin{equation}
\Ycal_n = \begin{pmatrix}
0 & \boldsymbol{1}  \\
\boldsymbol{1}^T & Y_n
\end{pmatrix} = 
\begin{pmatrix}
0 & 1 & 1 & \cdots & 1 \\
1 & 0 & Y_{12} & \cdots & Y_{1n} \\
1 & Y_{12} & 0 & \cdots & Y_{2n} \\
\vdots & \vdots & \vdots & \ddots & \vdots \\
1 & Y_{1n} & Y_{2n} & \cdots & 0
\end{pmatrix}.
\label{eq:modified-cayley-general}
\end{equation}
Its rows and columns are indexed by ${0,1,\cdots\hskip -1pt,n}$, where the index $0$ labels the border and $i,j\in\{1,\cdots\hskip -1pt,n\}$ label the propagators of the original $n$-gon. Writing $G(\boldsymbol{v}_1,\cdots\hskip -1pt,\boldsymbol{v}_m) \equiv \det(\boldsymbol{v}_i\cdot \boldsymbol{v}_j)$ for the \textit{Gram determinant} of $m$ vectors, the modified Cayley matrix $\Ycal_n$ has the following properties: 
\begin{align}
\text{rank}\,\Ycal_n &= \min (n+1,D+2)\,,
\label{eq:Y-rank}
\\
\det \Ycal_n &= -2^{n-1}\,G(\bk_1,\dots,\bk_{n-1}) \equiv -2^{n-1} G_n\,,
\label{eq:detY-gram}
\\
\Ycal_n[\begin{smallmatrix}0\\0\end{smallmatrix}] &= \det Y_n \equiv \Del_n = 2^n\hskip 1ptG(\boldsymbol{q}_1,\dots,\boldsymbol{q}_n)\big|_{\bq_i^2=0}\,.
\label{eq:cayley-minor}
\end{align}
The determinant of $\Ycal_n$ is thus proportional to the Gram determinant $G_n$ of any $(n{-}1)$ of the $n$ external momenta $\bk_i$, while the $(0,0)$ minor defines the \textit{leading Landau polynomial} $\Delta_n$ of the $n$-gon, which is proportional to the Gram determinant of the internal momenta restricted to be on-shell $\bq_i^2 = 0$. 
More generally, for multi-indices $I$ and $J$, $\Ycal_n[\begin{smallmatrix}I\\J\end{smallmatrix}]$ denotes the signed minor (or {\it cofactor}) obtained by deleting the rows $I$ and columns $J$. 
These minors are the basic kinematic quantities that enter the scalar $n$-gon reduction formulas below.

We also introduce the \emph{Cayley--Menger matrix} $\text{CM}_n$, defined by
\begin{equation}
       \text{CM}_n =
    \begin{pmatrix}
0 & 1 & 1 & \cdots & 1 & 1\\[1pt]
1 & 0 & Y_{12} & \cdots & Y_{1n} & -d_1\\[1pt]
1 & Y_{12} & 0 & \cdots & Y_{2n} & -d_2\\[1pt]
\vdots & \vdots & \vdots & \ddots & \vdots & \vdots\\[1pt]
1 & Y_{1n} & Y_{2n} & \cdots & 0 & -d_n\\[1pt]
1 & -d_1 & -d_2 & \cdots & -d_n & 0
\end{pmatrix}.
\label{eq:CM}
\end{equation}
The upper-left $(n{+}1)\times(n{+}1)$ block of $\text{CM}_n$ is the modified Cayley matrix $\Ycal_n$, while the additional final row and column of $\mathrm{CM}_n$ contain the entries $-d_i\equiv-\bq_i^2 = - x_{0i}^2$.
For generic kinematics, this matrix satisfies
\begin{align}\label{eq:rankcm}
    \text{rank}\,\text{CM}_n &= \min (n+2,D + 2)\,,\\
    \det\text{CM}_n&= -2^{n}\hskip 1pt G(\bm q_1,\cdots\hskip -1pt,\bm q_n)\,.
\end{align}
The second relation expresses the Gram determinant of the internal momenta in terms of signed squared distances. The rank deficiencies of the matrices $\Ycal_n$ and $\text{CM}_n$ drive the reduction mechanism discussed below. Table~\ref{tab:cayley} summarizes the three Cayley matrices along with the properties described above.

\begin{table}[t]
\centering
\renewcommand{\arraystretch}{1.2}
\setlength{\tabcolsep}{7.9pt}
\begin{tabular}{c l c c c}
\toprule
Matrix & Name & Order & Rank & Determinant \\
\midrule
$Y_n$ & Cayley & $n$ & $\min(n,D{+}2)$ & $\Delta_n=2^n\,G(\bq_1,\dots,\bq_n)|_{q_i^2=0}$  \\
$\Ycal_n$ & modified Cayley & $n+1$ & $\min(n{+}1,D{+}2)$ & $-2^{n-1}\,G(\bk_1,\dots,\bk_{n-1})$  \\
$\text{CM}_n$ & Cayley--Menger & $n+2$& $\min(n{+}2,D{+}2)$ & $-2^n\,G(\bq_1,\dots,\bq_n)$  \\
\bottomrule
\end{tabular}
\caption{The three Cayley matrices and their properties.}
\label{tab:cayley}
\end{table}

\subsubsection{Reduction Formulas} 

The reduction proceeds in two stages, each relying on a different algebraic consequence of the rank deficiencies described above. We summarize the resulting formulas here and defer their derivations to Appendix~\ref{app:mel-reductionderivation}.

\paragraph{Linear reduction.}

When $n>D+1$, the Cayley--Menger matrix $\text{CM}_n$ has order greater than $D+3$ but rank only $D+2$. Likewise, the submatrix $\text{CM}_{(l_1\cdots l_{D+2})}$ associated with any set of $D+2$ propagator labels $\{l_1,\dots,l_{D+2}\}\subset\{1,\dots,n\}$ has rank $D+2$.
Consequently, every $(D+3){\times}(D+3)$ minor containing the final propagator column vanishes, and expanding it along that column yields a {\it linear} relation among the denominators $d_{l_r}=\bq_{l_r}^2$, with coefficients given by signed minors of $\mathcal{Y}_{(l_1\cdots l_{D+2})}$. 
Substituting this relation into the integrand expresses the $n$-gon as a sum of $(n{-}1)$-gons, and the resulting single-step reduction reads \cite{Melrose:1965}
\begin{equation}
I_{12\cdots n} = \sum_{r=1}^{D+2} \frac{\Ycal_{(l_1\cdots l_{D+2})}[\begin{smallmatrix}0\\ l_r\end{smallmatrix}]}{\Ycal_{(l_1\cdots l_{D+2})}[\begin{smallmatrix}0\\0\end{smallmatrix}]}\,I^{(n)}_{l_r}\,,
\label{eq:single-step}
\end{equation}
where $I^{(n)}_{l_r}$ is the $(n{-}1)$-gon obtained by removing the propagator $d_{l_r}$.

Iterating the linear single-step relation~\eqref{eq:single-step} above until only $(D{+}1)$-gons remain gives
\begin{equation}\label{eq:complete-D+1}
  I_{12\cdots n} = \sum_{l_1 < l_2 < \cdots < l_{D+1}}
  f^{(n)}_{l_1 \cdots l_{D+1}}I_{l_1 \cdots l_{D+1}}\,,
\end{equation}
where $l_1, \cdots\hskip -1pt, l_{D+1}$ are the remaining $D+1$ internal momenta. 
The coefficient accumulated along the reduction chain is
\begin{equation}\label{eq:f-coeff}
 f^{(n)}_{l_1 \cdots l_{D+1}} = \prod_{\alpha = D+2}^{n}\frac{\Ycal_{(l_1\cdots l_{D+1}\,l_\alpha)}[\begin{smallmatrix}0\\ l_\alpha\end{smallmatrix}]}{\Ycal_{(l_1\cdots l_{D+1}\,l_\alpha)}[\begin{smallmatrix}0\\ 0\end{smallmatrix}]}\,,
\end{equation}
where $\Ycal_{(l_1 \cdots l_{D+1}\, l_\alpha)}$ denotes the modified Cayley matrix formed from the indicated subset of propagators.  
Crucially, this coefficient is independent of the order in which the intermediate reductions are performed. Any two reduction chains terminating in the same final set $\{l_1, \dots, l_{D+1}\}$ thus yield the same expression~\eqref{eq:f-coeff}.

\begin{figure}[t]
  \centering
  \begin{overpic}[
    width=\linewidth
  ]{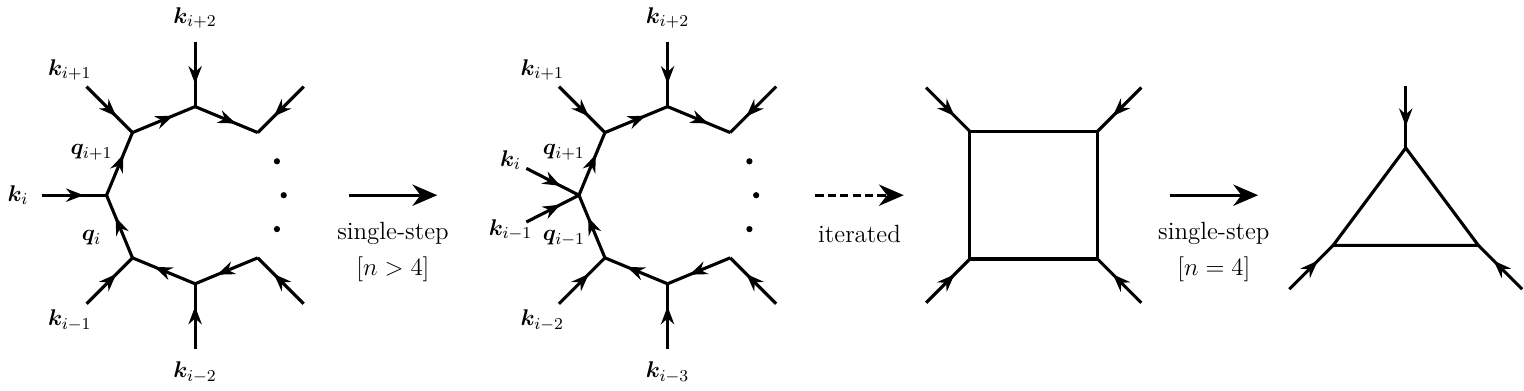}
    \put(25.7,16){%
      \makebox(0,0){\scriptsize\eqref{eq:single-step}}%
    }
    \put(56.1,16){%
      \makebox(0,0){\scriptsize\eqref{eq:complete-D+1}}%
    }
    \put(79.6,16){%
      \makebox(0,0){\scriptsize\eqref{eq:D+1-to-D}}%
    }
  \end{overpic}
  \caption{The Melrose reduction hierarchy in $D=3$.}
  \label{fig:melrose-reduction}
\end{figure}

\paragraph{Quadratic reduction.}

Each $(D{+}1)$-gon in~\eqref{eq:complete-D+1} sits at the critical case $n=D+1$, where $\Ycal_{D+1}$ has full rank and the linear constraint underlying the single-step reduction no longer applies.
Instead, one uses the quadratic Gram constraint for the internal momenta, which gives the reduction formula~\cite{Melrose:1965}
\begin{equation}
 I_{12\cdots (D{+}1)} = \sum_{i=1}^{D+1} \frac{\Ycal_{D{+}1}[\begin{smallmatrix}0\\ i\end{smallmatrix}]}{\Del_{D{+}1}}\, I^{(D{+}1)}_i\,,
\label{eq:D+1-to-D}
\end{equation}
where $I^{(D{+}1)}_i$ denotes the $D$-gon with the propagator at the $i$-th position removed, and $\Delta_{D{+}1} = \Ycal_{D{+}1}[\begin{smallmatrix}0\\0\end{smallmatrix}]$ is the leading Landau polynomial of the $(D{+}1)$-gon.
In Section~\ref{sec:fourpoint}, we apply this formula explicitly in $D=3$ to reduce the box integral to its four sub-triangles. 
Unlike the linear single-step relation~\eqref{eq:single-step}, this reduction admits no choice of reduction channel and is thus unique.

Combining the iterated linear relation~\eqref{eq:complete-D+1} with the quadratic relation~\eqref{eq:D+1-to-D} gives a complete reduction of the $n$-gon to its $\binom{n}{D}$ sub-$D$-gons. 
This process in $D=3$ is illustrated in Fig.~\ref{fig:melrose-reduction}. The resulting coefficients are products of modified Cayley minors and depend only on the external kinematics.
The explicit general formula is presented in Appendix~\ref{app:mel-reductionderivation}.

\subsubsection{The Triangle}
\label{sec:threepoint}
At the base of the reduction hierarchy in $D=3$ sits the triangle diagram, which is simple to evaluate. 
Using the standard star-triangle formula~\cite{Symanzik:1972wj}, one finds
\begin{equation}
I_{123} = \int_{\ell} \frac{1}{\ell^2(\bl + \bk_1)^2(\bl+\bk_1 +\bk_2)^2} =\frac{1}{k_1k_2k_3}\,,
\label{eq:triangle-eval}
\end{equation}
where $k_i \equiv |\bk_i|$ are the magnitudes of the external momenta. 
The corresponding three-point function is therefore\footnote{Upon shadow transforming the scalars, one obtains the three-point function of $\Delta=2$ boundary operators $\langle \tilde{\varphi}_{\bm k_1} \tilde{\varphi}_{\bm k_2} \tilde{\varphi}_{\bm k_3} \rangle = \frac{1}{N^2} \delta_{\bk_1{+}\bk_2{+}\bk_3}$, which in position space becomes a pure contact term proportional to a product of delta functions.
The contact form is consistent with the vanishing of the $J = 0$ single-trace OPE coefficient in the AdS higher-spin/vector-model duality~\cite{Petkou:2003,Sezgin:2003pt}, which precludes a nontrivial scalar exchange channel from contributing to higher-point correlators built from cubic vertices.} 
\begin{equation}
\langle \varphi_{\bm k_1}\varphi_{\bm k_2}\varphi_{\bm k_3} \rangle = \frac{1}{4 N^2}  I_{123} \, \delta_{\bk_1+\bk_2+\bk_3}\,.
\label{eq:3pt-final}
\end{equation}
This result matches the bulk computation and the form allowed by conformal symmetry. Genuinely nonlocal features of the higher-spin correlators first appear at four points, where the underlying box integral has nontrivial analytic structure that cannot be reproduced by a local bulk interaction.

\section{The Box}
\label{sec:fourpoint}

We now evaluate the scalar four-point function by explicitly computing the $D=3$ box integral. 
In \S\ref{sec:boxtotirangleredux}, we carry out the box-to-triangle reduction using the framework outlined in the previous section.
In \S\ref{sec:analstrucbox}, we discuss the Landau singularities and analytic structure of the box integral, showing in particular that three of the four Landau branches cancel algebraically in the final result.

\subsection{Reduction to Triangles}
\label{sec:boxtotirangleredux}

\begin{figure}[t]
\centering
\includegraphics{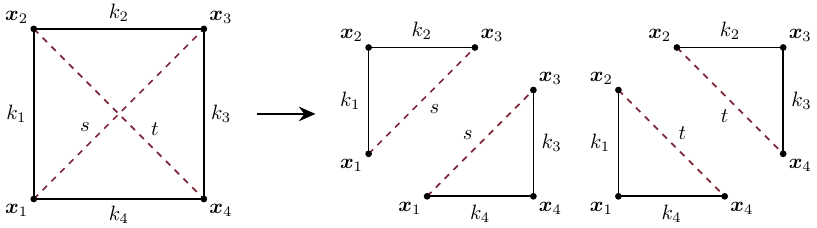}
\caption{The complete graph $K_4$ on the four dual vertices and its reduction to sub-triangles.
}
\label{fig:box-geometry}
\end{figure}

The box integral depends on a choice of cyclic ordering of the four external momenta. We take the ordering $(1234)$ as the reference box, while the remaining ordered contributions follow by permuting the external labels. 
For this ordering, the corresponding box integral is
\begin{align}
I_{1234}
&=
\int_\ell
\frac{1}{\ell^2(\bl+\bk_1)^2(\bl+\bk_1+\bk_2)^2
(\bl+\bk_1+\bk_2+\bk_3)^2}~,
\label{eq:box-def-int}
\end{align}
which is a function of six invariants specified by the edge set $E_4$ of the complete graph $K_4$
\begin{equation}
     E_4 = \{k_1,k_2,k_3,k_4,s,t\}\,,
     \label{eq:fourpointdisvar}
\end{equation}
as illustrated in Fig.~\ref{fig:box-geometry}.
Here $k_i=|\bk_i|$ are the four sides of the dual quadrilateral, while
\begin{equation}
s\equiv k_{12}=|\bk_1+\bk_2|\,,
\quad
t\equiv k_{23}=|\bk_2+\bk_3|
\end{equation}
are its two diagonals. 
These six variables specify the physical kinematics of the ordered box contribution $I_{1234}$.

Using the reduction formula \eqref{eq:D+1-to-D}, the ordered box integral reduces uniquely to its four triangle subgraphs 
\begin{equation}
I_{1234}
=
\frac{\Ycal_4[\begin{smallmatrix}0\\1\end{smallmatrix}]}{\Del_4}\,I_{234}
+\frac{\Ycal_4[\begin{smallmatrix}0\\2\end{smallmatrix}]}{\Del_4}\,I_{134}
+\frac{\Ycal_4[\begin{smallmatrix}0\\3\end{smallmatrix}]}{\Del_4}\,I_{124}
+\frac{\Ycal_4[\begin{smallmatrix}0\\4\end{smallmatrix}]}{\Del_4}\,I_{123}\,,
\label{eq:box-sum-triangles}
\end{equation}
as depicted in Fig.~\ref{fig:box-geometry}, with the individual triangle integrals given by 
\begin{align}
I_{234}=\frac{1}{k_2k_3t}\,,\quad
I_{134}=\frac{1}{k_3k_4s}\,,\quad I_{124}=\frac{1}{k_1 k_4 t}\,,\quad I_{123}=\frac{1}{k_1k_2s}\,.
\label{eq:triangle-values}
\end{align}
The reduction coefficients are the signed minors $\Ycal_4[\begin{smallmatrix}0\\i\end{smallmatrix}]$ of the modified Cayley matrix
\begin{equation}
\Ycal_4=
\begin{pmatrix}
0 & 1 & 1 & 1 & 1 \\
1 & 0 & -k_1^2 & -s^2 & -k_4^2 \\
1 & -k_1^2 & 0 & -k_2^2 & -t^2 \\
1 & -s^2 & -k_2^2 & 0 & -k_3^2 \\
1 & -k_4^2 & -t^2 & -k_3^2 & 0
\end{pmatrix},
\label{eq:Y4-box}
\end{equation}
obtained by deleting the zeroth row and $i$-th column.

\subsection{Analytic Structure}
\label{sec:analstrucbox}

We now extract the final form of the box integral from the reduction formulas~\eqref{eq:box-sum-triangles}--\eqref{eq:Y4-box}.
The leading Landau singularity of the box integral is given by $\Del_4 \equiv \det Y_4 = 0$, whose direct evaluation gives the K\"all\'{e}n function
\begin{align}
    \Delta_4 = \lambda(k_1^2 k_3^2, k_2^2 k_4^2, s^2 t^2)\,, \quad \lambda(a,b,c) = a^2 + b^2 + c^2 - 2ab - 2ac - 2bc\,. 
    \label{eq:kallen}
\end{align}
Written in terms of the squared invariants alone (as in a scattering amplitude), $\Delta_4$ is an irreducible quartic polynomial. But the box integral naturally lives over the Euclidean distance variables~\eqref{eq:fourpointdisvar}, in which case $\Delta_4$ factorizes completely into four quadratic factors\footnote{This four-point quartic is closely related to the rank-two Pl\"ucker relation. For massless four-particle kinematics, the quartic in flat-space Mandelstam variables is the \emph{squared} Pl\"ucker quadric and naturally belongs to the squared-Grassmannian viewpoint \cite{CalvoCortes:2025ks,Devriendt:2024twolives}. Passing to the Euclidean distance ring effectively provides a square-root lift of this quartic, so the rank-two relation splits into the four quadratic factors~$F_{\pm\pm}$.}
\begin{equation}\label{eq:Landaufact}
\Delta_4 = F_{++}F_{+-}F_{-+}F_{--}~, \quad
  F_{\varepsilon_1\varepsilon_2} \equiv
  st+\varepsilon_1 k_1k_3+\varepsilon_2 k_2k_4\,,\quad
  \varepsilon_1,\varepsilon_2=\pm1~.  
\end{equation}
The four quadratic factors $F_{\varepsilon_1\varepsilon_2}$ define the four branches of the leading Landau locus $\Delta_4 = 0$, distinguished by the relative signs of the two products of opposite edges of $K_4$, namely $k_1 k_3$ and $k_2 k_4$, relative to the diagonal product $st$. 

\begin{figure}[t]
\centering
\includegraphics{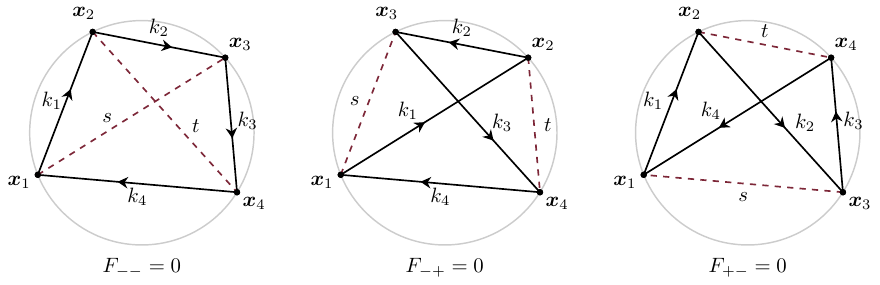}
\caption{The three Euclidean Ptolemy limits of the dual box graph $K_4$.}
\label{fig:euclidean-ptolemy}
\end{figure}

The three branches with at least one minus sign have a direct geometric interpretation: they are the Euclidean Ptolemy conditions for the three inequivalent cyclic orderings of the four dual vertices, as illustrated in Fig.~\ref{fig:euclidean-ptolemy}. 
For the reference ordering $\{\bx_1,\bx_2,\bx_3,\bx_4\}$, the quadrilateral has sides $(k_1,k_2,k_3,k_4)$ and diagonals $(s,t)$. The corresponding Ptolemy condition is
\begin{equation}
F_{--}=0
\quad \Leftrightarrow \quad
st=k_1k_3+k_2k_4\,.
\label{eq:euclidean-ptolemy}
\end{equation}
The two mixed-sign branches give the corresponding Ptolemy relations for the other two orderings. For the ordering $\{\bx_1,\bx_3,\bx_2,\bx_4\}$, the quadrilateral has sides $(s,k_2,t,k_4)$ and diagonals $(k_1,k_3)$, giving
\begin{equation}
F_{-+}=0
\quad \Leftrightarrow \quad
k_1k_3=st+k_2k_4\,.
\end{equation}
Similarly, for the ordering $\{\bx_1,\bx_2,\bx_4,\bx_3\}$, the quadrilateral has sides $(k_1,t,k_3,s)$ and diagonals $(k_2,k_4)$, leading to
\begin{equation}
F_{+-}=0
\quad \Leftrightarrow \quad
k_2k_4=st+k_1k_3\,.
\end{equation}
By contrast, the remaining factor
$F_{++}=st+k_1k_3+k_2k_4$
is strictly positive in the physical Euclidean region $k_i,s,t>0$ and its vanishing locus can be reached only after analytic continuation.

Bringing the four triangle contributions in~\eqref{eq:box-sum-triangles} over a common denominator, the numerator assembles into $F_{+-}F_{-+}F_{--}\,\mathcal{N}_4$, where
\begin{equation}\label{eq:N4def}
  \mathcal{N}_4 \equiv (k_1k_2+k_3k_4)s + (k_1k_4+k_2k_3)t\,.
\end{equation}
The ordered correlator therefore takes the form
\begin{align}
    I_{1234} = \frac{F_{+-}F_{-+}F_{--}\,\mathcal{N}_4}{k_1 k_2 k_3 k_4st \Del_4} = \frac{(k_1 k_2 + k_3 k_4)s + (k_1 k_4 + k_2 k_3)t}{k_1 k_2 k_3 k_4st(st+ k_1 k_3 + k_2 k_4)}\,.
    \label{eq:4gonfinal}
\end{align}
We see that three of the four Landau branches cancel against the numerator, leaving $F_{++}$ as the only physical singularity,\footnote{Here and throughout, we use ``physical singularity'' to mean a non-spurious singularity with nonzero residue. Such singularities lie in Lorentzian kinematics and are therefore inaccessible from the Euclidean region. In particular, the Ptolemy singularity of the box can be reached for real momenta in the physical $u$-channel process $1+3\to2+4$, where it corresponds to a box anomalous threshold~\cite{Mizera:2021fap}.} which we will call the \emph{Ptolemy singularity}, in addition to the soft and collapsed singularities associated with the explicit factors $k_1k_2k_3k_4st$.
This reproduces the known result of~\cite{Anninos:2017eib}, obtained using the methods of \cite{Bzowski:2011ab, Bzowski:2013sza}.

The full correlator is Bose symmetric, with the remaining two ordered correlators following from $I_{1234}$ by permuting the external momenta. 
The full connected scalar four-point function is then given by
\begin{equation}
\big\langle \varphi_{\bm k_1}\varphi_{\bm k_2}\varphi_{\bm k_3}\varphi_{\bm k_4}\big\rangle_{\mathrm{conn.}}
=\frac{1}{4N^{3}}(I_{1234}+I_{1324}+I_{1243}) \, 
\delta_{\bk_1+\cdots+\bk_4}\,.
\label{eq:fullfourpointfunction}
\end{equation}
Let us now make a few remarks on the final answer obtained above.

\paragraph{Graph symmetry.}
The numerator $\Ncal_4$ of~\eqref{eq:N4def} organizes into subgraphs of the dual graph $K_4$, whose six edges are the distance variables~\eqref{eq:fourpointdisvar}.
Recall from~\eqref{eq:ngondual} that the ordered box in dual space reads
\begin{align}
I_{1234} =\int_{x_0} \frac{1}{x_{01}^2x_{02}^2x_{03}^2x_{04}^2}\,.
\label{eq:box-dual-symmetry}
\end{align}
It follows directly that the ordered $I_{1234}$ is invariant under the relabeling of dual vertices, $x_i\leftrightarrow x_j$, or the action of the symmetric group $S_4$. 
The subgraphs appearing in $\mathcal{N}_4$ must therefore close under the orbits of $S_4$.

Each term in $\mathcal{N}_4$ is the product of the three edge lengths of a triangular subgraph of $K_4$. 
For example, $k_1k_2s$ corresponds to the triangle with vertices $\{\bx_1,\bx_2,\bx_3\}$ in Fig.~\ref{fig:box-geometry}. 
Its stabilizer is $S_3\times S_1$, giving an orbit of size $|S_4|/|S_3\times S_1|=24/6=4$, so the four triangles of $K_4$ form a single orbit. 
Under the action of the rotation generator $\sigma_4$ defined as
\begin{equation}
\sigma_4:\, \{k_1\to k_2\to k_3\to k_4\to k_1,\, s\leftrightarrow t \} \,,
\label{eq:4pt-rotation}
\end{equation}
the four triangles close into a single orbit
\begin{equation}
k_1 k_2\, s \;\xrightarrow{\sigma_4}\; k_2 k_3\, t \;\xrightarrow{\sigma_4}\; k_3 k_4\, s \;\xrightarrow{\sigma_4}\; k_1 k_4\, t \;\xrightarrow{\sigma_4}\; k_1 k_2\, s\,.
\label{eq:box-numerator-orbit}
\end{equation}
The numerator is thus the sum over the four triangle subgraphs of $K_4$.
Although this symmetry pattern is somewhat trivial at four points, we will see that a similar orbit decomposition becomes particularly useful at five points.

\paragraph{Ptolemy singularity.}

The fact that only $F_{++}$ survives in the physical denominator can also be understood from the Feynman-parameter representation of the ordered box
\begin{equation}
I_{1234} = \frac{\Gamma(5/2)}{\pi^{3/2}} \int_{\alpha_i\ge 0} \frac{\der^4\alpha\, \delta(1-\sum_{i=1}^{4}\alpha_i)}{[\mathcal{F}_4(\alpha)]^{5/2}}\,,
\label{eq:box-feynman-parameters}
\end{equation}
where the second Symanzik polynomial is given by
\begin{align}
\mathcal{F}_4 (\alpha) &= -\frac{1}{2}\alpha^T Y_4\, \alpha = \alpha_1\alpha_2 k_1^2  +\alpha_1\alpha_3 s^2 +\alpha_1\alpha_4 k_4^2  +\alpha_2\alpha_3 k_2^2 +\alpha_2\alpha_4 t^2 +\alpha_3\alpha_4 k_3^2\,.
\label{eq:F4-symanzik}
\end{align}
In the physical Euclidean region $k_i,s,t>0$, every coefficient is strictly positive. It follows that $\mathcal{F}_4(\alpha)>0$ throughout the interior of the simplex $\alpha_i>0$ and vanishes only on its boundary. The integrand therefore has no interior pinch, and the integral~\eqref{eq:box-feynman-parameters} is regular in the Euclidean domain.

This explains why the Euclidean branches $F_{--}$, $F_{-+}$, and $F_{+-}$ do not produce poles in the correlator. The leading Landau equations are
\begin{equation}
\mathcal{F}_4(\alpha)=0\,,
\quad
\nabla_{\alpha}\mathcal{F}_4(\alpha)=0\,.
\end{equation}
The second condition is equivalent to $Y_4\cdot\alpha=0$. Each of the three Euclidean branches is the Ptolemy locus associated with one of the cyclic orderings of the four dual vertices: $\{\bx_1,\bx_2,\bx_3,\bx_4\}$ gives $F_{--}=0$, while $\{\bx_1,\bx_3,\bx_2,\bx_4\}$ and $\{\bx_1,\bx_2,\bx_4,\bx_3\}$ give $F_{-+}=0$ and $F_{+-}=0$, respectively. Although these are genuine loci in the Euclidean distance domain, the corresponding $\alpha$ have components of mixed sign and therefore lie outside the positive Feynman simplex. Their cancellation in \eqref{eq:4gonfinal} is the algebraic reflection of this analyticity. 

The only non-collapsed branch that remains in the correlator is the Ptolemy singularity~$F_{++}=0$. This locus cannot be reached when all $k_i,s,t>0$, but becomes accessible after analytic continuation to Lorentzian kinematics, as expected for correlators in the Hartle--Hawking/Bunch--Davies vacuum.

\section{The Pentagon}
\label{sec:fivepoint}

The branch-by-branch description of spurious pole cancellation is special to four points. For the pentagon, the leading Landau polynomial $\Delta_5$ is irreducible in the distance variables, so there is no analogous decomposition into separate branches.
Moreover, physical five-point kinematics is defined on the Gram locus $\det\Ycal_5=0$, and the distance variables are no longer independent.
As a consequence, evaluating the pentagon using the reduction formula introduces a denominator with a spurious singularity. 
Although this spurious pole must cancel on the physical kinematic locus, its presence obscures the true analytic structure of the correlator and makes the resulting expression unsuitable.

In this section, we explicitly show that this spurious pole cancels after imposing the Gram constraint, and derive a manifestly spurious-free representation. Since $\Delta_5$ is irreducible, the cancellation cannot be analyzed branch by branch as in the box case. Instead, we parametrize the external momenta in components, establish the cancellation in this parametrization, and then reconstruct the physical answer.

\subsection{Reduction to Boxes}
\label{sec:pent-to-boxreduction}

\begin{figure}[t]
\centering
\includegraphics[scale=0.9]{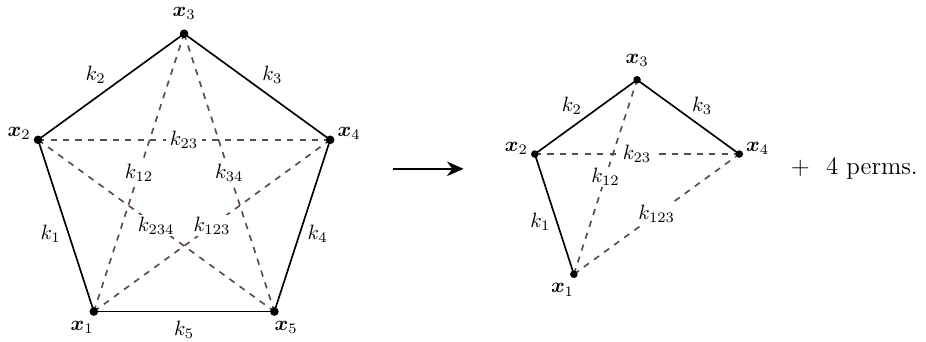}
\caption{The complete graph $K_5$ on the five dual vertices and its reduction to sub-boxes.
}
\label{fig:K5}
\end{figure}

We again consider a single ordered contribution with cyclic ordering $(12345)$. 
The full five-point correlator $I_5$ is obtained by summing over permutations of the external legs.
For this ordered correlator, the relevant scalar pentagon integral in $D=3$ is
\begin{align}
 I_{12345} &=   
 \!\int_\ell \frac{1}{\ell^2(\bl + \bk_1)^2(\bl + \bk_1+\bk_2)^2(\bl +\bk_1+\bk_2+\bk_3)^2 (\bl - \bk_5)^2} \,,
 \label{eq:5-gon}
\end{align}
which is a function of ten invariants given by the edge set $E_5$ of the complete graph $K_5$
\begin{equation}
E_5=\{k_1,k_2,k_3,k_4,k_5,k_{12},k_{23},k_{34},k_{123},k_{234}\}\,,
\label{eq:disvar5pt}
\end{equation}
where $k_{ij}\equiv |\boldsymbol{k}_i+\boldsymbol{k}_j|$ and $k_{ijl}\equiv |\boldsymbol{k}_i+\boldsymbol{k}_j+\boldsymbol{k}_l|$. These variables are illustrated in Fig.~\ref{fig:K5}.

Since the pentagon has $n = 5 >D+1$ in $D=3$ dimensions, we can directly apply the single-step reduction formula~\eqref{eq:single-step}.
Using the modified Cayley matrix at five points
\begin{equation}
\Ycal_5 = \begin{pmatrix}
0 & 1 & 1 & 1 & 1 & 1 \\
1 & 0 & -k_1^2 & -k_{12}^2 & -k_{123}^2 & -k_5^2 \\
1 & -k_1^2 & 0 & -k_2^2 & -k_{23}^2 & -k_{234}^2 \\
1 & -k_{12}^2 & -k_2^2 & 0 & -k_3^2 & -k_{34}^2 \\
1 & -k_{123}^2 & -k_{23}^2 & -k_3^2 & 0 & -k_4^2 \\
1 & -k_5^2 & -k_{234}^2 & -k_{34}^2 & -k_4^2 & 0
\end{pmatrix},
\label{eq:Y5-explicit}
\end{equation}
the pentagon reduces to a sum of five box integrals as
\begin{equation}
I_{12345} = \sum_{i=1}^{5}
\frac{\Ycal_5[\begin{smallmatrix}0\\i\end{smallmatrix}]}{\Ycal_5[\begin{smallmatrix}0\\0\end{smallmatrix}]}\, I^{(5)}_i\,,
\label{eq:cayley-pentagon-reduction}
\end{equation}
where $I_i^{(5)}$ denotes the box obtained by deleting the $i$-th propagator. Since the boxes were evaluated in Section~\ref{sec:fourpoint}, this expresses the ordered pentagon as a rational function of the ten distance variables.

There is, however, one important subtlety. From~\eqref{eq:Y-rank}, we see that $\Ycal_5$ is rank deficient as it is a $6\times6$ matrix but its rank is only $D + 2 = 5$. 
This leads to the Gram constraint 
\begin{equation}
G_5 \equiv G(\bk_1, \bk_2, \bk_3, \bk_4) \propto \det\Ycal_5 = 0\,,
\label{eq:gramconstraintfivepoint}
\end{equation}
which gives a polynomial relation between the ten distance variables.\footnote{Equivalently, five momentum vectors in $D{=}3$ contain $15$ components. Momentum conservation removes $3$ and rotational invariance removes another $3$, leaving $9$ independent degrees of freedom. The ten distance variables are thus overcomplete and obey one polynomial relation, namely the Gram constraint $G_5=0$.}
At the same time, the reduction~\eqref{eq:cayley-pentagon-reduction} introduces the leading Landau polynomial of the pentagon
\begin{equation}
\Delta_5 \equiv \det Y_5 = \Ycal_5[\begin{smallmatrix}0\\0\end{smallmatrix}]
\end{equation}
into the denominator.
At four points, we saw that the analogous Landau polynomial $\Delta_4$ factorized into four factors~\eqref{eq:Landaufact}, three of which were spurious and canceled by the numerator. 
At five points, $\Delta_5$ is an irreducible polynomial in the ten distance variables~\cite{DAndrea:2004}, so no factor-by-factor cancellation is possible. 
If $\Delta_5$ is spurious, it must cancel as a whole.\footnote{Similar spurious singularities have been studied for one-loop hexagons in $D>3$; see, e.g.,~\cite{Binoth:2002xh,Duplancic:2003tv,DelDuca:2011ne,Dixon:2011ng,Henn:2022ydo}.}

\subsection{Spurious Pole Cancellation}
\label{sec:pentagonresult}
We now show that the apparent pole at $\Delta_5=0$ is spurious and cancels upon restriction to the Gram locus~\eqref{eq:gramconstraintfivepoint}. 
After carrying out the reduction~\eqref{eq:cayley-pentagon-reduction}, the ordered pentagon integral takes the form
\begin{equation}
I_{12345} = \frac{\Ncal_5}{\Dphys\,\Delta_5}\,.
\label{eq:schematicfinalform}
\end{equation}
Here $\Dphys$ collects the physical singularities inherited from the five box subintegrals $I_i^{(5)}$.
It consists of the product of ten edge factors of $K_5$ coming from the triangle integrals at the bottom of the reduction, together with the five four-point Ptolemy factors $F_{++}^{(i)}$:
\begin{align}
\Dphys
&= \,\prod_{k_I \in E_5} k_I \prod_{i=1}^{5}F_{++}^{(i)} 
= k_1 k_2 k_3 k_4 k_5 k_{12} k_{23} k_{34} k_{123} k_{234} \nonumber \\
&\quad\times
\underbrace{\big(k_{23} k_{34} + k_3 k_{234} + k_2 k_4\big)}_{F^{(1)}_{++}}\,
\underbrace{\big(k_{34} k_{123} + k_4 k_{12} + k_3 k_5\big)}_{F^{(2)}_{++}}\,
\underbrace{\big(k_{123} k_{234} + k_1 k_4 + k_{23} k_5\big)}_{F^{(3)}_{++}} \nonumber \\
&\quad\times
\underbrace{\big(k_{12} k_{234} + k_1 k_{34} + k_2 k_5\big)}_{F^{(4)}_{++}}\,
\underbrace{\big(k_{12} k_{23} + k_2 k_{123} + k_1 k_3\big)}_{F^{(5)}_{++}}\,,
\label{eq:Dphys}
\end{align}
where $F_{++}^{(i)}$ is the Ptolemy singularity of the sub-box obtained by removing the $i$-th vertex from $K_5$, see Fig.~\ref{fig:K5}. 
We now show that the additional factor $\Delta_5$ is not a physical singularity and cancels against the numerator $\Ncal_5$ on the Gram locus.

If the ten distance variables $k_I\in E_5$ are temporarily treated as independent by relaxing the Gram constraint~\eqref{eq:gramconstraintfivepoint}, then $\Ncal_5$ is divisible by $\Delta_5$ only after restricting to the physical locus $G_5=0$.
Equivalently, there exist polynomials $h_1$ and $h_2$ in the distance variables such that
\begin{equation}
\Ncal_5=h_1\Delta_5+h_2G_5\,.
\label{eq:gram-decomposition}
\end{equation}
Mathematically, $\Ncal_5$ belongs to the ideal $\langle\Delta_5,\,G_5\rangle$ in the polynomial ring of the ten distance variables.
On the physical locus $G_5=0$,  $\Del_5$ cancels in the expression \eqref{eq:schematicfinalform}. The pentagon integral thus takes the form
\begin{equation}
I_{12345} = \frac{\Ncal_{\phys}}{\Dphys}\,,
\label{eq:pentagon-after-cancel}
\end{equation}
with the physical numerator $\Ncal_{\phys}=h_1$ given by a degree-13 polynomial in $k_I$.

We now describe the derivation of the physical representation. Since this analysis is somewhat technical, readers primarily interested in the final result may skip to~\S\ref{sec:physicalrep}.

\subsubsection{Gauge Fixing}

Constructing $\Ncal_{\phys}$ directly in terms of $k_I$ is possible in principle, but eliminating the Gram constraint leads to large intermediate expressions.
We instead solve the constraint from the outset by gauge-fixing the kinematics.\footnote{We thank Elia Mazzucchelli and Bernd Sturmfels for discussions on this point.} 
Using rotational invariance and homogeneity, we set
\begin{align}
\bk_1=(1,0,0)\,,\quad \bk_2=(y_1,y_2,y_3)\,,\quad \bk_3=(y_4,y_5,y_6)\,,\quad \bk_4=(y_7,y_8,y_9)\,,
\label{eq:componentmap}
\end{align}
with $\bk_5=-\sum_{i=1}^4\bk_i$. We denote evaluation on this parametrization by $|_{\rm gf}$ and restore the overall scale by homogeneity.
Since the momenta are physical three-vectors and obey momentum conservation, the Gram constraint is automatic in this parametrization.

A useful feature of this parametrization is that the gauge-fixed Landau polynomial $\Delta_5\big|_{\rm gf}$ becomes a perfect square. To see this, consider the dual vertices~\eqref{eq:dualvertdef}, which in this parametrization take the form
\begin{align}
\bx_1&=\boldsymbol 0\,,\quad \bx_2=(1,0,0)\,,\quad \bx_3=(1+y_1,\,y_2,\,y_3)\,,\quad \bx_4=(1+y_1+y_4,\,y_2+y_5,\,y_3+y_6)\,,\nonumber\\
\bx_5&=(1+y_1+y_4+y_7,\,y_2+y_5+y_8,\,y_3+y_6+y_9)\,.
\end{align}
It is convenient to introduce the embedding vector $P(\bx)=\big(x^2,\bx,1\big)^T$ and package the five vectors into the matrix $P_5=\big(P(\bx_1),\cdots\hskip -1pt,P(\bx_5)\big)$. The Cayley matrix then factorizes as 
\begin{equation}
Y_5=P_5^T\eta\hskip 1pt P_5\,,
\quad
\eta=
\begin{pmatrix}
0&0&0&0&-1\\[-2pt]
0&2&0&0&0\\[-2pt]
0&0&2&0&0\\[-2pt]
0&0&0&2&0\\[-2pt]
-1&0&0&0&0
\end{pmatrix},
\end{equation}
with its determinant given by
\begin{equation}
\Delta_5\big|_{\rm gf}
=
\det Y_5
=
-8\Omega_5^2\,,
\quad
\Omega_5
\equiv
\det P_5\,.
\label{eq:gaugefixed-delta5}
\end{equation}
Here $\Omega_5$ is a degree-five polynomial whose vanishing is equivalent to the five dual vertices lying on a common sphere.\footnote{The condition $\det P_5=0$ implies that there exists a nonzero vector $(a,\bm b,c)$ such that $(a,\bm b,c)\cdot P_5=0$ or  $a|\bx_i|^2+\bm b\cdot\bx_i+c=0$ for every $i$. Completing the square gives the equation of a common sphere.} It is an oriented quantity and changes sign under reflections, whereas $\Delta_5$ depends only on distances and is reflection invariant. This gives a geometric interpretation of why $\Omega_5$ appears through its square.

The reduction coefficients in~\eqref{eq:cayley-pentagon-reduction} are the modified Cayley-minor ratios
\begin{equation}
c_i=\frac{\Ycal_5[\begin{smallmatrix}0\\ i\end{smallmatrix}]}{\Ycal_5[\begin{smallmatrix}0\\ 0\end{smallmatrix}]}\,,
\label{eq:pentagon-reduction-coeffs}
\end{equation}
for $i=1,\cdots\hskip -1pt,5$. After gauge fixing, the denominator factorizes as $-8\Omega_5^2$, while each numerator contains a single factor of $\Omega_5$. Hence, one factor of $\Omega_5$ cancels, leaving
\begin{equation}
c_i\big|_{\rm gf}=\frac{\rho_i(y)}{\Omega_5}\,,
\label{eq:gf-simple-coeff}
\end{equation}
where $\rho_i(y)$ is a polynomial in the component variables. 
The remaining factor $\Omega_5$ does not cancel within each coefficient separately, but only after summing the five sub-box contributions. 
After substituting the explicit results for the box integrals, the gauge-fixed pentagon takes the form
\begin{equation}
I_{12345}\big|_{\rm gf}=\sum_{i=1}^{5}\frac{\rho_i(y)}{\Omega_5}\,I_i^{(5)}\big|_{\rm gf}=\frac{\Ncal_{\rm gf}}{\Omega_5\Dphys}\,.
\label{eq:gf-simple-pole-form}
\end{equation}
The cancellation of the spurious pole is thus equivalent to showing that $\Ncal_{\rm gf}$ is divisible by~$\Omega_5$.

To establish this divisibility, let us define
\begin{equation}
    \kappa_I(y) = k_I^2\big|_{\rm gf}\quad \Leftrightarrow\quad k_I\big|_{\rm gf} = \sqrt{\kappa_I(y)}\,.
\end{equation}
The even and odd powers of each $k_I$ then take the form
\begin{equation}
    k_I^n\big|_{\rm gf} =\begin{cases}
         \kappa_I(y)^{n/2}  & n\ \text{even}\,,\\
\kappa_I(y)^{(n-1)/2}\sqrt{\kappa_I(y)}  & n \ \text{odd}\,.\\
    \end{cases}
\end{equation}
The powers of $\kappa_I(y)$ can be absorbed into coefficients polynomial in $y$, whereas the remaining $\sqrt{\kappa_I(y)}$ is identified with the corresponding distance $k_I$. 
The products of the remaining linear $k_I$ factors are naturally associated with cycles of $K_5$. For an {\it Eulerian cycle}\footnote{An Eulerian cycle is a closed path in a graph that traverses each edge exactly once. This differs from a {\it simple cycle} which restricts each vertex to be visited only once.} $\gamma=(i_1i_2\cdots i_r)$ in $K_5$, we define the {\it cycle monomial} $X_\gamma$  as 
\begin{equation}
X_\gamma \equiv X_{i_1i_2\cdots i_r}
=\prod_{a=1}^{r}x_{i_a i_{a+1}}\,,
\end{equation}
with $i_{r+1}\equiv i_1$, and $x_{ij}$ the dual distance assigned to the edge $(ij)$ as usual.
For example, $X_{123}
=x_{12}x_{23}x_{31} = k_1k_2k_{12}$ and $X_{12345}
=x_{12}x_{23}x_{34}x_{45}x_{51}
=k_1k_2k_3k_4k_5$. 
We may thus expand the numerator $\Ncal_{\rm gf}$ as
\begin{equation}
\Ncal_{\rm gf}
=
\sum_\gamma r_\gamma(y)\,X_\gamma\,,
\label{eq:norm-monomial-expansion}
\end{equation}
where each coefficient $r_\gamma(y)$ is a polynomial in $y$. 

Since the distinct cycle monomials in this expansion are independent, the spurious pole must cancel separately in the coefficient of each $X_\gamma$. This means that every $r_\gamma$ is proportional to $\Omega_5$:
\begin{equation}
r_\gamma(y)=\Omega_5\hskip 1pt h_\gamma(y)\,,
\label{eq:gf-coeff-div}
\end{equation}
with $h_\gamma(y)$ polynomial. It follows that
\begin{equation}
I_{12345}\big|_{\rm gf}=\frac{\Ncal_{\phys}|_{\rm gf}}{\Dphys}\,,\quad\text{with}\quad \Ncal_{\phys}\big|_{\rm gf}=\sum_\gamma h_\gamma(y)\hskip 1pt X_\gamma\,.
\label{eq:gf-after-cancel}
\end{equation}
Although the intermediate polynomials are large, divisibility by
$\Omega_5$ can be verified explicitly for the coefficient of every $X_\gamma$.

\subsubsection{Reconstruction}

Finally, we undo the gauge fixing and reconstruct each $h_\gamma$ as a polynomial in the squared distances. Let $n_\gamma$ denote its homogeneous degree in the variables $k_I^2$. We start with the ansatz
\begin{equation}
h_\gamma
= \sum_{\beta_I\geq0}
b_{\gamma,\boldsymbol{\beta}}
\prod_{I}
k_I^{2\beta_I}\,,\quad\text{with}\quad |\boldsymbol{\beta}| = \sum_{I}\beta_I=n_\gamma\,,
\label{eq:ansatzorbit}
\end{equation}
where $\beta_I$ are non-negative integers and the sum is restricted by the stabilizer symmetry of~$\gamma$ and $I$ labels the elements ${k_I \in E_5}$. 
The coefficients $b_{\gamma,\boldsymbol{\beta}}$ are then determined by evaluating the ansatz at generic values of the component variables~$y$ and matching it to the known gauge-fixed polynomial $h_\gamma(y)$. 
Each sample point gives one linear equation, and using slightly more equations than unknowns avoids accidental rank deficiency.

We solve the resulting linear system over a finite field to avoid the large intermediate fractions produced by Gaussian elimination. For example, working modulo the prime $p=32003$ determines the coefficients in $\mathbb F_p$. 
The resulting modular coefficients have small integer lifts, which we verify at additional kinematic points. In practice, only nine or ten coefficients are nonzero among the several thousand monomials allowed by the initial ansatz. 
Repeating this reconstruction for every independent $\gamma$ determines the physical numerator $\Ncal_{\rm phys}$ in terms of the distance variables. 

\newpage

\subsection{Physical Representation}\label{sec:physicalrep}

The method described above reconstructs the pentagon integral in a form without any spurious singularities.
Before presenting the final result, we first describe the symmetries that organize the answer.

\begin{figure}[t]
\centering
\includegraphics{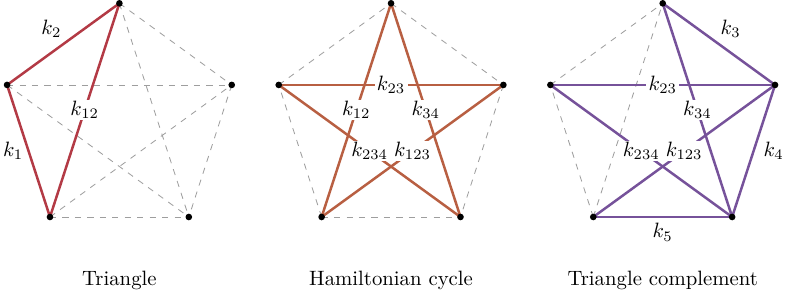}

\caption{Representative subgraphs in the orbit decomposition of the
five-point numerator $\Ncal_{\phys}$.}
\label{fig:pentagon-orbits}
\end{figure}

\paragraph{Graph symmetry.}

The cycle monomials $X_\gamma$ appearing in the numerator of $I_{12345}$ are naturally associated with subgraphs $\gamma\subset K_5$. 
Their symmetry is most transparent in dual variables. 
Recall from~\eqref{eq:ngondual} that the ordered pentagon may be written in dual space as
\begin{align}
I_{12345}
=\int_{x_0}
\frac{1}{x_{01}^2x_{02}^2x_{03}^2x_{04}^2x_{05}^2}\,.
\label{eq:pentagon-dual-symmetry}
\end{align}
Because the five propagators are identical, this representation is invariant under arbitrary permutations of the dual vertices corresponding to the action of $S_5$.
Consequently, the subgraphs contributing to the numerator are naturally organized into complete $S_5$ orbits of subgraphs of the dual graph $K_5$.

We find that the cycle monomials of $I_{12345}$ fall into three distinct $S_5$ orbit types:
\begin{itemize}
\item \textit{Triangle (3-cycle):} A triangle consists of three edges forming a closed loop, such as $\{k_1,k_2,k_{12}\}$ shown in Fig.~\ref{fig:pentagon-orbits}. 
Its stabilizer is $S_3\times S_2$, giving an orbit of size $|S_5|/|S_3\times S_2|=120/12=10$.

\item \textit{Hamiltonian cycle  (5-cycle):} A Hamiltonian cycle consists of five edges visiting every vertex exactly once before closing, such as the external cycle $\{k_1,k_2,k_3,k_4,k_5\}$ or the pentagram
$\{k_{12},k_{123},k_{23},k_{234},k_{34}\}$ shown in Fig.~\ref{fig:pentagon-orbits}. Its stabilizer is $D_5$, so the corresponding orbit contains $|S_5|/|D_5|=120/10=12$ elements.

\item \textit{Triangle complement (7-cycle):} The complement of a triangle consists of the seven edges of $K_5$ that remain after deleting the triangle. For example, removing $\{k_1,k_2,k_{12}\}$ leaves $\{k_{123},k_{23},k_{234},k_3,k_{34},k_4,k_5\}$ shown in Fig.~\ref{fig:pentagon-orbits}. 
It inherits the stabilizer $S_3\times S_2$ of the deleted triangle, again giving $10$ elements.
\end{itemize}
These three orbits consist of
$32$ cycle monomials in total, as summarized in Table~\ref{tab:orbit-classes}.

\newpage
\paragraph{Result.}
It is sufficient to specify one representative coefficient for each of the three $S_5$ orbits, since all remaining coefficients are obtained by the group action. We choose the following representatives:
\begin{itemize}
\item For the triangle with $X_{123} = k_1k_2k_{12}$, the coefficient is
\begin{align}
h_{123}
&= 2\ksq{4}(\ksq{3}\ksq{23}\ksq{34}\ksq{234} + \ksq{3}\ksq{34}\ksq{123}\ksq{5} + \ksq{23}\ksq{234}\ksq{123}\ksq{5})\label{eq:seed3}\\[3pt]
&\quad + \ksq{3}\ksq{23}\ksq{123}(\ksq{1}\ksq{34} + \ksq{2}\ksq{5} + \ksq{12}\ksq{234}) + \ksq{5}\ksq{34}\ksq{234}(\ksq{1}\ksq{3} + \ksq{2}\ksq{123} + \ksq{12}\ksq{23})\,. \nonumber
\end{align}
The three groups of terms are generated by cyclic relabeling of the three vertices of the triangle. They are the cyclic orbits of the seed monomials $k_3^2 k_{23}^2 k_{34}^2 k_{234}^2$, $k_1^2 k_{34}^2$, and $k_1^2 k_3^2$ with coefficients $2$, $1$, and $1$, respectively.

\item For the Hamiltonian cycle with $X_{12345} = k_1k_2k_3k_4k_5$, the coefficient is given by
\begin{align}
h_{12345}
=
\sum_{i=0}^{4} \sigma_5^{i}\bigl[\ksq{1}\ksq{12}\ksq{234}\ksq{34}
+ 2\ksq{123}\ksq{12}\ksq{234}\ksq{23}\bigr] \,,
\label{eq:seed-hamiltonian}
\end{align}
where $\sigma_5$ is the rotation generator
\begin{equation}
    \sigma_5 : \ \{k_1\hskip -1pt \to\hskip -1pt k_2\hskip -1pt \to\hskip -1pt k_3 \hskip -1pt\to\hskip -1pt k_4 \hskip -1pt\to\hskip -1pt k_5 \hskip -1pt\to\hskip -1pt k_1,\, k_{12} \hskip -1pt\to\hskip -1pt k_{23} \hskip -1pt\to\hskip -1pt k_{34} \hskip -1pt\to\hskip -1pt k_{123} \hskip -1pt\to\hskip -1pt k_{234} \hskip -1pt\to\hskip -1pt k_{12}\}\,.
\end{equation}
The weight-$1$ orbit pairs one external invariant with three diagonals, while the weight-$2$ orbit collects four of the five diagonals.

\begin{table}[t!]
\centering
\begin{tabular}{lcccc}
\toprule
Subgraph of $K_5$ & Count & $X_\gamma$ degree & $h_\gamma$ degree & Stabilizer \\
\midrule
 Triangle & $10$ & $3$ & $10$ & $S_3\times S_2$ \\
 Hamiltonian cycle & $12$ & $5$ & $8$ & $D_5$ \\
 Triangle complement & $10$ & $7$ & $6$ & $S_3\times S_2$ \\
\bottomrule
\end{tabular}
\caption{Orbit decomposition of the physical five-point numerator $\Ncal_{\phys}$.}
\label{tab:orbit-classes}
\end{table}

\item We take as the representative $X_{\overline{123}} = k_{123}k_{23}k_{234}k_3k_{34}k_4k_5$, which is the complement of the triangle $X_{123}$. Its coefficient is
\begin{align}
h_{\overline{123}}
&= 4\ksq{1}\ksq{2}\ksq{12} + 2\ksq{4}(\ksq{1}\ksq{2} + \ksq{2}\ksq{12} + \ksq{1}\ksq{12}) \label{eq:seed-complement}\\[3pt]
&\quad + \ksq{12}\ksq{34}\ksq{123} + \ksq{2}\ksq{3}\ksq{234} + \ksq{2}\ksq{23}\ksq{34} + \ksq{12}\ksq{3}\ksq{5} + \ksq{1}\ksq{123}\ksq{234} + \ksq{1}\ksq{23}\ksq{5}\, .\nonumber
\end{align}
The first two structures depend only on the three edges $k_1$, $k_2$, and $k_{12}$ of the deleted triangle: their product appears with coefficient $4$, while the symmetric sum of pairwise products is multiplied by $2\ksq{4}$. Both structures are invariant under the stabilizer $S_3\times S_2$. The remaining six monomials form a single $S_3\times S_2$ orbit with unit coefficient.
\end{itemize}
\newpage
Combining the three sectors, the ordered five-point function takes the compact form
\begin{align}
\boxed{I_{12345}
=
\frac{
\mathcal N_{\triangle}
+\mathcal N_{H}^{\phantom{1}}
+\mathcal N_{\bar\triangle}
}{
\mathcal D_{\rm phys}
}\,,}
\label{eq:finalfivepointordered}
\end{align}
where 
\begin{align}
\mathcal N_{\triangle}
&=
\sum_{\gamma\in \triangle}
h_{\gamma}X_{\gamma}\,,\quad
\mathcal N_H
=
\sum_{\gamma\in H}
h_{\gamma}X_{\gamma}\,,\quad 
\mathcal N_{\bar\triangle}=
\sum_{\gamma\in \bar \triangle}
h_{\gamma}X_{\gamma}\,,
\end{align}
denote the $S_5$ orbits of triangles $\triangle$, Hamiltonian cycles $H$, and triangle complements $\bar\triangle$, respectively, and their representative coefficients are given in \eqref{eq:seed3}, \eqref{eq:seed-hamiltonian} and \eqref{eq:seed-complement}. 
We have also numerically verified that our analytic result agrees with the integral \eqref{eq:5-gon}. 
The full expression is provided in an ancillary \textsc{Mathematica} notebook.

\section{Symmetries and Singularities}
\label{sec:symmandsing}
In this section, we discuss the conformal symmetry and singularity structure of general $n$-point ordered scalar correlators $I_{12\cdots n}$, using the four- and five-point ordered correlators derived in the previous sections as representative examples.
In \S\ref{sec:confsym}, we show that they obey the conformal Ward identities expected of a boundary scalar correlator. 
In \S\ref{sec:softandcollapsed}, we describe their soft and collapsed limits, together with the residues at the Ptolemy singularities.

\subsection{Conformal Symmetry}
\label{sec:confsym}

Conformal symmetry imposes nontrivial constraints on correlators. 
Dilatations determine their overall homogeneity, while special conformal transformations (SCTs) impose second-order differential constraints. 
For a scalar operator of dimension $\Delta$ in $D$ dimensions, the momentum-space SCT generator acting on the $i$-th leg is
\begin{equation}
\bm{K}_i
=
\bk_i\hskip 1pt\partial_i^2
-2(\bk_i\!\cdot\!\bm\partial_i)\bm\partial_i+2(\Delta-D)\bm\partial_i\,,
\quad\text{with}\quad
\bm\partial_i\equiv \frac{\partial}{\partial \bk_i}\, .
\label{eq:Kcartesianmain}
\end{equation}
An $n$-point function of a scalar operator ${\cal O}$ with scaling dimension $\Delta$ satisfies
\begin{equation}
    \sum_{i=1}^n \bm{K}_i \,\langle {\cal O}_1\cdots {\cal O}_n\rangle'=0\,,
\end{equation}
where ${\cal O}_i \equiv {\cal O}(\bk_i)$ and the prime indicates that the momentum-conserving delta function has been removed.

For correlators written in terms of distance variables $k_I$, these vector equations can be converted into scalar differential constraints. As derived in Appendix~\ref{app:distance-cwi}, they take the form
\begin{equation}
\mathbb K_i^{(n)}\langle {\cal O}_1\cdots {\cal O}_n\rangle'=0\,, \label{eq:generalSCToperatormain}
\end{equation}
for $i\in\{1,\cdots\hskip -1pt,n-1\}$, where
\begin{align}
\mathbb K_i^{(n)}&\equiv
\sum_{j=1}^{n-1}
G_{ij}
\bigl(D_j^{(n)}-D_n^{(n)}\bigr)\,,\\[-3pt]
D_i^{(n)}
&=
2\sum_{I\ni i}\sum_J
\frac{\bk_{I\cap J}\!\cdot\!\hat{\bk}_J}{k_I}\,
\partial_I\partial_J
-\sum_{I,J\ni i}
\hat{\bk}_I\!\cdot\!\hat{\bk}_J\,
\partial_I\partial_J
-\sum_{I\ni i}
\frac{2(\Delta-D)|I|+D-1}{k_I}\,\partial_I\, ,
\label{eq:full-Dcnmain}
\end{align}
with $G_{ij}=\bk_i\cdot\bk_j$ the Gram matrix and $\partial_I \equiv \partial_{k_I}$. Here $I$ and $J$ label the momentum invariants $k_I$ and $k_J$ associated with edges in $E_n$, while $|I|$ denotes the number of external momenta entering $\bk_I$. The momentum sum associated with their overlap is $\bk_{I\cap J}\equiv \sum_{i\in I\cap J}\bk_i$.
For example, if $\bk_{12}=\bk_1+\bk_2$ and
$\bk_{234}=\bk_2+\bk_3+\bk_4$, then $\bk_{12\cap 234}=\bk_2$.

We can verify these conformal Ward identities directly for the box and pentagon integrals. For four scalar operators of dimension $\Delta=1$, the correlator $I_{1234}$ in~\eqref{eq:4gonfinal} is homogeneous of degree $-5$. Its SCT constraints are given by
\begin{equation}
\label{eq:box-special-ward}
\mathbb{K}_i^{(4)} I_{1234}=0\,.
\end{equation}
In fact, the four-point case admits a stronger equivalent form. The three independent external momenta generically span $\mathbb{R}^3$, so their Gram matrix is invertible. The projected equations~\eqref{eq:box-special-ward} can thus be unprojected to give
\begin{equation}
\label{eq:box-pairwise-ward}
\bigl(D_i^{(4)}-D_4^{(4)}\bigr)I_{1234}=0\,.
\end{equation}
These equations are equivalent to those derived in~\cite{Arkani-Hamed:2018kmz}.
It is enough to verify one of these equations, since the remaining two follow by cyclic permutations of the momenta.

At five points, four external momenta cannot be linearly independent in three spatial dimensions, and their Gram determinant 
vanishes, $G_5=0$. Restricting to this physical kinematic locus and acting with the SCT generators on the expression for $I_{12345}$ in~\eqref{eq:finalfivepointordered}, we find
\begin{equation}
\label{eq:pent-special-Ward}
\left[\mathbb{K}_i^{(5)} I_{12345}\right]_{G_5=0}
=0\,,
\end{equation}
for $i=1,\cdots\hskip -1pt,4$.
Together with the required homogeneity, this confirms that the pentagon result satisfies the conformal Ward identities.

\subsection{Singularities}
\label{sec:softandcollapsed}

We now turn to the singularities of the ordered correlators $I_{12\cdots n}$. 
Their singularities occur on the loci where the physical denominator vanishes. 
As illustrated by the box in Section~\ref{sec:fourpoint} and the pentagon in Section~\ref{sec:fivepoint}, once the spurious leading Landau singularity has canceled, the scalar $n$-gon admits a physical representation of the form
\begin{equation}
I_{12\cdots n}
=
\frac{\Ncal_{\rm phys}}{\Dcal_{\rm phys}}\,,
\quad\text{with}\quad
\Dphys
=
\prod_{k_I\in E_n} k_I
\prod_i F_{++}^{(i)}\,.
\label{eq:genericdenom}
\end{equation}
The first product runs over the edge set $E_n$ of the complete graph $K_n$, as defined in~\eqref{eq:invariantsetEn}. 
The second product runs over its $\binom{n}{4}$ box subgraphs, each of which contributes a Ptolemy factor $F_{++}^{(i)}$ of the form encountered in Section~\ref{sec:fourpoint}.

Every factor in $\Dcal_{\rm phys}$ corresponds to a physical singularity of the correlator. The side lengths $k_i$ produce soft singularities, the diagonal lengths $k_{i\cdots j}$ produce collapsed singularities, and the factors $F_{++}^{(i)}$ produce Ptolemy singularities.

The soft and collapsed singularities have a common origin. Any two propagators in the $n$-gon integral~\eqref{eq:ngon-def} differ by a partial sum of consecutive external momenta. Consequently, every invariant $k_I\in E_n$ is associated with a pair of propagators: side lengths correspond to adjacent propagators, while diagonal lengths correspond to nonadjacent propagators. In the limit $k_I\to0$, the momenta carried by the corresponding pair become equal, and the two propagators coincide. The leading singular behavior is then governed by the resulting bubble subintegral, with the remaining $n-2$ propagators evaluated at the common singular point of the coincident pair.
We discuss these two cases in turn before computing the residues at the Ptolemy singularities.

\paragraph{Soft limit.}

Consider the soft limit $\bk_i\to0$. The two adjacent propagators whose momenta differ by $\bk_i$ then coincide. The leading singular region factorizes into the three-dimensional bubble integral
\begin{equation}
\int_{\ell}\frac{1}{\bl^2(\bl+\bk_i)^2}=\frac{1}{k_i}\,,
\label{eq:softbubble}
\end{equation}
multiplied by the remaining $n-2$ hard propagators evaluated at the common singular point of the coincident pair,
$\bl=-(\bk_1+\cdots+\bk_i)$.
These hard propagators carry partial sums beginning at the $(i+1)$-th leg. Hence, for $n\geq4$, with all indices understood modulo $n$, the leading soft behavior is
\begin{equation}
\lim_{k_i \to 0} I_{12\cdots n} = \frac{1}{k_i} \frac{1}{k_{i+1}^2 k_{i-1}^2} \prod_{a=i+2}^{i+n-3} \frac{1}{k_{(i+1)\cdots a}^2}\,,
\label{eq:softgen}
\end{equation}
where momentum conservation sets $k_{(i+1)\cdots(i+n-2)} = k_{i-1}$ in the soft limit. 

This factorization has a simple interpretation in the dual graph. As $k_i=x_{i,i+1}\to0$, the vertices $\bx_i$ and $\bx_{i+1}$ merge, contracting one side of the $n$-gon and reducing it to an $(n-1)$-gon. The factors in~\eqref{eq:softgen} are the inverse squared distances from the merged vertex to the remaining $n-2$ vertices. Two of these correspond to the neighboring sides, $k_{i+1}$ and $k_{i-1}$, while the remaining $n-4$ correspond to the diagonals $k_{(i+1)\cdots a}$.
For the box and the pentagon, this gives
\begin{align}
\lim_{k_1\to 0} I_{1234} = \frac{1}{k_1}  \frac{1}{k_2^2k_4^2}\,, \quad
\lim_{k_1\to 0} I_{12345} = \frac{1}{k_1}  \frac{1}{k_2^2k_{23}^2\,k_5^2}\,,
\label{eq:soft-limit}
\end{align}
with the remaining soft limits following by cyclic relabeling.

\begin{figure}[t]
\centering
\includegraphics{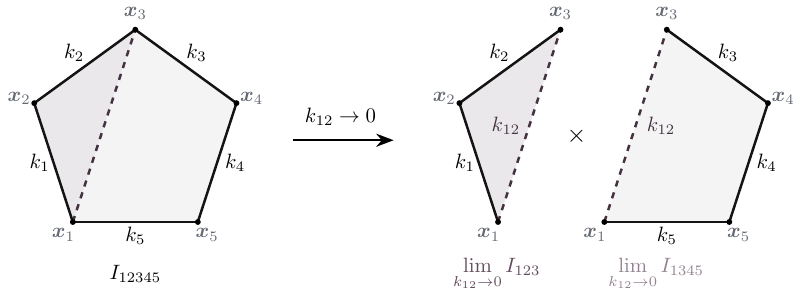}
\caption{
Factorization of the pentagon integral $I_{12345}$
in the collapsed channel $k_{12}\to0$.
}
\label{fig:collapsedlimitpentagon}
\end{figure}


\paragraph{Collapsed limit.}
A collapsed limit is one in which a partial sum vanishes, $k_{i\cdots j}=|\bk_i+\cdots+\bk_j|\to 0$.
The mechanism is basically the same as in the soft limit. The two propagators whose momenta differ by $\bk_i+\cdots+\bk_j$ coincide, and the resulting bubble subintegral produces a simple pole $1/k_{i\cdots j}$. Evaluating the hard propagators at their common pole $\bl = -(\bm{k}_1 + \cdots + \bm{k}_j)$ gives
\begin{equation}
\lim_{k_{i\cdots j} \to 0} I_{12\cdots n} = \frac{1}{k_{i\cdots j}} \prod_{a=i+1}^{j} \frac{1}{k_{a\cdots j}^2} \prod_{b=j+1}^{i-2} \frac{1}{k_{(j+1)\cdots b}^2}\,,
\label{eq:collgen}
\end{equation}
where the indices and the second product are understood cyclically modulo $n$.

This factorization is manifest in the dual graph. The diagonal $k_{i\cdots j}=x_{i,j+1}$ joins the vertices $\bx_i$ and $\bx_{j+1}$. As this diagonal shrinks, the two vertices merge and the original polygon separates into two smaller polygons that meet at the merged vertex. Writing
$m=j-i+1$,
one obtains an $(m+1)$-gon containing the legs $k_i,\cdots\hskip -1pt,k_j$ and an $(n-m+1)$-gon containing the remaining legs. In both polygons, the vanishing diagonal appears as an additional soft leg.

The two products in~\eqref{eq:collgen} are precisely the soft residues of these lower-point polygons. The collapsed singularity therefore factorizes as
\begin{equation}
\lim_{k_{i\cdots j} \to 0}\, I_{12\cdots n} = \frac{1}{k_{i \cdots j}} \Big[ \Res_{k_e = 0} I_{i\cdots j e} \Big] \Big[ \Res_{k_e = 0} I_{e (j+1)\cdots(i-1)} \Big]\,,
\label{eq:collres}
\end{equation}
where $e$ labels the additional leg carrying the exchanged momentum  $\bm{k}_e \equiv \bm{k}_i + \cdots + \bm{k}_j$. 
These collapsed limits correspond to OPE channels of the boundary correlator. The leading, angle-independent residue is the $J=0$ contribution to the momentum-space conformal partial wave expansion. From the bulk perspective, this is the factorization associated with scalar exchange. The angular dependence of the subleading collapsed expansion organizes the contributions from the infinite tower of higher-spin fields.\footnote{The partial wave coefficients for arbitrary spin-$J$ exchange were explicitly computed for the four-point function $I_4$ in Grassmannian space~\cite{De:2026shn}. We leave a direct conformal partial wave analysis in momentum space to future work.}

At four points, the collapsed limit $s\to0$ separates the dual box into the two triangles $I_{123}$ and $I_{134}$. Equation~\eqref{eq:collres} then gives
\begin{equation} \lim_{s \to 0} I_{1234} = \frac{1}{s} \Big[\Res_{s= 0} I_{123}\Big]  \Big[\Res_{s= 0} I_{134}\Big] = \frac{1}{s}  \frac{1}{k_2^2}  \frac{1}{k_3^2}\,. \label{eq:collapsed4pt} \end{equation}
At five points, the limit $k_{12}\to0$ separates the pentagon into the triangle $I_{123}$ and the sub-box $I_{1345}$, as shown in Fig.~\ref{fig:collapsedlimitpentagon}. Their soft limit gives
\begin{equation} 
\lim_{k_{12} \to 0} I_{12345} = \frac{1}{k_{12}} \Big[\Res_{k_{12}= 0} I_{123}\Big] \Big[\Res_{k_{12}= 0} I_{1345}\Big] = \frac{1}{k_{12}}  \frac{1}{k_2^2}  \frac{1}{k_3^2 k_5^2}\,. \label{eq:collapsed5pt} 
\end{equation}
The remaining collapsed channels follow by cyclic relabeling. 
More generally, the factorization~\eqref{eq:collres} holds at arbitrary multiplicity: the leading singularity on every collapsed pole is the product of the soft residues of two lower-point ordered correlators.

\begin{figure}[t]
\centering
\includegraphics{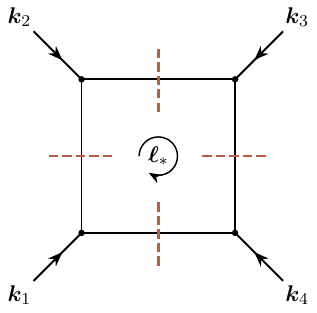}
\caption{The maximal four-line cut of the box in $D=3$.}
\end{figure}

\paragraph{Ptolemy limit.}

The remaining factors of $\Dphys$ are the Ptolemy singularities $F^{(i)}_{++}$, one for each four-point subgraph of the ordered correlator $I_{12 \cdots n}$.
Each Ptolemy factor $F^{(i)}_{++}$ is strictly positive in the Euclidean region and is reached only after analytic continuation.

At four points, there is a single Ptolemy factor, $F_{++}=st+k_1k_3+k_2k_4$. The ordered correlator $I_{1234}$ in~\eqref{eq:4gonfinal} becomes singular on the locus $F_{++}=0$. The origin of this singularity can be seen directly from the maximal cut of the box integral. Recall that
\begin{align}
I_{1234}
&=
\int_\ell
\frac{1}{\ell^2(\bl+\bk_1)^2(\bl+\bk_1+\bk_2)^2
(\bl+\bk_1+\bk_2+\bk_3)^2}\,.
\label{eq:box-def-int2}
\end{align}
Writing the four propagator denominators as $d_i=\bq_i^2$, the maximal cut requires
\begin{equation} 
d_1=d_2=d_3=d_4=0\,. 
\end{equation}
Since the loop momentum has only three components, these four conditions are over-constrained for generic external kinematics. Instead, three of the conditions determine the loop momentum, while the fourth imposes a constraint on the external data.

More explicitly, the differences $d_i-d_1$, for $i=2,3,4$, are linear in $\bl$. Setting them to zero fixes a unique solution $\bl_*$: \begin{equation} 2\,\bl_*\!\cdot\bk_1=-k_1^2\,, \quad 2\,\bl_*\!\cdot(\bk_1+\bk_2)=-s^2\,, \quad 2\,\bl_*\!\cdot(\bk_1+\bk_2+\bk_3)=-k_4^2\,. \label{eq:boxloopmomsol} \end{equation} The Jacobian associated with these three linear conditions is \begin{equation} J=8\det(\bk_1,\bk_2,\bk_3)\,, \quad J^2=-8\det\Ycal_4\,. \end{equation} On the solution~\eqref{eq:boxloopmomsol}, all four propagators take the same value. The remaining cut condition is therefore $\bl_*^2=0$. This condition depends only on the external kinematics and is related to the leading Landau polynomial by 
\begin{equation} 
\Delta_4=-\frac14\,J^2\bl_*^2\,. \label{eq:leadinglandauboxcutmom} \end{equation} 
For generic external momenta, $J\neq0$, and hence the final cut condition $\bl_*^2=0$ is equivalent to $\Delta_4=0$. The four-line maximal cut therefore exists only on the leading Landau locus. The Ptolemy condition $F_{++}=0$ selects the corresponding branch of this locus, on which the localized loop momentum becomes null.\footnote{In three-dimensional Euclidean space, a real null vector must vanish. The relevant on-shell configuration is therefore necessarily complex, in agreement with the strict positivity of $F_{++}$ in the physical Euclidean region.}

The residue on the Ptolemy pole is the leading singularity obtained by fully localizing the box integrand on this on-shell configuration. 
It can be computed algebraically by solving the equation $F_{++}=0$ for any one of the four external energies. 
Solving for $k_i$ gives \begin{equation} k_i^* = -\frac{k_{i+1}k_{i-1}+st}{k_{i+2}}\,, \end{equation} where the indices are understood modulo four. 
Since $\partial_{k_i}F_{++}=k_{i+2}$, the residue in the variable $k_i$ is \begin{equation} 
\Res_{k_i=k_i^*} I_{1234} =-\frac{\Ncal_4|_{k_i=k_i^*}}{st(k_{i+1} k_{i-1}+st) \prod_{j \neq i} k_j}\,,
\end{equation}
where the product in the denominator runs over the three unresolved energies. 
Resolving the pole at an odd and at an even vertex, for example at energies $k_1$ and $k_2$, gives
\begin{align} 
  \Res_{k_1 = k_1^*}\,I_{1234}
  &= \frac{2}{k_3^2 (k_2 k_4+st)}\,
  \bigl(\hat{\bk}_2\cdot\hat{\boldsymbol{t}}-\hat{\bk}_4\cdot\hat{\boldsymbol{s}} \bigr)\,, \label{eq:ptolemyresk1}\\
  \Res_{k_2 = k_2^*}\,I_{1234}
  &= -\frac{2}{k_4^2 (k_1 k_3+st)}\,
  \bigl(\hat{\bk}_1\cdot\hat{\boldsymbol{t}} + \hat{\bk}_3\cdot\hat{\boldsymbol{s}}\bigr)\,,
  \label{eq:ptolemyresk2}
\end{align}
where 
\begin{align}\label{eq:unitvectors4pt}
  \hat{\bk}_i = \frac{\bk_i}{k_i}\,,\quad \hat{\boldsymbol{s}} = \frac{\bk_1+\bk_2}{s}\,,\quad
  \hat{\boldsymbol{t}} = \frac{\bk_2+\bk_3}{t}\,.
\end{align}
The residues in the four energies $k_{i}$ form a single $\sigma_4$ orbit.
The $\sigma_4$ generator~\eqref{eq:4pt-rotation} acts on the unit vectors~\eqref{eq:unitvectors4pt} as $\hat{\bk}_i\mapsto\hat{\bk}_{i+1}$, $\hat{\boldsymbol{s}}\mapsto\hat{\boldsymbol{t}}$ and
$\hat{\boldsymbol{t}}\mapsto-\hat{\boldsymbol{s}}$, and maps~\eqref{eq:ptolemyresk1} to~\eqref{eq:ptolemyresk2}.
The residues at $k_3$ and $k_4$ follow by applying $\sigma_4$ successively.
In each case the prefactor is the reciprocal of the square of the norm diagonally opposite the resolved vertex, divided by the part of $F_{++}$ that does not contain the resolved energy.
The remaining factor contracts the two adjacent external unit vectors with the two diagonal directions, as a difference at the odd vertices and as a sum at the even ones.

At five points, the same singularities reappear, but now there are five Ptolemy factors, as is manifest from the physical denominator $\Dphys$ in~\eqref{eq:Dphys}. Each factor in $\prod_{i=1}^5 F_{++}^{(i)}$ is associated with the box subgraph obtained by deleting vertex $i$ from $K_5$. The corresponding pole lies on the Ptolemy locus of that sub-box, and its residue is determined by the same maximal-cut analysis as at four points. The four propagators of the sub-box are localized on the cut solution, while the remaining propagator is evaluated on that configuration. Ptolemy singularities are therefore not unique to the four-point correlator. At higher multiplicity, they are inherited from the box subgraphs that arise in the reduction of the $n$-gon, with each box subgraph contributing its own Ptolemy pole to the ordered correlator~$I_{12\cdots n}$.

\section{Conclusions and Outlook}
\label{sec:conclusions}

Cosmological correlators on the future boundary of de Sitter space encode the late-time structure of quantum fields in inflationary backgrounds. Understanding their analytic structure from symmetry, locality, and unitarity is the central goal of the cosmological bootstrap program. So far, this approach has been most powerful for perturbative bulk theories with a finite number of weakly coupled fields, where boundary correlators are organized by their energy singularities and factorization limits. A natural question is how this structure gets modified when the bulk spectrum contains an infinite tower of higher-spin fields that must be treated collectively. In ordinary local bulk theories, the total-energy singularity at $E= 0$ is a universal feature of boundary correlators, and its leading behavior reproduces the high-energy limit of the corresponding flat-space scattering amplitude. It was argued in~\cite{Arkani-Hamed:2018kmz} that a stringy ultraviolet completion should instead make the boundary correlator regular in this limit, replacing the total-energy singularity with a different analytic structure.

The higher-spin/vector-model duality in $\ds_4$ provides a controlled setting in which to address this question. The perturbative bulk spectrum consists of an infinite tower of massless even-spin fields, while the boundary correlators can be computed exactly using the dual vector model. In the $Q$-model formulation~\cite{Anninos:2017eib}, the connected scalar $n$-point function is generated by cyclic Wick contractions and reduces to one-loop $n$-gon integrals in three Euclidean dimensions. Using the reduction mechanism described in Section~\ref{sec:qformalism}, we have shown that these momentum-space boundary correlators are rational functions of the external momenta at all multiplicities and obtained their physical representations for~$n\leq 5$.

The four-point scalar correlator~\eqref{eq:fullfourpointfunction} already displays several remarkable features. 
Notably, it is devoid of any total- and partial-energy singularities, and its only nontrivial singular locus is the Ptolemy singularity $st+ k_1 k_3 + k_2 k_4 = 0$, which is the unique physical Landau singularity of the one-loop box in three dimensions. 
At five points, we showed that the leading Landau polynomial $\Del_5$ is in fact a spurious pole of the correlator. 
Moreover, the resulting numerator is not a generic polynomial but instead organizes into three orbits of subgraphs of the complete graph $K_5$, giving the five-point answer an elegant graph-theoretic structure shown in \eqref{eq:finalfivepointordered}. 
This provides another example of a broader theme in modern approaches to cosmological observables, in which complicated answers can often be reorganized in terms of hidden combinatorial or geometric data, as exemplified by cosmological polytopes~\cite{Arkani-Hamed:2017fdk,Benincasa:2019vqr,Benincasa:2024leu}, cosmohedra~\cite{Arkani-Hamed:2024jbp,Ardila-Mantilla:2026cbo,Glew:2025otn}, kinematic flow~\cite{Arkani-Hamed:2023kig,Arkani-Hamed:2023bsv,Hang:2024xas, Baumann:2024mvm,Baumann:2025qjx,Glew:2025ypb,Baumann:2026atn}, cluster algebra~\cite{Mazloumi:2025pmx,Capuano:2025myy,Capuano:2026pgq,Paranjape:2026htn,Ferro:2026oph}, and hypergeometric structures~\cite{Liu:2024str,Grafe:2026avi}. In the present case, this organization emerges from the exact resummation of an infinite higher-spin tower, and points toward a graph-theoretic description of holographic correlators.

Taken together, these results suggest that the simplicity of the final correlators reflects a deeper organizing principle rather than an accident of low multiplicity. Clarifying its origin and exploring its broader implications is a natural direction for future work.

\begin{itemize}[leftmargin=*]
 
\item \textbf{Combinatorial bootstrap.} The most immediate next step is to extend the computation to six points and beyond. 
By iterating the reduction formula~\eqref{eq:complete-D+1}, any $n$-gon with $n\geq 6$ can be written as a sum of boxes. The central challenge is not to simply evaluate the integral, but to make the cancellation of spurious poles manifest and derive a physical representation of the correlator. 
More ambitiously, the five-point result suggests a purely combinatorial construction in which conformal invariance and the expected soft and collapsed limits emerge as outputs, rather than being imposed as fundamental inputs. The hexagon provides the first nontrivial test of whether this structure extends to arbitrary multiplicity. Such a construction could also provide a useful framework for computing higher-point CFT correlators in momentum space more generally~\cite{Bzowski:2019kwd,Coriano:2019nkw,Jain:2020rmw,Jain:2020puw,Jain:2022ajd}.

\item \textbf{Graviton correlators.} The spin-2 bilinear operator dual to the bulk graviton is given by the convolution~\cite{Anninos:2017eib}
\begin{equation}
 \int\!\frac{\text{d}^3p}{(2\pi)^3}  :\!Q^\alpha_\bp Q^\alpha_{\bk-\bp}\!: \left[(\bm z \cdot \bp)^2 + (\bm z \cdot (\bk-\bp))^2 + 6 (\bm z \cdot \bp) \bm z \cdot (\bp-\bk) \right],
\label{eq:B2-momentum}
\end{equation}
with a null vector $\bm z$.
As in the scalar case~\eqref{eq:B0-momentum}, its correlators reduce to $n$-gon integrals, but now with nontrivial tensor numerators. A direct tensor reduction is cumbersome because the individual tensor integrals diverge even when their sum is finite~\cite{Anninos:2019nib}. A more efficient approach would be to generate the spin-two correlators by acting on the scalar answers with cosmological weight-shifting operators~\cite{Baumann:2019oyu,Baumann:2020dch, Lee:2022fgr,Lee:2023qqx}. This would isolate the universal scalar integral from the spin-dependent numerator and provide a systematic route to graviton correlators at higher multiplicity.

\item \textbf{Scattering amplitudes in $D=3$.}
The same formulas also give compact expressions for scalar one-loop $n$-gon amplitudes in three-dimensional Lorentzian kinematics, with massless internal propagators and external legs of masses $m_i$. Starting from the Euclidean result for $I_{12\cdots n}$, the continuation to scattering kinematics is obtained through
\begin{equation}
k_i^2 \to m_i^2\,, \quad
k_{i\cdots j}^2 \to s_{i\cdots j}^{\rm flat}+i0\,,
\end{equation}
where $s_{i\cdots j}^{\rm flat}=(\bp_i+\cdots+\bp_j)^2$ are the usual Lorentzian Mandelstam invariants. At four points, the result~\eqref{eq:4gonfinal} maps to the three-dimensional four-mass box integral and agrees numerically with the expression in~\cite{Lipstein:2012kd}. Expressed in distance variables, the answer takes a compact rational form.
Upon continuation to Lorentzian kinematics, the resulting one-loop amplitudes thus have only square-root branch points, rather than the polylogs familiar from other representations of the four-mass box.

\item \textbf{Other holographic models.} 
The boundary $Q$-model~\cite{Anninos:2017eib} is one realization of a vector model dual to a higher-spin theory in $\ds_4$, building on the $\mathrm{Sp}(N)$ proposal of~\cite{Anninos:2011ui}. 
We expect the framework applied here to extend to a broader family of holographic setups.
These include the type-B Vasiliev theory dual to a free-fermion vector model~\cite{Sezgin:2003pt,Leigh:2003gk}, parity-broken Vasiliev theories with Chern-Simons matter
duals~\cite{Aharony:2011jz,Maldacena:2012sf,Chang:2012kt}, the (A)dS counterparts of these dualities~\cite{Klebanov:2002ja,Sezgin:2002rt,Giombi:2009wh,Chang:2013afa,Anninos:2014hia}, and their supersymmetric extensions~\cite{Hertog:2017ymy}. Another useful class of examples is provided by the free $\Box^k$ scalar theories~\cite{Brust:2016gjy,Brust:2016zns}, whose partially conserved currents are dual to partially massless fields in dS space. These models may offer useful clues to how boundary correlators are deformed by the presence of massive bulk fields.
It would also be interesting to study higher-point correlators in broader holographic approaches to cosmology~\cite{McFadden:2009fg,McFadden:2010na,Bzowski:2011ab,Bzowski:2013sza}.

\item \textbf{A Grassmannian formulation.}
In~\cite{De:2026shn}, we showed that the four-point scalar correlator~\eqref{eq:fullfourpointfunction} can be written in a remarkably simple way as an integral over the orthogonal Grassmannian $\mathrm{OGr}(4,8)$~\cite{Arundine:2026fbr}. In suitable Grassmannian variables, the full crossing-symmetric answer reduces to the form $(S^2+T^2+U^2)/(STU)$. This expression mirrors the field-theory $(\alpha'\to0)$ limit of the Veneziano amplitude, even though the higher-spin theory should instead be viewed as a tensionless $(\alpha'\to\infty)$ limit with an infinite tower of massless fields. It remains to understand whether this resemblance is accidental or points to a deeper physical principle, and whether the five-point correlator obtained here admits a similarly simple Grassmannian formulation.
\end{itemize}

More broadly, our results point toward a description of de Sitter higher-spin theory in which the full spectrum is treated as a single object, rather than as an infinite collection of individual exchanges. In this picture, the boundary correlator would be primary, while the bulk interpretation would emerge from its analytic and geometric structure. Identifying the principles that govern this description may offer useful clues toward a nonperturbative understanding of quantum gravity in de Sitter space.

\paragraph{Acknowledgements.}

We thank Dionysios Anninos, Mattia Arundine, Vivek Chakrabhavi, Jordan Cotler, Frederik Denef, Harry Goodhew, Kurt Hinterbichler, Yu-tin Huang, Javier Huenupi, Elia Mazzucchelli, Shani Meynet, Sebastian Mizera, Evgeny Skvortsov, Bernd Sturmfels, and Zimo Sun for insightful discussions. 
HL~thanks the participants of the MIAPbP workshop ``Primordial Cosmology: Novel Perspectives from Scattering Amplitudes, Holography and the Bootstrap'' for stimulating discussions and the Korea Institute for Advanced Study for its hospitality while the work was completed.
The work of SD~was completed at the Aspen Center for Physics, which is supported by National Science Foundation grant PHY-2210452.
SD~is supported by funds provided by the Center for Particle Cosmology at the University
of Pennsylvania. 
HL~is supported by the U.S.~Department of Energy under grant DE-SC0013528.

\newpage
\appendix
\section{The Melrose Reduction}
\label{app:mel-reductionderivation}

In this appendix, we derive the reduction formulas presented in Section~\ref{sec:qformalism}, following the construction of~\cite{Melrose:1965}.  
We treat the two cases separately:
\begin{itemize}
    \item For $n > D+1$, the rank deficiency of the Cayley--Menger matrix $\text{CM}_n$ of~\eqref{eq:CM} yields a \emph{linear} identity among propagator denominators. This removes one propagator at a time and reduces an $n$-gon to $(n-1)$-gons.
\item For $n = D+1$, the linear identity is exhausted. A remaining {\it quadratic} Gram identity then reduces the critical $(D+1)$-gon to $D$-gons.
\end{itemize}
For the three-dimensional correlators in this work, this means that pentagons and higher $n$-gons can be reduced to boxes via linear identities, and subsequently to triangles using the quadratic identity.
We first derive the single-step reduction formulas~\eqref{eq:single-step} and~\eqref{eq:D+1-to-D} in \S\ref{sec:linear}, and then present the full iterated reduction in \S\ref{sec:iteration}.

\subsection{Single-Step Reductions}
\paragraph{Linear reduction ($n > D+1$)}
\label{sec:linear}
When $n > D + 1$, the Cayley--Menger matrix~$\text{CM}_n$ has order $n + 2 > D + 3$ but rank only $D + 2$.  
Every minor of order $\geq D + 3$ therefore vanishes identically.  
The strategy is to construct a particular vanishing minor of order $D + 3$, expand it along the propagator column, and read off a linear identity among the propagators~$d_i$. 
Deleting the propagator row $(n{+}1)$ at the outset is what makes the identity linear.
The entries $-d_i$ then survive solely in the propagator column, as a result of which each term of the expansion carries exactly one factor of $d$.
Setting $m=n-D-2$, we make the following deletions in the $\text{CM}_n$ matrix:
\begin{itemize}
  \item among rows: the propagator row~$(n{+}1)$ together with $m$~bulk rows $\{a_1, \cdots\hskip -1pt, a_m\} \subset \{1, \cdots\hskip -1pt, n\}$,
  \item among columns: any $m{+}1$ columns
    $\{b_1, \cdots\hskip -1pt, b_{m+1}\} \subset \{0, \cdots\hskip -1pt, n\}$, excluding the propagator column~$n{+}1$.
\end{itemize}
The surviving submatrix has order $(n + 2) - (m+1) = D + 3$ and retains the propagator column. 
Because it is a minor of a rank-$(D{+}2)$ matrix, its determinant vanishes.

We now perform a Laplace expansion of this vanishing minor along the propagator column~$(n{+}1)$. 
Each term of the expansion deletes column $(n{+}1)$ as well, so the accompanying minors lie entirely inside the modified Cayley block $\Ycal_n$ of~\eqref{eq:modified-cayley-general} and hence carry no label $(n{+}1)$. 
The expansion index instead runs over the surviving rows $\{0,l_1,\cdots,l_{D+2}\}$, with the border row contributing unity and each bulk row $i$ contributing $-d_i$.
The expansion then reads 
\begin{equation}\label{eq:raw_expansion}
  \Ycal_n[\begin{smallmatrix} a_1 \ \cdots \ a_m & 0 \\ b_1\ \cdots \ b_m & b_{m+1}\end{smallmatrix}]
  \;-\; \sum_{i=1}^{n} \Ycal_n[\begin{smallmatrix} a_1 \ \cdots \ a_m & i \\ b_1 \ \cdots \ b_m & b_{m+1}\end{smallmatrix}]\, d_i = 0\,.
\end{equation}
These minors vanish when $i=a_1,\dots,a_m$, since a deleted row index is then repeated. 
Thus \eqref{eq:raw_expansion} can be written in the form 
\begin{equation}\label{eq:clean_expansion}
    \Ycal_n[\begin{smallmatrix} a_1 \ \cdots \ a_m & 0 \\ b_1\ \cdots \ b_m & b_{m+1} \end{smallmatrix}]
  \;=\; \sum_{r=1}^{D+2}   \Ycal_n[\begin{smallmatrix} a_1 \ \cdots \ a_m & l_r \\ b_1\ \cdots \ b_m & b_{m+1} \end{smallmatrix}] \, d_{l_r} \,,
\end{equation}
where $\{l_{1}, \cdots\hskip -1pt, l_{D+2}\} = \{1, \cdots\hskip -1pt, n\}\setminus\{a_1, \cdots\hskip -1pt, a_m\}$ denotes the set of surviving bulk indices. 
This is a linear, off-shell identity valid for all~$\bl$, where the left side and each minor on the right depend only on the external kinematics, while the~$d_i$ carry all the loop-momentum dependence.

We now divide both sides of~\eqref{eq:clean_expansion} by $\prod_{j=1}^{n} d_j$ and integrate over~$\bl$.
Each $d_{l_r}$ in the numerator cancels one
propagator, producing the $(n{-}1)$-gon $I^{(n)}_{l_r}$ with the propagator $d_{l_r}$ removed
\begin{equation}\label{eq:pre218}
   \Ycal_n[\begin{smallmatrix}  a_1 \ \cdots \ a_m & 0 \\ b_1\ \cdots \ b_m & b_{m+1}\end{smallmatrix}] \,
  I_{1 2\cdots n}
  \;=\; \sum_{r=1}^{D+2}
   \Ycal_n[\begin{smallmatrix} a_1 \ \cdots \ a_m & l_r \\ b_1\ \cdots \ b_m & b_{m+1}\end{smallmatrix}] \, I^{(n)}_{l_r}.
\end{equation}
While the column labels $\{b_1,\cdots\hskip -1pt,b_{m+1}\}$ are freely chosen, it can be shown that the ratio of minors
$ \Ycal_n[\begin{smallmatrix} a_1 \ \cdots \ a_m & l_r \\ b_1\ \cdots \ b_m & b_{m+1}\end{smallmatrix}]\big/ \Ycal_n[\begin{smallmatrix} a_1 \ \cdots \ a_m & 0 \\ b_1\ \cdots \ b_m & b_{m+1}\end{smallmatrix}]$ is independent of this choice \cite{Melrose:1965}.  
We may therefore set $b_1 = a_1, \cdots\hskip -1pt, b_m = a_m, b_{m+1} = 0$, after which the minors reduce to those of the modified Cayley matrix $\Ycal_{(l_1\cdots l_{D+2})}$ formed from the $D+2$ surviving propagators alone.  
Exchanging the two index rows, which is permitted because $\Ycal_{(l_1\cdots l_{D+2})}$ is symmetric, gives the reduction formula \eqref{eq:single-step} quoted in the main text
\begin{equation}
I_{12\cdots n} = \sum_{r=1}^{D+2} \frac{\Ycal_{(l_1\cdots l_{D+2})}[\begin{smallmatrix}0\\ l_r\end{smallmatrix}]}{\Ycal_{(l_1\cdots l_{D+2})}[\begin{smallmatrix}0\\ 0\end{smallmatrix}]}\,I^{(n)}_{l_r}~.
\label{eq:rederivedredux}
\end{equation}
The $D+2$ propagators $\{l_1,\cdots\hskip -1pt,l_{D+2}\}$ are
chosen freely, with different choices giving algebraically distinct but numerically equivalent reductions.

\paragraph{Quadratic reduction ($n = D+1$)}
\label{sec:quadratic}

For $n = D + 1$, the Cayley--Menger matrix~$\text{CM}_n$ yields a \emph{quadratic} relation among the propagators $d_i$,  which is not directly useful as a starting point.  
We instead follow~\cite{Melrose:1965} in returning to the Gram determinant of the internal momenta $\bq_i = \bl + \bP_i$ where $\bP_i = \sum_{a=1}^{i-1} \bk_a$ denotes the sum of the first $i-1$ external momenta.
Since $\bq_1 = \bl$ and $\bq_i = \bl + \bP_i$ for $i \geq 2$, the Gram determinant of the internal momenta $G(\boldsymbol{q}_1, \cdots\hskip -1pt, \boldsymbol{q}_n)$ can be expressed as
\begin{equation}\label{eq:gram_exposed}
G(\bq_1, \dots,\bq_n) = G(\bl, \bP_2, \cdots\hskip -1pt, \bP_{n})=\det
  \begin{pmatrix}
    \bl^2            & \bl\cdot \bP_2        & \cdots & \bl\cdot \bP_{n} \\[2pt]
    \bl\cdot \bP_2     & \bP_2^2                 & \cdots & \bP_2\cdot \bP_{n}  \\
    \vdots            &      \vdots                   & \ddots & \vdots            \\
    \bl\cdot \bP_{n}   & \bP_2\cdot \bP_{n}      & \cdots & \bP_{n}^2
  \end{pmatrix}
  \;=\; 0\,.
\end{equation}
The determinant vanishes because the $n=D+1$ vectors $\bl, \bP_2, \cdots \bP_n$ are linearly dependent in $D$ dimensions.
Its only dependence on the loop momentum enters through the first row and column, where $\bl^2 = d_1$ and $2\bl\cdot \bP_i = d_i - d_1 - \bP_i^2$. 
The determinant is therefore quadratic in the propagator denominators. 
The entry $\bl^2$ contributes linearly through $d_1$, while each $\bl\cdot\bP_i$ in the first row multiplies a cofactor that is itself linear in $\bl$, and these cross terms are then quadratic in the $d_i$.

To organize the cofactor expansion of~\eqref{eq:gram_exposed}, one can introduce an order-$n$ auxiliary matrix $G^\mu$ whose first row is promoted to carry the partial sums $\bP_i$ of external momenta such that
\begin{equation}\label{eq:vecgram}
  G^\mu =
  \det
  \begin{pmatrix}
    0                 & P_2^\mu             & P_3^\mu        & \cdots & P_{n}^\mu      \\[2pt]
    1                 & \bP_2^2               & \bP_2\cdot \bP_3   & \cdots & \bP_2\cdot \bP_{n} \\
    1                 & \bP_2\cdot \bP_3        & \bP_3^2          & \cdots & \bP_3\cdot \bP_{n} \\
    \vdots            &   \vdots                  &     \vdots           & \ddots & \vdots           \\
    1                 & \bP_2\cdot \bP_{n}      & \bP_3\cdot \bP_{n}  & \cdots & \bP_{n}^2
  \end{pmatrix}.
\end{equation}
Its rows are labeled $\{1, 2, \cdots\hskip -1pt, n\}$, with the first row carrying the vector entries and the remaining rows forming the bordered external Gram matrix.
The $(1,1)$ minor of $G^\mu$ is the scalar Gram determinant of the $n-1 = D$ independent external momenta 
\begin{equation}\label{eq:Gext}
  G_n \equiv G(\bk_1, \cdots\hskip -1pt, \bk_{n-1}) = G(\bP_2, \cdots\hskip -1pt, \bP_{n})\,.
\end{equation}
For $i \in \{2, \cdots\hskip -1pt, n\}$, the minor $G[\begin{smallmatrix}i\\ 1\end{smallmatrix}]^\mu$ is a spatial vector (inherited from the surviving first row).  
Cramer's rule applied to the external Gram system gives the orthogonality relation
\begin{equation}\label{eq:dual}
  G[\begin{smallmatrix}i\\ 1\end{smallmatrix}]^\mu  P_{j,\mu} = -\delta_{ij}\,G_n\,,
  \quad i,\, j \in \{2, \cdots\hskip -1pt, n\}\,.
\end{equation}
The minors $-G[\begin{smallmatrix}i\\ 1\end{smallmatrix}]^\mu / G_n$ therefore form a dual basis to $\{\bP_2, \cdots\hskip -1pt, \bP_{n}\}$.

Expanding~\eqref{eq:gram_exposed} along its first row in terms of the minors~\eqref{eq:vecgram} and~\eqref{eq:dual} yields the following identity
\begin{equation}\label{eq:quad_ident}
  G_n\, \bl^2 + \sum_{i=2}^{n}G[\begin{smallmatrix}i\\ 1\end{smallmatrix}]^\mu P_{i}^\nu
  \ell_\mu \ell_\nu = 0\,.
\end{equation}
Since $\bl^2 = d_1$ and $2  \bl\cdot \bP_i = d_i - d_1 - \bP_i^2$, this is a polynomial of degree two in the propagator denominators~$d_i$, thereby confirming the quadratic nature of the $n = D+1$ reduction.
Defining the tensor Feynman integrals
\begin{equation}\label{eq:tensor_ints}
  F^{(n)}_\mu = \int_\ell \frac{\ell_\mu}{\prod_{j=1}^{n}d_j}\,,
  \quad
  F^{(n)}_{\mu\nu} = \int_\ell \frac{\ell_\mu\ell_\nu}{\prod_{j=1}^{n}d_j}\,,
  \quad
  F^{(n)}_{i,\mu} = \int_\ell
    \frac{\ell_\mu}{\prod_{j=1, j\neq i}^{n}d_j}\,,
\end{equation}
and dividing~\eqref{eq:quad_ident} by $\prod_{j=1}^{n} d_j$ and integrating leads to the tensor relation
\begin{equation}\label{eq:tensor_id}
  G_n {F^{(n)\,\mu}}_\mu
  + \sum_{i=2}^{n}\, G[\begin{smallmatrix}i\\ 1\end{smallmatrix}]^\mu P_{i}^\nu
  F^{(n)}_{\mu\nu} = 0\,.
\end{equation}
Applying standard tensor integral reduction techniques, one can first reduce the trace ${F^{(n)\,\mu}}_\mu = \int d_1/\prod d_j$
to the $(n{-}1)$-gon~$I^{(n)}_1$.
The contractions $F^{(n)}_{\mu\nu}P_i^{\nu}$ are then reduced using $2\bl\cdot\bP_i = d_i - d_1 - \bP_i^2$, which produces the vector integrals $F^{(n)}_{i,\mu}$, $F^{(n)}_{1,\mu}$ and $F^{(n)}_{\mu}$ of~\eqref{eq:tensor_ints}. 
These in turn reduce to combinations of the $n$-gon and $(n{-}1)$-gons, arriving at a result purely in terms of scalar integrals.  
The final result can be expressed in a determinant form as
\begin{equation}\label{eq:det_form}
  \det
  \begin{pmatrix}
    I_{12\cdots n}    & -I^{(n)}_1   & -I^{(n)}_2   & \cdots & -I^{(n)}_{n} \\[2pt]
    1      & Y_{11}       & Y_{12}       & \cdots & Y_{1n}         \\
    1      & Y_{12}       & Y_{22}       & \cdots & Y_{2n}         \\
    \vdots &    \vdots          &     \vdots         & \ddots & \vdots         \\
    1      & Y_{1n}       & Y_{2n}       & \cdots & Y_{nn}
  \end{pmatrix}
  \;=\; 0\,,
\end{equation}
where $I_{12\cdots n}$ is the $n$-gon and $I^{(n)}_a$ denotes the $(n{-}1)$-gon with the propagator at the $a$-th position removed.

The matrix~\eqref{eq:det_form} is $\Ycal_n$ with its border row replaced by $(I_{12\cdots n},\, -I^{(n)}_1,\, \cdots\hskip -1pt,\, -I^{(n)}_{n})$.  
Expanding along this row gives
\begin{equation}\label{eq:expand_det}
  \Ycal_n[\begin{smallmatrix}0\\ 0\end{smallmatrix}] \, I_{12\cdots n} - \sum_{a=1}^{n} \Ycal_n[\begin{smallmatrix}0\\ a\end{smallmatrix}] \, I^{(n)}_a = 0\,.
\end{equation}
Since $\Ycal_n\left[\begin{smallmatrix}0\\ 0\end{smallmatrix}\right] = \Delta_n$ is the leading Landau polynomial of the $n{=}(D{+}1)$-gon, we obtain
\begin{equation}\label{eq:221}
  I_{12\cdots (D{+}1)} = \sum_{a=1}^{D+1} \frac{\Ycal_{D{+}1}[\begin{smallmatrix}0\\ a\end{smallmatrix}]}{\Del_{D{+}1}}\, I^{(D{+}1)}_a\,,
\end{equation}
which is the reduction formula~\eqref{eq:D+1-to-D} presented in the main paper.  
In contrast to~\eqref{eq:rederivedredux}, this reduction
is unique and there are no free choices, because $\Ycal_n$ at $n = D + 1$ has full rank and the cofactor expansion is completely determined.

\subsection{Iteration}
\label{sec:iteration}

The linear step~\eqref{eq:rederivedredux} can be repeated until only critical $(D+1)$-gons remain. 
This gives
\begin{equation}\label{eq:219}
  I_{12\cdots n} = \sum_{l_1 < l_2 < \cdots < l_{D+1}}
  f^{(n)}_{l_1 \cdots l_{D+1}} I_{l_1 \cdots l_{D+1}}\,,
\end{equation}
where $l_1, \cdots\hskip -1pt, l_{D+1}$ are the $D+1$ internal momenta remaining. 
The coefficient accumulated along the reduction chain is
\begin{equation}\label{eq:220}
 f^{(n)}_{l_1 \cdots l_{D+1}} = \prod_{\alpha = D+2}^{n}\;\frac{\Ycal_{(l_1\cdots l_{D+1}\,l_\alpha)}[\begin{smallmatrix}0\\ l_\alpha\end{smallmatrix}]}{\Ycal_{(l_1\cdots l_{D+1}\,l_\alpha)}[\begin{smallmatrix}0\\ 0\end{smallmatrix}]}\,,
\end{equation}
where $\Ycal_{(l_1 \cdots l_{D+1}\, l_\alpha)}$ denotes the modified Cayley matrix formed from the indicated subset of propagators.  
Crucially, these coefficients are independent of the order in which the intermediate reductions are performed; for any two chains of choices leading to the same final set $\{l_1, \cdots\hskip -1pt, l_{D+1}\}$, the accumulated product~\eqref{eq:220} is the same.
This is the content of the lemma~(A.20) of~\cite{Melrose:1965}, whose proof rests on the minor identities~(A.11)--(A.13) of the bordered determinant in the same reference.

Each $(D{+}1)$-gon in~\eqref{eq:219} is further reduced by the unique step~\eqref{eq:221}.  
Combining gives a decomposition into $\binom{n}{D}$ sub-$D$-gons:
\begin{equation}\label{eq:222}
  I_{12\cdots n} =\sum_{l_1 < \cdots < l_D}
  a^{(n)}_{l_1 \cdots l_D} I_{l_1 \cdots l_D}\,.
\end{equation}
The coefficient is obtained by combining the iterated linear step~\eqref{eq:220} with the unique quadratic step~\eqref{eq:221}.
Each $D$-gon $I_{l_1\cdots l_D}$ appears inside every $(D{+}1)$-gon of the form $I_{l_1\cdots l_D\, l_\beta}$, and summing over the extra propagator gives
\begin{equation}\label{eq:a_coeff}
  a^{(n)}_{l_1 \cdots l_D}
  \;=\; \sum_{\substack{l_\beta = 1 \\[1pt] l_\beta \notin \{l_1,\cdots\hskip -1pt, l_D\}}}^{n}\;
  f^{(n)}_{l_1 \cdots l_D\, l_\beta}
  \frac{\Ycal_{(l_1\cdots l_{D}\,l_\beta)}[\begin{smallmatrix}0\\ l_\beta \end{smallmatrix}]}{\Ycal_{(l_1\cdots l_{D}\,l_\beta)}[\begin{smallmatrix}0\\ 0 \end{smallmatrix}]}\,,
\end{equation}
where $f^{(n)}_{l_1\cdots l_{D+1}}$ is given by~\eqref{eq:220} and $\Ycal_{(l_1\cdots l_D\, l_\beta)}\minor{0}{0} = \Del_{(l_1\cdots l_D\, l_\beta)}$ is the Landau polynomial of the sub-$(D{+}1)$-gon.

\newpage
\section{van Neerven--Vermaseren Reduction}
\label{app:vv-reduction}

In this appendix, we present a parallel reduction of the $n$-gon integrals studied in this work based on the van Neerven--Vermaseren basis~\cite{vanNeerven:1984} (see~\cite{Jain:2020rmw} for a similar application).
The main text uses the Melrose reduction procedure \cite{Melrose:1965} because the $\Ycal$-matrix machinery of Section~\ref{sec:qformalism} provides a more geometrical and combinatorial framework for solving the $n$-gon integrals. 
The van Neerven--Vermaseren basis is nevertheless useful as it gives a conceptually independent derivation of the reduction formulas used in our work, and also provides a strong numerical cross-check on the results obtained using the formulas presented in Section~\ref{sec:qformalism}.
While we present the essential details here, the interested reader is referred to \cite{Ellis:2011cr} for a detailed overview. 

In $D$~dimensions, the generalized Kronecker symbol is defined as the determinant \cite{vanOldenborgh:1989wn}
\begin{equation}\label{eq:gKd}
\delta^{\mu_1\mu_2\cdots\mu_n}_{\nu_1\nu_2\cdots\nu_n} =
  \begin{vmatrix}
    \delta^{\mu_1}_{\nu_1} & \delta^{\mu_1}_{\nu_2} & \cdots & \delta^{\mu_1}_{\nu_n}\\[2pt]
    \delta^{\mu_2}_{\nu_1} & \delta^{\mu_2}_{\nu_2} & \cdots & \delta^{\mu_2}_{\nu_n}\\
    \vdots & \vdots & \ddots & \vdots \\
    \delta^{\mu_n}_{\nu_1} & \delta^{\mu_n}_{\nu_2} & \cdots & \delta^{\mu_n}_{\nu_n}
  \end{vmatrix}\,.
\end{equation}
It vanishes identically for $n \geq D+1$, and for $n=D$ it factorizes into the product of two Levi-Civita tensors.
Contracting pairs of indices with vectors gives the $n$-particle Gram determinant
\begin{equation}\label{eq:Gram-delta}
  G(\bk_1,\cdots\hskip -1pt,\bk_n)
  = \delta^{\bk_1 \bk_2 \cdots \bk_n}_{\bk_1 \bk_2 \cdots \bk_n}
  = \det(\bk_i \cdot \bk_j)\,.
\end{equation}
For the $n$-gon in $D=3$ with $n\geq 4$, the physical space is spanned by the three independent partial sums $\bP_2,\bP_3,\bP_4$.
The van Neerven--Vermaseren basis vectors are
\begin{equation}\label{eq:vnv-dual}
  v_1^{\mu}
  = \frac{\delta^{\mu\, \bP_3\, \bP_4}_{\bP_2\, \bP_3\, \bP_4}}{G_4}\,,\quad
  v_2^{\mu}
  = \frac{\delta^{\bP_2\, \mu\, \bP_4}_{\bP_2\, \bP_3\, \bP_4}}{G_4}\,,\quad
  v_3^{\mu}
  = \frac{\delta^{\bP_2\, \bP_3\, \mu}_{\bP_2\, \bP_3\, \bP_4}}{G_4}\,,
\end{equation}
where $G_4 \equiv G(\bP_2, \bP_3, \bP_4) = \delta^{\bP_2\,\bP_3\,\bP_4}_{\bP_2\,\bP_3\,\bP_4}$ is the Gram determinant of the independent external momenta as defined in \eqref{eq:Gext}. 
They satisfy the duality relation
\begin{equation}\label{eq:vnv-orth}
  \bm v_i \cdot \bP_{j+1} = \delta_{ij}
  \qquad (i,j = 1,2,3)\,,
\end{equation}
and the loop momentum $\bl$ can be expanded as
\begin{equation}\label{eq:vnv-expand}
  \ell^{\mu}
  = \sum_{i=1}^{3}(\bl \cdot \bP_{i+1})\,v_i^{\mu}\,.
\end{equation}
The scalar products $\bl \cdot  \bP_{i+1}$ can be written in terms of inverse propagators using
\begin{equation}\label{eq:vnv-identity}
  2\,\bl \cdot  \bP_{i+1}
  = d_{i+1} - d_1 - \bP_{i+1}^2\,,
  \quad i = 1,2,3\,,
\end{equation}
which follows directly from $d_{i+1} = (\bl + \bP_{i+1})^2$ and $d_1 = \bl^2$.
Substituting into~\eqref{eq:vnv-expand} gives
\begin{equation}\label{eq:vnv-ldecomp}
  \ell^{\mu}
  = -\,\frac{w^{\mu}}{2}
    + \frac{1}{2}\sum_{i=1}^{3}
      (d_{i+1} - d_1)\,v_i^{\mu}\,,
\end{equation}
where
\begin{equation}\label{eq:vnv-w}
  w^{\mu}
  = \sum_{i=1}^{3} v_i^{\mu}\bP_{i+1}^2~,
\end{equation}
is a fixed kinematic vector.
Each propagator difference $d_{i+1} - d_1$
in~\eqref{eq:vnv-ldecomp} cancels one propagator from the denominator upon partial-fractioning, thereby producing integrals with one fewer propagator.

\paragraph{The box.}~A key result of~\cite{vanNeerven:1984} states that a $(D{+}1)$-gon scalar integral in $D$~dimensions can be reduced to a sum of $(D{+}1)$ scalar $D$-gon integrals.
In three dimensions, the four propagators $d_1,\cdots\hskip -1pt,d_4$ of the box overdetermine the loop momentum.
Squaring~\eqref{eq:vnv-ldecomp} and imposing
$\bl^2 = d_1$ yields a quadratic relation among the propagators
with purely kinematic coefficients, whose partial-fraction decomposition gives
\begin{equation}\label{eq:vnv-box}
  I_{1234} = \frac{1}{\bm w^2}\bigg[
      (\bm w\!\cdot\! \bm v_1) I_{134}
    + (\bm w\!\cdot\! \bm v_2) I_{124}
    + (\bm w\!\cdot\! \bm v_3) I_{123}
    + \bigl(2 - \bm w\!\cdot\!(\bm v_1{+} \bm v_2{+}\bm v_3)\bigr) I_{234}
  \bigg]\,.
\end{equation}
When specialized to the ordered box kinematics, this reproduces the same rational answer \eqref{eq:4gonfinal} as obtained via the Melrose reduction \eqref{eq:box-sum-triangles}.

\paragraph{The pentagon.}~For the pentagon, the propagator factor
$d_5 = (\bl + \bP_5)^2$ is not among the four propagators $d_1,\cdots\hskip -1pt,d_4$ that already fix $\bl$ in $D=3$.
The van Neerven--Vermaseren expansion expresses $d_5$ as a linear function of
$d_1,\cdots\hskip -1pt,d_4$ using the relation
\begin{equation}\label{eq:vnv-d5}
  d_5 = d_1 + 2\,\bl \cdot \bP_5 + \bP_5^2
  = \sum_{i=1}^{3}(\bm v_i\!\cdot\! \bP_5)\,(d_{i+1} - d_1 - \bP_{i+1}^2)
    + d_1 + \bP_5^2\,,
\end{equation}
where we have substituted~\eqref{eq:vnv-identity} in the second relation.
Dividing the pentagon integrand by this relation and partial-fractioning produces the reduction
\begin{align}
    I_{12345} = \frac{1}{\bm w \cdot \bP_5 - \bP_5^2} \Big[(\bm v_1 \cdot \bP_5)I_{1345} + (\bm  v_2 \cdot \bP_5) I_{1245} + (\bm  v_3 \cdot \bP_5) I_{1235}& \nonumber \\
    +\, (1- (\bm v_1 + \bm v_2 + \bm v_3)\cdot \bP_5) I_{2345} - I_{1234}&\Big]\,. 
    \label{eq:vvpentagon}
\end{align}

\paragraph{$\boldsymbol{n}$-gons.}~The same pattern is completely recursive. Once the three independent partial sums $\bP_2, \bP_3,\bP_4$ are chosen, every additional propagator $d_j$ with $j \geq 5$ is a linear combination of $d_1,\cdots\hskip -1pt,d_4$ and can be eliminated by the same partial-fraction step.
For example, in the case of the hexagon one finds
\begin{align}\label{eq:vnv-hex}
  I_{123456} = \frac{1}{\bm w\cdot \bP_5 - \bP_5^2}\Big[
      (\bm v_1\cdot \bP_5) I_{13456}
    + (\bm v_2\cdot \bP_5) I_{12456}
    + (\bm v_3\cdot \bP_5) I_{12356}&
  \nonumber \\
    + \bigl(1 - (\bm v_1+\bm v_2+\bm v_3)\cdot \bP_5\bigr) I_{23456} - I_{12346}&
  \Big]\,,
\end{align}
and similarly for the 7-gon $I_{1234567}$.
The recursive structure means the reduction process remains tractable beyond five points, even though the resulting formulas are less adapted to the final symmetry than the presentation in terms of the minors of the $\Ycal$-matrix.

\newpage
\section{Conformal Ward Identities}
\label{app:distance-cwi}

The scalar correlators constructed in the main text are naturally functions of the distance variables associated with the edge set $E_n$ of $K_n$. The conformal generators, however, act on the external momentum vectors. 
In this appendix, we translate the conformal generators into these distance variables. Our index conventions are:
\begin{itemize}
\item $i,j=1,\cdots\hskip -1pt,n$ denote the external legs of an $n$-point function.
\item $I,J=1,\cdots\hskip -1pt,|E_n|$ denote the distance variables of a momentum $n$-gon.
\end{itemize}
For an external scalar of dimension $\Delta$ in $D$ dimensions, the special conformal generator acting on the momentum $\bk_i$ is
\begin{equation}
\bm{K}_i
=
\bk_i\hskip 1pt\partial_i^2
-2(\bk_i\!\cdot\!\bm\partial_i)\bm\partial_i+2(\Delta-D)\bm\partial_i\,,
\quad\text{with}\quad
\bm\partial_i\equiv \frac{\partial}{\partial \bk_i}\, .
\label{eq:Kcartesian}
\end{equation}
Each distance $k_I$ is the norm of a consecutive sum of external momenta. Momentum conservation gives two complementary representations; we choose one and denote its set of labels by $I$. We then define
\begin{equation}
\bk_I\equiv \sum_{i\in I}\bk_i\,,
\quad
k_I\equiv |\bk_I|\,,
\quad
\hat{\bk}_I\equiv \frac{\bk_I}{k_I}\,,
\quad
\partial_I\equiv \frac{\partial}{\partial k_I}\, .
\label{eq:kI-def}
\end{equation}
This representative is held fixed under differentiation. For example, choosing $k_{12}=|\bk_1+\bk_2|$ sets $I=\{1,2\}$ and implies $\partial k_{12}/\partial\bk_3=0$, even though momentum conservation also gives $k_{12}=|\bk_3+\cdots+\bk_n|$.

Using the chain rule, the first derivative then takes the form
\begin{equation}
\bm\partial_i 
=\sum_{I\ni\, i}\hat{\bk}_I\hskip 1pt\partial_I\, .
\label{eq:kI-chain}
\end{equation}
The second-derivative structures appearing in~\eqref{eq:Kcartesian} are given by
\begin{align}
\partial_i^2
&=
\sum_{I,J\ni i}
\bigl(\hat{\bk}_I\!\cdot\!\hat{\bk}_J\bigr)
\partial_I\partial_J
+(D-1)\sum_{I\ni i}\frac{1}{k_I}\,\partial_I\,,
\label{eq:lap-kI}
\\
(\bk_i\!\cdot\!\bm\partial_i)\bm\partial_i
&=
\sum_{I,J\ni i}
\bigl(\bk_i\!\cdot\!\hat{\bk}_J\bigr)\hat{\bk}_I\,
\partial_I\partial_J
+\sum_{I\ni i}\frac{1}{k_I}
\big(
\bk_i-\bigl(\bk_i\!\cdot\!\hat{\bk}_I\bigr)\hat{\bk}_I\big)\partial_I\, .
\label{eq:dhess-kI}
\end{align}
Substituting these relations into~\eqref{eq:Kcartesian} and collecting the external momentum vectors gives
\begin{equation}
\sum_{i=1}^{n}\bm{K}_i
=
-\sum_{i=1}^{n}\bk_i\hskip 1pt D_i^{(n)}\, ,
\label{eq:Da-defining}
\end{equation}
where
\begin{equation}
D_i^{(n)}
=
2\sum_{I\ni i}\sum_J
\frac{\bk_{I\cap J}\!\cdot\!\hat{\bk}_J}{k_I}\,
\partial_I\partial_J
-\sum_{I,J\ni i}
\hat{\bk}_I\!\cdot\!\hat{\bk}_J\,
\partial_I\partial_J
-\sum_{I\ni i}
\frac{2(\Delta-D)|I|+D-1}{k_I}\,\partial_I\, .
\label{eq:full-Dcn}
\end{equation}
Here $\bk_{I\cap J}\equiv \sum_{i\in I\cap J}\bk_i$ is the overlap momentum, which vanishes when the chosen representatives are disjoint, and $|I|$ counts the number of momenta defining $k_I$.

Eliminating $\bk_n=-\sum_{i=1}^{n-1}\bk_i$, conformal invariance requires
\begin{equation}
\sum_{i=1}^{n-1}\bk_i
\bigl(D_i^{(n)}-D_n^{(n)}\bigr)
\langle {\cal O}_1\cdots {\cal O}_n\rangle'=0\, .
\label{eq:diffward}
\end{equation}
Contracting this vector equation with a basis of linearly independent external momenta gives an equivalent set of scalar Ward identities. Contracting with each of the $n-1$ external momenta gives the projected equations, of which only three are independent at generic kinematics in $D=3$. We have
\begin{equation}
\mathbb K_i^{(n)}
\equiv
\sum_{j=1}^{n-1}
\bk_i\cdot\bk_j
\bigl(D_j^{(n)}-D_n^{(n)}\bigr)\, ,
\label{eq:generalSCToperator}
\end{equation}
for $i=1,\cdots\hskip -1pt,n-1$. Conformal invariance is then equivalent to
\begin{equation}
\mathbb K_i^{(n)}
\langle{\cal O}_1\cdots{\cal O}_n\rangle'=0\, .
\end{equation}
We now apply this to the $n=4,5$ cases.

\paragraph{The box.}

The ordered box depends on the six distance variables
\begin{equation}
E_4=\{k_1,k_2,k_3,k_4,k_{12}= s,k_{23}= t\}\, ,
\label{eq:box-vars-cwi}
\end{equation}
introduced in~\eqref{eq:fourpointdisvar}. Since $\bk_1$ appears only in $k_1$ and $k_{12}$, the general expression~\eqref{eq:full-Dcn} reduces to
\begin{equation}
\begin{aligned}
D^{(4)}_1={}&
\partial_1^2+\partial_{12}^2
-\frac{2\Delta-D-1}{k_1}\partial_1
-\frac{4\Delta-3D-1}{k_{12}}\partial_{12}
\\
&+\frac{2k_1}{k_{12}}\partial_1\partial_{12}
+\frac{2k_2}{k_{12}}\partial_2\partial_{12}
+\frac{k_2^2+k_{23}^2-k_3^2}{k_{12}k_{23}}\partial_{12}\partial_{23}\,.
\end{aligned}
\label{eq:D4-1}
\end{equation}
This agrees with the operator in~\cite{Arkani-Hamed:2018kmz}. The remaining operators $D_i^{(4)}$ are obtained by cyclic permutations.

At four points in $D=3$, the Ward identity takes a stronger form. After eliminating one momentum by momentum conservation, the remaining momenta are generically linearly independent. The vector constraint~\eqref{eq:diffward} therefore forces each coefficient to vanish separately, giving
\begin{equation}
(D_i^{(4)}-D_j^{(4)})\langle{\cal O}_1\cdots{\cal O}_4\rangle'=0\, ,
\end{equation}
for any $i,j$. This pairwise form is special to four points. Acting on the box result~\eqref{eq:4gonfinal}, it is straightforward to check that $(D_i^{(4)}-D_j^{(4)})I_{1234}=0$.

\newpage
\paragraph{The pentagon.}

Five-point functions depend on the ten distances
\begin{equation}
E_5=\{k_1,k_2,k_3,k_4,k_5,k_{12},k_{23},k_{34},k_{123},k_{234}\}\, ,
\label{eq:E5-cwi}
\end{equation}
introduced in~\eqref{eq:disvar5pt}. Unlike the four-point case, these ten variables are not independent in three dimensions. We construct the SCT operators by treating the ten variables as independent and impose the Gram constraint afterward.

For leg $1$, the relevant distances are $k_1$, $k_{12}$, $k_{123}$, and the corresponding operator is
\begin{align}
D_1^{(5)}={}&
\Bigl[
\partial_1^2+\partial_{12}^2+\partial_{123}^2
-\frac{2\Delta-D-1}{k_1}\partial_1
-\frac{4\Delta-3D-1}{k_{12}}\partial_{12}
-\frac{6\Delta-5D-1}{k_{123}}\partial_{123}
\Bigr]
\nonumber\\
&+\frac{2}{k_{12}}
(k_1\partial_1+k_2\partial_2)\partial_{12}
+\frac{2}{k_{123}}
(k_1\partial_1+k_2\partial_2+k_3\partial_3
+k_{12}\partial_{12}+k_{23}\partial_{23})\partial_{123}
\nonumber\\[1mm]
&+\Bigl[
\frac{k_2^2+k_{23}^2-k_3^2}{k_{12}k_{23}}\partial_{23}
+\frac{k_2^2+k_{234}^2-k_{34}^2}{k_{12}k_{234}}\partial_{234}
\Bigr]\partial_{12}
\nonumber\\[1mm]
&+\Bigl[
\frac{k_3^2+k_{34}^2-k_4^2}{k_{34}k_{123}}\partial_{34}
+\frac{k_{23}^2+k_{234}^2-k_4^2}{k_{123}k_{234}}\partial_{234}
\Bigr]\partial_{123}\,.
\label{eq:D5-1}
\end{align}
Recall that the physical five-point kinematics obey $G_5=0$. As shown in \eqref{eq:pent-special-Ward}, we find
\begin{equation}
\left[
\mathbb K_i^{(5)}I_{12345}
\right]_{G_5=0}
=0\, .
\label{eq:check5}
\end{equation}
Since the Gram matrix has rank three, one of these four equations is redundant.

\newpage
\bibliographystyle{JHEP}
\bibliography{refs}

@article{Hang:2024xas,
    author = "Hang, Yanfeng and Shen, Cong",
    title = "{A note on kinematic flow and differential equations for two-site one-loop graph in FRW spacetime}",
    eprint = "2410.17192",
    archivePrefix = "arXiv",
    primaryClass = "hep-th",
    doi = "10.1007/JHEP09(2025)209",
    journal = "JHEP",
    volume = "09",
    pages = "209",
    year = "2025"
}

@article{Baumann:2024mvm,
    author = "Baumann, Daniel and Goodhew, Harry and Lee, Hayden",
    title = "{Kinematic flow for cosmological loop integrands}",
    eprint = "2410.17994",
    archivePrefix = "arXiv",
    primaryClass = "hep-th",
    doi = "10.1007/JHEP07(2025)131",
    journal = "JHEP",
    volume = "07",
    pages = "131",
    year = "2025"
}

@article{Henn:2022ydo,
    author = "Henn, Johannes M. and Matija{\v{s}}i{\'c}, Antonela and Miczajka, Julian",
    title = "{One-loop hexagon integral to higher orders in the dimensional regulator}",
    eprint = "2210.13505",
    archivePrefix = "arXiv",
    primaryClass = "hep-th",
    reportNumber = "MPP-2022-131",
    doi = "10.1007/JHEP01(2023)096",
    journal = "JHEP",
    volume = "01",
    pages = "096",
    year = "2023"
}

@article{Dixon:2011ng,
    author = "Dixon, Lance J. and Drummond, James M. and Henn, Johannes M.",
    title = "{The one-loop six-dimensional hexagon integral and its relation to MHV amplitudes in N=4 SYM}",
    eprint = "1104.2787",
    archivePrefix = "arXiv",
    primaryClass = "hep-th",
    reportNumber = "HU-EP-11-17, CERN-PH-TH-2011-075, SLAC-PUB-14434, LAPTH-013-11, CERN--PH--TH-2011-075, SLAC--PUB--14434",
    doi = "10.1007/JHEP06(2011)100",
    journal = "JHEP",
    volume = "06",
    pages = "100",
    year = "2011"
}

@article{DelDuca:2011ne,
    author = "Del Duca, Vittorio and Duhr, Claude and Smirnov, Vladimir A.",
    title = "{The massless hexagon integral in D = 6 dimensions}",
    eprint = "1104.2781",
    archivePrefix = "arXiv",
    primaryClass = "hep-th",
    reportNumber = "DCPT-11-34, IPPP-11-17",
    doi = "10.1016/j.physletb.2011.07.079",
    journal = "Phys. Lett. B",
    volume = "703",
    pages = "363--365",
    year = "2011"
}

@article{Duplancic:2003tv,
    author = "Duplancic, G. and Nizic, B.",
    title = "{Reduction method for dimensionally regulated one loop N point Feynman integrals}",
    eprint = "hep-ph/0303184",
    archivePrefix = "arXiv",
    reportNumber = "IRB-TH-2-03",
    doi = "10.1140/epjc/s2004-01723-7",
    journal = "Eur. Phys. J. C",
    volume = "35",
    pages = "105--118",
    year = "2004"
}

@article{Binoth:2002xh,
    author = "Binoth, T. and Heinrich, G. and Kauer, N.",
    title = "{A Numerical evaluation of the scalar hexagon integral in the physical region}",
    eprint = "hep-ph/0210023",
    archivePrefix = "arXiv",
    reportNumber = "EDINBURGH-2002-16, IPPP-02-56, DCPT-02-112",
    doi = "10.1016/S0550-3213(03)00052-X",
    journal = "Nucl. Phys. B",
    volume = "654",
    pages = "277--300",
    year = "2003"
}

@article{Mizera:2021fap,
    author = "Mizera, Sebastian",
    title = "{Crossing symmetry in the planar limit}",
    eprint = "2104.12776",
    archivePrefix = "arXiv",
    primaryClass = "hep-th",
    doi = "10.1103/PhysRevD.104.045003",
    journal = "Phys. Rev. D",
    volume = "104",
    number = "4",
    pages = "045003",
    year = "2021"
}

@article{Chakraborty:2025myb,
    author = "Chakraborty, Priyesh",
    title = "{Primordial non-Gaussianity from light compact scalars}",
    eprint = "2501.07672",
    archivePrefix = "arXiv",
    primaryClass = "hep-th",
    doi = "10.1007/JHEP11(2025)023",
    journal = "JHEP",
    volume = "11",
    pages = "023",
    year = "2025"
}

@article{Green:2013rd,
    author = "Green, Daniel and Lewandowski, Matthew and Senatore, Leonardo and Silverstein, Eva and Zaldarriaga, Matias",
    title = "{Anomalous Dimensions and Non-Gaussianity}",
    eprint = "1301.2630",
    archivePrefix = "arXiv",
    primaryClass = "hep-th",
    reportNumber = "SLAC-PUB-15334, SU-ITP-12-42",
    doi = "10.1007/JHEP10(2013)171",
    journal = "JHEP",
    volume = "10",
    pages = "171",
    year = "2013"
}

@article{Aoki:2026yrb,
    author = "Aoki, Shuntaro",
    title = "{Primordial Correlators from a Kaluza-Klein Graviton Continuum}",
    eprint = "2608.01762",
    archivePrefix = "arXiv",
    primaryClass = "hep-th",
    reportNumber = "RIKEN-iTHEMS-Report-26",
    month = "8",
    year = "2026"
}

@article{Kumar:2025anx,
    author = "Kumar, Soubhik and Nee, Michael",
    title = "{Warped dimensions at the cosmological collider}",
    eprint = "2510.19900",
    archivePrefix = "arXiv",
    primaryClass = "hep-ph",
    doi = "10.1007/JHEP04(2026)035",
    journal = "JHEP",
    volume = "04",
    pages = "035",
    year = "2026"
}

@article{Kumar:2018jxz,
    author = "Kumar, Soubhik and Sundrum, Raman",
    title = "{Seeing Higher-Dimensional Grand Unification In Primordial Non-Gaussianities}",
    eprint = "1811.11200",
    archivePrefix = "arXiv",
    primaryClass = "hep-ph",
    reportNumber = "UMD-PP-018-09",
    doi = "10.1007/JHEP04(2019)120",
    journal = "JHEP",
    volume = "04",
    pages = "120",
    year = "2019"
}

@article{Hubisz:2024xnj,
    author = "Hubisz, Jay and Lee, Seung J. and Li, He and Sambasivam, Bharath",
    title = "{Cosmological quasiparticles and the cosmological collider}",
    eprint = "2408.08951",
    archivePrefix = "arXiv",
    primaryClass = "astro-ph.CO",
    doi = "10.1103/PhysRevD.111.023543",
    journal = "Phys. Rev. D",
    volume = "111",
    number = "2",
    pages = "023543",
    year = "2025"
}

@article{Pimentel:2025rds,
    author = "Pimentel, Guilherme L. and Yang, Chen",
    title = "{Strongly coupled sectors in inflation: gapless theories and unparticles}",
    eprint = "2503.17840",
    archivePrefix = "arXiv",
    primaryClass = "hep-th",
    doi = "10.1007/JHEP04(2026)146",
    journal = "JHEP",
    volume = "04",
    pages = "146",
    year = "2026"
}

@article{Jiang:2025mlm,
    author = "Jiang, Yikun and Pimentel, Guilherme L. and Yang, Chen",
    title = "{Strongly coupled sectors in inflation: gapped theories of unparticles}",
    eprint = "2512.23796",
    archivePrefix = "arXiv",
    primaryClass = "hep-th",
    doi = "10.1007/JHEP06(2026)125",
    journal = "JHEP",
    volume = "06",
    pages = "125",
    year = "2026"
}

@article{Aoki:2023tjm,
    author = "Aoki, Shuntaro",
    title = "{Continuous spectrum on cosmological collider}",
    eprint = "2301.07920",
    archivePrefix = "arXiv",
    primaryClass = "hep-th",
    doi = "10.1088/1475-7516/2023/04/002",
    journal = "JCAP",
    volume = "04",
    pages = "002",
    year = "2023"
}

@article{Vasiliev:2003ev,
    author = "Vasiliev, M. A.",
    title = "{Nonlinear equations for symmetric massless higher spin fields in (A)dS(d)}",
    eprint = "hep-th/0304049",
    archivePrefix = "arXiv",
    reportNumber = "FIAN-TD-07-03",
    doi = "10.1016/S0370-2693(03)00872-4",
    journal = "Phys. Lett. B",
    volume = "567",
    pages = "139--151",
    year = "2003"
}

@article{Mazloumi:2025pmx,
    author = "Mazloumi, Pouria and Xu, Xiaofeng",
    title = "{Cluster algebras for cosmological correlators}",
    eprint = "2512.14854",
    archivePrefix = "arXiv",
    primaryClass = "hep-th",
    reportNumber = "MITP-25-081",
    doi = "10.1007/JHEP03(2026)256",
    journal = "JHEP",
    volume = "03",
    pages = "256",
    year = "2026"
}

@article{Glew:2025ypb,
    author = "Glew, Ross and Pokraka, Andrzej",
    title = "{Kinematic flow from the flow of cuts}",
    eprint = "2508.11568",
    archivePrefix = "arXiv",
    primaryClass = "hep-th",
    doi = "10.1007/JHEP06(2026)158",
    journal = "JHEP",
    volume = "06",
    pages = "158",
    year = "2026"
}

@article{Chowdhury:2026upp,
    author = "Chowdhury, Chandramouli and Jazayeri, Sadra and Lipstein, Arthur and Marshall, Joe and Mei, Jiajie and Sachs, Ivo",
    title = "{Cosmological Correlator Discontinuities from Scattering Amplitudes}",
    eprint = "2602.03841",
    archivePrefix = "arXiv",
    primaryClass = "hep-th",
    month = "2",
    year = "2026"
}

@article{Jazayeri:2023xcj,
    author = "Jazayeri, Sadra and Renaux-Petel, S{\'e}bastien and Werth, Denis",
    title = "{Shapes of the cosmological low-speed collider}",
    eprint = "2307.01751",
    archivePrefix = "arXiv",
    primaryClass = "hep-th",
    doi = "10.1088/1475-7516/2023/12/035",
    journal = "JCAP",
    volume = "12",
    pages = "035",
    year = "2023"
}

@article{Bodas:2025vpb,
    author = "Bodas, Arushi and Broadberry, Edward and Sundrum, Raman and Xu, Zhaohui",
    title = "{Charged loops at the cosmological collider with chemical potential}",
    eprint = "2507.22978",
    archivePrefix = "arXiv",
    primaryClass = "hep-ph",
    reportNumber = "FERMILAB-PUB-25-0519-V",
    doi = "10.1007/JHEP01(2026)083",
    journal = "JHEP",
    volume = "01",
    pages = "083",
    year = "2026"
}

@article{You:2026xoq,
    author = "You, Jingtao and Song, Linghao and Han, Chengcheng and He, Hong-Jian and Chen, Xingang and Xianyu, Zhong-Zhi",
    title = "{Cosmological Collider Signatures from Right-Handed Neutrino Loop}",
    eprint = "2605.21419",
    archivePrefix = "arXiv",
    primaryClass = "hep-ph",
    month = "5",
    year = "2026"
}

@article{Coriano:2019nkw,
    author = "Corian{\`o}, Claudio and Maglio, Matteo Maria and Theofilopoulos, Dimosthenis",
    title = "{Four-Point Functions in Momentum Space: Conformal Ward Identities in the Scalar/Tensor case}",
    eprint = "1912.01907",
    archivePrefix = "arXiv",
    primaryClass = "hep-th",
    doi = "10.1140/epjc/s10052-020-8089-1",
    journal = "Eur. Phys. J. C",
    volume = "80",
    number = "6",
    pages = "540",
    year = "2020"
}

@article{Jain:2022ajd,
    author = "Jain, Prabhav and Jain, Sachin and Sahoo, Bibhut and Dhruva, K. S. and Zade, Aashna",
    title = "{Mapping Large N Slightly Broken Higher Spin (SBHS) theory correlators to free theory correlators}",
    eprint = "2207.05101",
    archivePrefix = "arXiv",
    primaryClass = "hep-th",
    doi = "10.1007/JHEP12(2023)173",
    journal = "JHEP",
    volume = "12",
    pages = "173",
    year = "2023"
}

@article{Jain:2020puw,
    author = "Jain, Sachin and John, Renjan Rajan and Malvimat, Vinay",
    title = "{Constraining momentum space correlators using slightly broken higher spin symmetry}",
    eprint = "2008.08610",
    archivePrefix = "arXiv",
    primaryClass = "hep-th",
    doi = "10.1007/JHEP04(2021)231",
    journal = "JHEP",
    volume = "04",
    pages = "231",
    year = "2021"
}

@article{Caron-Huot:2020bkp,
    author = {Caron-Huot, Simon and Dixon, Lance J. and Drummond, James M. and Dulat, Falko and Foster, Jack and G{\"u}rdo{\u{g}}an, {\"O}mer and von Hippel, Matt and McLeod, Andrew J. and Papathanasiou, Georgios},
    title = "{The Steinmann Cluster Bootstrap for $N$ = 4 Super Yang-Mills Amplitudes}",
    eprint = "2005.06735",
    archivePrefix = "arXiv",
    primaryClass = "hep-th",
    reportNumber = "DESY-20-087",
    doi = "10.22323/1.376.0003",
    journal = "PoS",
    volume = "CORFU2019",
    pages = "003",
    year = "2020"
}

@article{Baumann:2025qjx,
    author = "Baumann, Daniel and Goodhew, Harry and Joyce, Austin and Lee, Hayden and Pimentel, Guilherme L. and Westerdijk, Tom",
    title = "{Geometry of kinematic flow}",
    eprint = "2504.14890",
    archivePrefix = "arXiv",
    primaryClass = "hep-th",
    doi = "10.1007/JHEP05(2026)211",
    journal = "JHEP",
    volume = "05",
    pages = "211",
    year = "2026"
}

@article{Baumann:2026atn,
    author = "Baumann, Daniel and Joyce, Austin and Lee, Hayden and Salehi Vaziri, Kamran",
    title = "{Differential Equations for Massive Correlators}",
    eprint = "2604.08658",
    archivePrefix = "arXiv",
    primaryClass = "hep-th",
    month = "4",
    year = "2026"
}

@article{Liu:2024str,
    author = "Liu, Haoyuan and Xianyu, Zhong-Zhi",
    title = "{Massive inflationary amplitudes: differential equations and complete solutions for general trees}",
    eprint = "2412.07843",
    archivePrefix = "arXiv",
    primaryClass = "hep-th",
    reportNumber = "USTC-ICTS/PCFT-24-56",
    doi = "10.1007/JHEP09(2025)183",
    journal = "JHEP",
    volume = "09",
    pages = "183",
    year = "2025"
}

@article{Benincasa:2019vqr,
    author = "Benincasa, Paolo",
    title = "{Cosmological Polytopes and the Wavefuncton of the Universe for Light States}",
    eprint = "1909.02517",
    archivePrefix = "arXiv",
    primaryClass = "hep-th",
    month = "9",
    year = "2019"
}

@article{Grafe:2026avi,
    author = {Gr{\"a}fe, Jonathan and Werth, Denis},
    title = "{All Tree-Level Massive Cosmological Correlators via Spectral Gluing}",
    eprint = "2607.18223",
    archivePrefix = "arXiv",
    primaryClass = "hep-th",
    month = "7",
    year = "2026"
}

@article{Jazayeri:2022kjy,
    author = "Jazayeri, Sadra and Renaux-Petel, S{\'e}bastien",
    title = "{Cosmological bootstrap in slow motion}",
    eprint = "2205.10340",
    archivePrefix = "arXiv",
    primaryClass = "hep-th",
    doi = "10.1007/JHEP12(2022)137",
    journal = "JHEP",
    volume = "12",
    pages = "137",
    year = "2022"
}

@article{Melville:2021lst,
    author = "Melville, Scott and Pajer, Enrico",
    title = "{Cosmological Cutting Rules}",
    eprint = "2103.09832",
    archivePrefix = "arXiv",
    primaryClass = "hep-th",
    doi = "10.1007/JHEP05(2021)249",
    journal = "JHEP",
    volume = "05",
    pages = "249",
    year = "2021"
}

@article{Goodhew:2020hob,
    author = "Goodhew, Harry and Jazayeri, Sadra and Pajer, Enrico",
    title = "{The Cosmological Optical Theorem}",
    eprint = "2009.02898",
    archivePrefix = "arXiv",
    primaryClass = "hep-th",
    doi = "10.1088/1475-7516/2021/04/021",
    journal = "JCAP",
    volume = "04",
    pages = "021",
    year = "2021"
}

@article{Sleight:2021plv,
    author = "Sleight, Charlotte and Taronna, Massimo",
    title = "{From dS to AdS and back}",
    eprint = "2109.02725",
    archivePrefix = "arXiv",
    primaryClass = "hep-th",
    doi = "10.1007/JHEP12(2021)074",
    journal = "JHEP",
    volume = "12",
    pages = "074",
    year = "2021"
}

@article{Sleight:2019hfp,
    author = "Sleight, Charlotte and Taronna, Massimo",
    title = "{Bootstrapping Inflationary Correlators in Mellin Space}",
    eprint = "1907.01143",
    archivePrefix = "arXiv",
    primaryClass = "hep-th",
    reportNumber = "PUPT-2590",
    doi = "10.1007/JHEP02(2020)098",
    journal = "JHEP",
    volume = "02",
    pages = "098",
    year = "2020"
}

@article{Glew:2025otn,
    author = "Glew, Ross and Lukowski, Tomasz",
    title = "{Amplitubes: graph cosmohedra}",
    eprint = "2502.17564",
    archivePrefix = "arXiv",
    primaryClass = "hep-th",
    doi = "10.1007/JHEP09(2025)074",
    journal = "JHEP",
    volume = "09",
    pages = "074",
    year = "2025"
}

@article{Capuano:2026pgq,
    author = "Capuano, Mattia and Ferro, Livia and Lukowski, Tomasz and Palazio, Alessandro and Zhang, Yao-Qi",
    title = "{Generalised Cluster Adjacency for Cosmology}",
    eprint = "2603.09965",
    archivePrefix = "arXiv",
    primaryClass = "hep-th",
    month = "3",
    year = "2026"
}

@article{Capuano:2025myy,
    author = "Capuano, Mattia and Ferro, Livia and Lukowski, Tomasz and Palazio, Alessandro",
    title = "{Cosmology meets cluster algebra}",
    eprint = "2512.14859",
    archivePrefix = "arXiv",
    primaryClass = "hep-th",
    month = "12",
    year = "2025"
}

@article{Ferro:2026oph,
    author = "Ferro, Livia and Lukowski, Tomasz and Ren, Lecheng and Spradlin, Marcus and Volovich, Anastasia and Weng, He-Chen and Zhang, Yao-Qi",
    title = "{de Sitter Wavefunction from Quadrangular Polylogarithms: Chain Graphs}",
    eprint = "2605.06542",
    archivePrefix = "arXiv",
    primaryClass = "hep-th",
    month = "5",
    year = "2026"
}

@article{Paranjape:2026htn,
    author = "Paranjape, Shruti and Skowronek, Marcos and Spradlin, Marcus and Volovich, Anastasia and Weng, He-Chen",
    title = "{Cluster Bootstrap for Cosmological Correlators}",
    eprint = "2603.08670",
    archivePrefix = "arXiv",
    primaryClass = "hep-th",
    month = "3",
    year = "2026"
}

@article{Brust:2016gjy,
    author = "Brust, Christopher and Hinterbichler, Kurt",
    title = "{Free {\ensuremath{\square}}$^{k}$ scalar conformal field theory}",
    eprint = "1607.07439",
    archivePrefix = "arXiv",
    primaryClass = "hep-th",
    doi = "10.1007/JHEP02(2017)066",
    journal = "JHEP",
    volume = "02",
    pages = "066",
    year = "2017"
}

@article{Brust:2016zns,
    author = "Brust, Christopher and Hinterbichler, Kurt",
    title = "{Partially Massless Higher-Spin Theory}",
    eprint = "1610.08510",
    archivePrefix = "arXiv",
    primaryClass = "hep-th",
    doi = "10.1007/JHEP02(2017)086",
    journal = "JHEP",
    volume = "02",
    pages = "086",
    year = "2017"
}

@article{Anninos:2019nib,
    author = "Anninos, D. and De Luca, V. and Franciolini, G. and Kehagias, A. and Riotto, A.",
    title = "{Cosmological Shapes of Higher-Spin Gravity}",
    eprint = "1902.01251",
    archivePrefix = "arXiv",
    primaryClass = "hep-th",
    doi = "10.1088/1475-7516/2019/04/045",
    journal = "JCAP",
    volume = "04",
    pages = "045",
    year = "2019"
}

@article{Lee:2023qqx,
    author = "Lee, Hayden and Wang, Xinkang",
    title = "{Amplitude basis for conformal correlators}",
    eprint = "2312.17312",
    archivePrefix = "arXiv",
    primaryClass = "hep-th",
    doi = "10.1007/JHEP03(2024)147",
    journal = "JHEP",
    volume = "03",
    pages = "147",
    year = "2024"
}

@article{Baumann:2021fxj,
    author = "Baumann, Daniel and Chen, Wei-Ming and Duaso Pueyo, Carlos and Joyce, Austin and Lee, Hayden and Pimentel, Guilherme L.",
    title = "{Linking the singularities of cosmological correlators}",
    eprint = "2106.05294",
    archivePrefix = "arXiv",
    primaryClass = "hep-th",
    doi = "10.1007/JHEP09(2022)010",
    journal = "JHEP",
    volume = "09",
    pages = "010",
    year = "2022"
}

@article{Baumann:2019oyu,
    author = "Baumann, Daniel and Duaso Pueyo, Carlos and Joyce, Austin and Lee, Hayden and Pimentel, Guilherme L.",
    title = "{The cosmological bootstrap: weight-shifting operators and scalar seeds}",
    eprint = "1910.14051",
    archivePrefix = "arXiv",
    primaryClass = "hep-th",
    doi = "10.1007/JHEP12(2020)204",
    journal = "JHEP",
    volume = "12",
    pages = "204",
    year = "2020"
}

@article{Lee:2022fgr,
    author = "Lee, Hayden and Wang, Xinkang",
    title = "{Cosmological double-copy relations}",
    eprint = "2212.11282",
    archivePrefix = "arXiv",
    primaryClass = "hep-th",
    doi = "10.1103/PhysRevD.108.L061702",
    journal = "Phys. Rev. D",
    volume = "108",
    number = "6",
    pages = "L061702",
    year = "2023"
}

@article{De:2025bmf,
    author = "De, Shounak and Paranjape, Shruti and Pokraka, Andrzej and Spradlin, Marcus and Volovich, Anastasia",
    title = "{Hidden zeros of the cosmological wavefunction}",
    eprint = "2503.23579",
    archivePrefix = "arXiv",
    primaryClass = "hep-th",
    doi = "10.1007/JHEP07(2025)174",
    journal = "JHEP",
    volume = "07",
    pages = "174",
    year = "2025"
}

@article{Ardila-Mantilla:2026cbo,
    author = "Ardila{\nobreakdash-}Mantilla, Federico and Arkani{\nobreakdash-}Hamed, Nima and Figueiredo, Carolina and Vaz{\~a}o, Francisco",
    title = "{Combinatorics of the Cosmohedron}",
    eprint = "2603.03425",
    archivePrefix = "arXiv",
    primaryClass = "math.CO",
    month = "3",
    year = "2026"
}

@article{Arkani-Hamed:2024jbp,
    author = "Arkani-Hamed, Nima and Figueiredo, Carolina and Vaz{\~a}o, Francisco",
    title = "{Cosmohedra}",
    eprint = "2412.19881",
    archivePrefix = "arXiv",
    primaryClass = "hep-th",
    doi = "10.1007/JHEP11(2025)029",
    journal = "JHEP",
    volume = "11",
    pages = "029",
    year = "2025"
}

@article{Arkani-Hamed:2023bsv,
    author = "Arkani-Hamed, Nima and Baumann, Daniel and Hillman, Aaron and Joyce, Austin and Lee, Hayden and Pimentel, Guilherme L.",
    title = "{Kinematic Flow and the Emergence of Time}",
    eprint = "2312.05300",
    archivePrefix = "arXiv",
    primaryClass = "hep-th",
    doi = "10.1103/dsjm-tckw",
    journal = "Phys. Rev. Lett.",
    volume = "135",
    number = "3",
    pages = "031602",
    year = "2025"
}

@article{Arkani-Hamed:2023kig,
    author = "Arkani-Hamed, Nima and Baumann, Daniel and Hillman, Aaron and Joyce, Austin and Lee, Hayden and Pimentel, Guilherme L.",
    title = "{Differential equations for cosmological correlators}",
    eprint = "2312.05303",
    archivePrefix = "arXiv",
    primaryClass = "hep-th",
    doi = "10.1007/JHEP09(2025)009",
    journal = "JHEP",
    volume = "09",
    pages = "009",
    year = "2025"
}

@article{Elvang:2013cua,
    author = "Elvang, Henriette and Huang, Yu-tin",
    title = "{Scattering Amplitudes}",
    eprint = "1308.1697",
    archivePrefix = "arXiv",
    primaryClass = "hep-th",
    month = "8",
    year = "2013"
}

@inproceedings{Cheung:2017pzi,
    author = "Cheung, Clifford",
    title = "{TASI lectures on scattering amplitudes.}",
    booktitle = "{Theoretical Advanced Study Institute in Elementary Particle Physics}: {Anticipating the Next Discoveries in Particle Physics}",
    eprint = "1708.03872",
    archivePrefix = "arXiv",
    primaryClass = "hep-ph",
    reportNumber = "CALT-TH-2017-041",
    doi = "10.1142/9789813233348_0008",
    pages = "571--623",
    year = "2018"
}

@inproceedings{Dixon:2013uaa,
    author = "Dixon, Lance J.",
    title = "{A brief introduction to modern amplitude methods}",
    booktitle = "{Theoretical Advanced Study Institute in Elementary Particle Physics}: {Particle Physics: The Higgs Boson and Beyond}",
    eprint = "1310.5353",
    archivePrefix = "arXiv",
    primaryClass = "hep-ph",
    reportNumber = "SLAC-PUB-15775",
    doi = "10.5170/CERN-2014-008.31",
    pages = "31--67",
    year = "2014"
}

@inproceedings{Giombi:2016ejx,
    author = "Giombi, Simone",
    title = "{Higher Spin {\textemdash} CFT Duality}",
    booktitle = "{Theoretical Advanced Study Institute in Elementary Particle Physics}: {New Frontiers in Fields and Strings}",
    eprint = "1607.02967",
    archivePrefix = "arXiv",
    primaryClass = "hep-th",
    doi = "10.1142/9789813149441_0003",
    pages = "137--214",
    year = "2017"
}

@article{Farren:2026hao,
    author = "Farren, Alexander and McCulloch, Ciaran and Pajer, Enrico and Tong, Xi",
    title = "{All-Loop Renormalization and the Phase of the de Sitter Wavefunction}",
    eprint = "2603.08794",
    archivePrefix = "arXiv",
    primaryClass = "hep-th",
    month = "3",
    year = "2026"
}

@article{Chowdhury:2026dwm,
    author = "Chowdhury, Chandramouli and He, Song and Su, Yong-Xiang and Yang, Dongyu",
    title = "{On the simplicity of de Sitter correlators}",
    eprint = "2604.26421",
    archivePrefix = "arXiv",
    primaryClass = "hep-th",
    month = "4",
    year = "2026"
}

@article{Arundine:2026fbr,
    author = "Arundine, Mattia and Baumann, Daniel and Lee, Mang Hei Gordon and Pimentel, Guilherme L. and Rost, Facundo",
    title = "{The Cosmological Grassmannian}",
    eprint = "2602.07117",
    archivePrefix = "arXiv",
    primaryClass = "hep-th",
    month = "2",
    year = "2026"
}

@article{Cheung:2022mkw,
    author = "Cheung, Clifford and Remmen, Grant N.",
    title = "{Veneziano Variations: How Unique are String Amplitudes?}",
    eprint = "2210.12163",
    archivePrefix = "arXiv",
    primaryClass = "hep-th",
    reportNumber = "CALT-TH 2022-037",
    doi = "10.1007/JHEP01(2023)122",
    journal = "JHEP",
    volume = "01",
    pages = "122",
    year = "2023"
}

@article{Cheung:2023adk,
    author = "Cheung, Clifford and Remmen, Grant N.",
    title = "{Stringy Dynamics from an Amplitudes Bootstrap}",
    eprint = "2302.12263",
    archivePrefix = "arXiv",
    primaryClass = "hep-th",
    reportNumber = "CALT-TH 2023-006",
    doi = "10.1103/PhysRevD.108.026011",
    journal = "Phys. Rev. D",
    volume = "108",
    number = "2",
    pages = "026011",
    year = "2023"
}

@article{Cheung:2023uwn,
    author = "Cheung, Clifford and Remmen, Grant N.",
    title = "{Bespoke Dual Resonance}",
    eprint = "2308.03833",
    archivePrefix = "arXiv",
    primaryClass = "hep-th",
    reportNumber = "CALT-TH 2023-026",
    doi = "10.1103/PhysRevD.108.086009",
    journal = "Phys. Rev. D",
    volume = "108",
    number = "8",
    pages = "086009",
    year = "2023"
}

@article{Cheung:2024uhn,
    author = "Cheung, Clifford and Hillman, Aaron and Remmen, Grant N.",
    title = "{Bootstrap Principle for the Spectrum and Scattering of Strings}",
    eprint = "2406.02665",
    archivePrefix = "arXiv",
    primaryClass = "hep-th",
    reportNumber = "CALT-TH 2024-022",
    doi = "10.1103/PhysRevLett.133.251601",
    journal = "Phys. Rev. Lett.",
    volume = "133",
    number = "25",
    pages = "251601",
    year = "2024"
}

@article{Cheung:2024xuo,
    author = "Cheung, Clifford and Hillman, Aaron and Remmen, Grant N.",
    title = "{Uniqueness Criteria for the Virasoro-Shapiro Amplitude}",
    eprint = "2408.03362",
    archivePrefix = "arXiv",
    primaryClass = "hep-th",
    reportNumber = "CALT-TH 2024-030",
    doi = "10.1103/PhysRevD.111.086034",
    journal = "Phys. Rev. D",
    volume = "111",
    number = "8",
    pages = "086034",
    year = "2025"
}

@article{Arkani-Hamed:2025mce,
    author = "Arkani-Hamed, Nima and Glew, Ross and Vaz{\~a}o, Francisco",
    title = "{Correlators are simpler than wavefunctions}",
    eprint = "2512.23795",
    archivePrefix = "arXiv",
    primaryClass = "hep-th",
    month = "12",
    year = "2025"
}

@article{Arkani-Hamed:2015bza,
    author = "Arkani-Hamed, Nima and Maldacena, Juan",
    title = "{Cosmological Collider Physics}",
    eprint = "1503.08043",
    archivePrefix = "arXiv",
    primaryClass = "hep-th",
    year = "2015"
}

@article{Arkani-Hamed:2017fdk,
    author = "Arkani-Hamed, Nima and Benincasa, Paolo and Postnikov, Alexander",
    title = "{Cosmological Polytopes and the Wavefunction of the Universe}",
    eprint = "1709.02813",
    archivePrefix = "arXiv",
    primaryClass = "hep-th",
    year = "2017"
}

@article{Arkani-Hamed:2018kmz,
    author = "Arkani-Hamed, Nima and Baumann, Daniel and Lee, Hayden and Pimentel, Guilherme L.",
    title = "{The Cosmological Bootstrap: Inflationary Correlators from Symmetries and Singularities}",
    eprint = "1811.00024",
    archivePrefix = "arXiv",
    primaryClass = "hep-th",
    reportNumber = "CALT-TH-2018-45",
    doi = "10.1007/JHEP04(2020)105",
    journal = "JHEP",
    volume = "04",
    pages = "105",
    year = "2020"
}

@article{Baumann:2020dch,
    author = "Baumann, Daniel and Duaso Pueyo, Carlos and Joyce, Austin and Lee, Hayden and Pimentel, Guilherme L.",
    title = "{The cosmological bootstrap: spinning correlators from symmetries and factorization}",
    eprint = "2005.04234",
    archivePrefix = "arXiv",
    primaryClass = "hep-th",
    doi = "10.21468/SciPostPhys.11.3.071",
    journal = "SciPost Phys.",
    volume = "11",
    number = "3",
    pages = "071",
    year = "2021"
}

@article{Baumann:2022jpr,
    author = "Baumann, Daniel and others",
    title = "{Snowmass White Paper: The Cosmological Bootstrap}",
    eprint = "2203.08121",
    archivePrefix = "arXiv",
    primaryClass = "hep-th",
    reportNumber = "Snowmass 2021",
    year = "2022"
}

@article{Vasiliev:1990en,
    author = "Vasiliev, M. A.",
    title = "{Consistent equations for interacting gauge fields of all spins in 3+1 dimensions}",
    doi = "10.1016/0370-2693(90)91200-6",
    journal = "Phys. Lett. B",
    volume = "243",
    pages = "378--382",
    year = "1990"
}

@article{Klebanov:2002ja,
    author = "Klebanov, Igor R. and Polyakov, Alexander M.",
    title = "{AdS dual of the critical O(N) vector model}",
    eprint = "hep-th/0210114",
    archivePrefix = "arXiv",
    doi = "10.1016/S0370-2693(02)02980-5",
    journal = "Phys. Lett. B",
    volume = "550",
    pages = "213--219",
    year = "2002"
}

@article{Sezgin:2002rt,
    author = "Sezgin, E. and Sundell, P.",
    title = "{Massless higher spins and holography}",
    eprint = "hep-th/0205131",
    archivePrefix = "arXiv",
    doi = "10.1016/S0550-3213(02)00739-3",
    journal = "Nucl. Phys. B",
    volume = "644",
    pages = "303--370",
    year = "2002"
}

@article{Giombi:2009wh,
    author = "Giombi, Simone and Yin, Xi",
    title = "{Higher Spin Gauge Theory and Holography: The Three-Point Functions}",
    eprint = "0912.3462",
    archivePrefix = "arXiv",
    primaryClass = "hep-th",
    doi = "10.1007/JHEP09(2010)115",
    journal = "JHEP",
    volume = "09",
    pages = "115",
    year = "2010"
}

@article{Maldacena:2012sf,
    author = "Maldacena, Juan and Zhiboedov, Alexander",
    title = "{Constraining conformal field theories with a slightly broken higher spin symmetry}",
    eprint = "1204.3882",
    archivePrefix = "arXiv",
    primaryClass = "hep-th",
    doi = "10.1088/0264-9381/30/10/104003",
    journal = "Class. Quant. Grav.",
    volume = "30",
    pages = "104003",
    year = "2013"
}

@article{Anninos:2011ui,
    author = "Anninos, Dionysios and Hartman, Thomas and Strominger, Andrew",
    title = "{Higher Spin Realization of the dS/CFT Correspondence}",
    eprint = "1108.5735",
    archivePrefix = "arXiv",
    primaryClass = "hep-th",
    doi = "10.1088/1361-6382/34/1/015009",
    journal = "Class. Quant. Grav.",
    volume = "34",
    number = "1",
    pages = "015009",
    year = "2017"
}

@article{Anninos:2017eib,
  author = {Anninos, Dionysios and Denef, Frederik and Monten, Ruben and Sun, Zimo},
  title = {{Higher Spin de Sitter Hilbert Space}},
  eprint = {1711.10037},
  archivePrefix = {arXiv},
  primaryClass = {hep-th},
  doi = {10.1007/JHEP10(2019)071},
  journal = {JHEP},
  volume = {10},
  pages = {071},
  year = {2019},
}

@article{Melrose:1965,
  author = {Melrose, D. B.},
  title = {{Reduction of Feynman diagrams}},
  doi = {10.1007/BF02832919},
  journal = {Nuovo Cim. A},
  volume = {40},
  pages = {181--213},
  year = {1965}
}

@article{vanNeerven:1984,
  author = {van Neerven, W. L. and Vermaseren, J. A. M.},
  title = {{Large loop integrals}},
  doi = {10.1016/0370-2693(84)90237-5},
  journal = {Phys. Lett. B},
  volume = {137},
  pages = {241--244},
  year = {1984}
}

@article{DAndrea:2004,
  author = {D'Andrea, Carlos and Sombra, Martin},
  title = {{The Cayley-Menger determinant is irreducible for $n \geq 3$}},
  eprint = {math/0406359},
  archivePrefix = {arXiv},
  primaryClass = {math.AG},
  doi = {10.1007/s11202-005-0007-0},
  journal = {Siberian Math. J.},
  volume = {46},
  number = {1},
  pages = {71--76},
  year = {2005}
}

@article{Bzowski:2019kwd,
  author = {Bzowski, Adam and McFadden, Paul and Skenderis, Kostas},
  title = {{Conformal $n$-point functions in momentum space}},
  eprint = {1910.10162},
  archivePrefix = {arXiv},
  primaryClass = {hep-th},
  doi = {10.1103/PhysRevLett.124.131602},
  journal = {Phys. Rev. Lett.},
  volume = {124},
  pages = {131602},
  year = {2020}
}

@article{Hogervorst:2021uvp,
  author = {Hogervorst, Matthijs and Penedones, Jo\~ao and Vaziri, Kevin S.},
  title = {{Towards the non-perturbative cosmological bootstrap}},
  eprint = {2107.13871},
  archivePrefix = {arXiv},
  primaryClass = {hep-th},
  doi = {10.1007/JHEP02(2023)162},
  journal = {JHEP},
  volume = {02},
  pages = {162},
  year = {2023}
}

@article{Chowdhury:2023ssc,
    author = "Chowdhury, Chandramouli and Lipstein, Arthur and Mei, Jiajie and Sachs, Ivo and Vanhove, Pierre",
    title = "{The Subtle Simplicity of Cosmological Correlators}",
    eprint = "2312.13803",
    archivePrefix = "arXiv",
    primaryClass = "hep-th",
    doi = "10.1007/JHEP03(2025)007",
    journal = "JHEP",
    volume = "03",
    pages = "007",
    year = "2025"
}

@article{Giombi:2011kc,
  author = {Giombi, Simone and Yin, Xi},
  title = {{Higher Spin Gauge Theory and the Critical $O(N)$ Model}},
  eprint = {1105.4011},
  archivePrefix = {arXiv},
  primaryClass = {hep-th},
  doi = {10.1103/PhysRevD.85.086005},
  journal = {Phys. Rev. D},
  volume = {85},
  pages = {086005},
  year = {2012}
}

@article{CalvoCortes:2025ks,
  author = {Calvo Cortes, Veronica and Frost, Hadleigh and Sturmfels, Bernd},
  title = {{Kinematic Stratifications}},
  eprint = {2503.09571},
  archivePrefix = {arXiv},
  primaryClass = {math.CO},
  year = {2025}
}

@article{Devriendt:2024twolives,
  author = {Devriendt, Karel and Friedman, Hannah and Reinke, Bernhard and Sturmfels, Bernd},
  title = {{The Two Lives of the Grassmannian}},
  eprint = {2401.03684},
  archivePrefix = {arXiv},
  primaryClass = {math.AG},
  doi = {10.1007/s44426-025-00007-x},
  journal = {Acta Univ. Sapientiae Math.},
  volume = {17},
  pages = {8},
  year = {2025}
}

@article{De:2026shn,
    author = "De, Shounak and Lee, Hayden",
    title = "{The Vasiliev Grassmannian}",
    eprint = "2603.24656",
    archivePrefix = "arXiv",
    primaryClass = "hep-th",
    month = "3",
    year = "2026"
}

@article{Bzowski:2013sza,
    author = "Bzowski, Adam and McFadden, Paul and Skenderis, Kostas",
    title = "{Implications of conformal invariance in momentum space}",
    eprint = "1304.7760",
    archivePrefix = "arXiv",
    primaryClass = "hep-th",
    doi = "10.1007/JHEP03(2014)111",
    journal = "JHEP",
    volume = "03",
    pages = "111",
    year = "2014"
}

@article{Bzowski:2011ab,
    author = "Bzowski, Adam and McFadden, Paul and Skenderis, Kostas",
    title = "{Holographic predictions for cosmological 3-point functions}",
    eprint = "1112.1967",
    archivePrefix = "arXiv",
    primaryClass = "hep-th",
    doi = "10.1007/JHEP03(2012)091",
    journal = "JHEP",
    volume = "03",
    pages = "091",
    year = "2012"
}

@article{Symanzik:1972wj,
    author = "Symanzik, K.",
    title = "{On Calculations in conformal invariant field theories}",
    reportNumber = "DESY-72-6",
    doi = "10.1007/BF02824349",
    journal = "Lett. Nuovo Cim.",
    volume = "3",
    pages = "734--738",
    year = "1972"
}

@article{Petkou:2003,
    author = "Petkou, Anastasios C.",
    title = "{Evaluating the AdS dual of the critical O(N) vector model}",
    eprint = "hep-th/0302063",
    archivePrefix = "arXiv",
    reportNumber = "CERN-TH-2003-021",
    doi = "10.1088/1126-6708/2003/03/049",
    journal = "JHEP",
    volume = "03",
    pages = "049",
    year = "2003"
}

@article{Sezgin:2003pt,
    author = "Sezgin, E. and Sundell, P.",
    title = "{Holography in 4D (super) higher spin theories and a test via cubic scalar couplings}",
    eprint = "hep-th/0305040",
    archivePrefix = "arXiv",
    reportNumber = "MIFP-03-09, UU-07-03",
    doi = "10.1088/1126-6708/2005/07/044",
    journal = "JHEP",
    volume = "07",
    pages = "044",
    year = "2005"
}

@article{Ellis:2011cr,
    author = "Ellis, R. Keith and Kunszt, Zoltan and Melnikov, Kirill and Zanderighi, Giulia",
    title = "{One-loop calculations in quantum field theory: from Feynman diagrams to unitarity cuts}",
    eprint = "1105.4319",
    archivePrefix = "arXiv",
    primaryClass = "hep-ph",
    reportNumber = "FERMILAB-PUB-11-195-T",
    doi = "10.1016/j.physrep.2012.01.008",
    journal = "Phys. Rept.",
    volume = "518",
    pages = "141--250",
    year = "2012"
}

@article{vanOldenborgh:1989wn,
    author = "van Oldenborgh, G. J. and Vermaseren, J. A. M.",
    title = "{New Algorithms for One Loop Integrals}",
    reportNumber = "NIKHEF-H/89-17",
    doi = "10.1007/BF01621031",
    journal = "Z. Phys. C",
    volume = "46",
    pages = "425--438",
    year = "1990"
}

@article{Veneziano:1968yb,
    author = "Veneziano, G.",
    title = "{Construction of a crossing-symmetric, Regge behaved amplitude for linearly rising trajectories}",
    doi = "10.1007/BF02824451",
    journal = "Nuovo Cim. A",
    volume = "57",
    pages = "190--197",
    year = "1968"
}

@article{Caron-Huot:2016icg,
    author = "Caron-Huot, Simon and Komargodski, Zohar and Sever, Amit and Zhiboedov, Alexander",
    title = "{Strings from Massive Higher Spins: The Asymptotic Uniqueness of the Veneziano Amplitude}",
    eprint = "1607.04253",
    archivePrefix = "arXiv",
    primaryClass = "hep-th",
    doi = "10.1007/JHEP10(2017)026",
    journal = "JHEP",
    volume = "10",
    pages = "026",
    year = "2017"
}

@article{Chen:2009zp,
    author = "Chen, Xingang and Wang, Yi",
    title = "{Large non-Gaussianities with Intermediate Shapes from Quasi-Single Field Inflation}",
    eprint = "0909.0496",
    archivePrefix = "arXiv",
    primaryClass = "astro-ph.CO",
    doi = "10.1103/PhysRevD.81.063511",
    journal = "Phys. Rev. D",
    volume = "81",
    pages = "063511",
    year = "2010"
}

@article{Chen:2009we,
    author = "Chen, Xingang and Wang, Yi",
    title = "{Quasi-Single Field Inflation and Non-Gaussianities}",
    eprint = "0911.3380",
    archivePrefix = "arXiv",
    primaryClass = "hep-th",
    doi = "10.1088/1475-7516/2010/04/027",
    journal = "JCAP",
    volume = "04",
    pages = "027",
    year = "2010"
}

@article{Noumi:2012vr,
    author = "Noumi, Toshifumi and Yamaguchi, Masahide and Yokoyama, Daisuke",
    title = "{Effective field theory approach to quasi-single field inflation and effects of heavy fields}",
    eprint = "1211.1624",
    archivePrefix = "arXiv",
    primaryClass = "hep-th",
    doi = "10.1007/JHEP06(2013)051",
    journal = "JHEP",
    volume = "06",
    pages = "051",
    year = "2013"
}

@article{Chen:2015lza,
    author = "Chen, Xingang and Namjoo, Mohammad Hossein and Wang, Yi",
    title = "{Quantum Primordial Standard Clocks}",
    eprint = "1509.03930",
    archivePrefix = "arXiv",
    primaryClass = "astro-ph.CO",
    doi = "10.1088/1475-7516/2016/02/013",
    journal = "JCAP",
    volume = "02",
    pages = "013",
    year = "2016"
}

@article{Chen:2016uwp,
    author = "Chen, Xingang and Wang, Yi and Xianyu, Zhong-Zhi",
    title = "{Standard Model Background of the Cosmological Collider}",
    eprint = "1610.06597",
    archivePrefix = "arXiv",
    primaryClass = "hep-th",
    doi = "10.1007/JHEP04(2017)058",
    journal = "JHEP",
    volume = "04",
    pages = "058",
    year = "2017"
}

@article{Pimentel:2022fsc,
    author = "Pimentel, Guilherme L. and Wang, Dong-Gang",
    title = "{Boostless Cosmological Collider Bootstrap}",
    eprint = "2205.00013",
    archivePrefix = "arXiv",
    primaryClass = "hep-th",
    doi = "10.1007/JHEP10(2022)177",
    journal = "JHEP",
    volume = "10",
    pages = "177",
    year = "2022"
}

@article{Bodas:2020yho,
    author = "Bodas, Arushi and Kumar, Soubhik and Sundrum, Raman",
    title = "{The Scalar Chemical Potential in Cosmological Collider Physics}",
    eprint = "2010.04727",
    archivePrefix = "arXiv",
    primaryClass = "hep-th",
    doi = "10.1007/JHEP02(2021)079",
    journal = "JHEP",
    volume = "02",
    pages = "079",
    year = "2021"
}

@article{Pajer:2020wxk,
    author = "Pajer, Enrico",
    title = "{Building a Boostless Bootstrap for the Bispectrum}",
    eprint = "2010.12818",
    archivePrefix = "arXiv",
    primaryClass = "hep-th",
    doi = "10.1088/1475-7516/2021/01/023",
    journal = "JCAP",
    volume = "01",
    pages = "023",
    year = "2021"
}

@article{Jazayeri:2021fvk,
    author = "Jazayeri, Sadra and Pajer, Enrico and Stefanyszyn, David",
    title = "{From Locality and Unitarity to Cosmological Correlators}",
    eprint = "2103.08649",
    archivePrefix = "arXiv",
    primaryClass = "hep-th",
    doi = "10.1007/JHEP10(2021)065",
    journal = "JHEP",
    volume = "10",
    pages = "065",
    year = "2021"
}

@article{Benincasa:2024leu,
    author = "Benincasa, Paolo and Dian, Gabriele",
    title = "{The geometry of cosmological correlators}",
    eprint = "2401.05207",
    archivePrefix = "arXiv",
    primaryClass = "hep-th",
    reportNumber = "MPP-2023-150, DESY-24-006",
    doi = "10.21468/SciPostPhys.18.3.105",
    journal = "SciPost Phys.",
    volume = "18",
    number = "3",
    pages = "105",
    year = "2025"
}

@article{Bonifacio:2021azc,
    author = "Bonifacio, James and Pajer, Enrico and Wang, Dong-Gang",
    title = "{From Amplitudes to Contact Cosmological Correlators}",
    eprint = "2106.15468",
    archivePrefix = "arXiv",
    primaryClass = "hep-th",
    doi = "10.1007/JHEP10(2021)001",
    journal = "JHEP",
    volume = "10",
    pages = "001",
    year = "2021"
}

@article{Baumann:2011nk,
    author = "Baumann, Daniel and Green, Daniel",
    title = "{Signatures of Supersymmetry from the Early Universe}",
    eprint = "1109.0292",
    archivePrefix = "arXiv",
    primaryClass = "hep-th",
    doi = "10.1103/PhysRevD.85.103520",
    journal = "Phys. Rev. D",
    volume = "85",
    pages = "103520",
    year = "2012"
}

@article{Assassi:2012zq,
    author = "Assassi, Valentin and Baumann, Daniel and Green, Daniel",
    title = "{On Soft Limits of Inflationary Correlation Functions}",
    eprint = "1204.4207",
    archivePrefix = "arXiv",
    primaryClass = "hep-th",
    doi = "10.1088/1475-7516/2012/11/047",
    journal = "JCAP",
    volume = "11",
    pages = "047",
    year = "2012"
}

@article{Wang:2019gbi,
    author = "Wang, Lian-Tao and Xianyu, Zhong-Zhi",
    title = "{In Search of Large Signals at the Cosmological Collider}",
    eprint = "1910.12876",
    archivePrefix = "arXiv",
    primaryClass = "hep-ph",
    doi = "10.1007/JHEP02(2020)044",
    journal = "JHEP",
    volume = "02",
    pages = "044",
    year = "2020"
}

@article{Qin:2022xrs,
    author = "Qin, Zhehan and Xianyu, Zhong-Zhi",
    title = "{Helical Inflation Correlators: Partial Mellin-Barnes and Bootstrap Equations}",
    eprint = "2208.13790",
    archivePrefix = "arXiv",
    primaryClass = "hep-th",
    doi = "10.1007/JHEP04(2023)059",
    journal = "JHEP",
    volume = "04",
    pages = "059",
    year = "2023"
}

@article{Arkani-Hamed:2023jry,
    author = "Arkani-Hamed, Nima and Cheung, Clifford and Figueiredo, Carolina and Remmen, Grant N.",
    title = "{Multiparticle Factorization and the Rigidity of String Theory}",
    eprint = "2312.07652",
    archivePrefix = "arXiv",
    primaryClass = "hep-th",
    doi = "10.1103/PhysRevLett.132.091601",
    journal = "Phys. Rev. Lett.",
    volume = "132",
    pages = "091601",
    year = "2024"
}

@article{Benincasa:2024lpy,
    author = "Benincasa, Paolo and Brunello, Giacomo and Mandal, Manoj K. and Mastrolia, Pierpaolo and Vaz\~ao, Francisco",
    title = "{On one-loop corrections to the Bunch-Davies wavefunction of the universe}",
    eprint = "2408.16386",
    archivePrefix = "arXiv",
    primaryClass = "hep-th",
    doi = "10.1103/PhysRevD.111.085016",
    journal = "Phys. Rev. D",
    volume = "111",
    pages = "085016",
    year = "2025"
}

@article{Pimentel:2026abc,
    author = "Pimentel, Guilherme L. and Westerdijk, Tom",
    title = "{On Cosmological Correlators at One Loop}",
    eprint = "2601.00952",
    archivePrefix = "arXiv",
    primaryClass = "hep-th",
    year = "2026"
}

@article{Sundborg:2000wp,
    author = "Sundborg, Bo",
    title = "{Stringy gravity, interacting tensionless strings and massless higher spins}",
    eprint = "hep-th/0103247",
    archivePrefix = "arXiv",
    reportNumber = "USITP-01-07",
    year = "2001"
}

@article{Mikhailov:2002bp,
    author = "Mikhailov, Andrei",
    title = "{Notes on higher spin symmetries}",
    eprint = "hep-th/0201019",
    archivePrefix = "arXiv",
    year = "2002"
}

@article{McFadden:2009fg,
    author = "McFadden, Paul and Skenderis, Kostas",
    title = "{Holography for cosmology}",
    eprint = "0907.5542",
    archivePrefix = "arXiv",
    primaryClass = "hep-th",
    doi = "10.1103/PhysRevD.81.021301",
    journal = "Phys. Rev. D",
    volume = "81",
    pages = "021301",
    year = "2010"
}

@article{McFadden:2010na,
    author = "McFadden, Paul and Skenderis, Kostas",
    title = "{Holographic Non-Gaussianity}",
    eprint = "1011.0452",
    archivePrefix = "arXiv",
    primaryClass = "hep-th",
    doi = "10.1088/1475-7516/2011/05/013",
    journal = "JCAP",
    volume = "05",
    pages = "013",
    year = "2011"
}

@article{Lee:2016vti,
    author = "Lee, Hayden and Baumann, Daniel and Pimentel, Guilherme L.",
    title = "{Non-Gaussianity as a Particle Detector}",
    eprint = "1607.03735",
    archivePrefix = "arXiv",
    primaryClass = "hep-th",
    doi = "10.1007/JHEP12(2016)040",
    journal = "JHEP",
    volume = "12",
    pages = "040",
    year = "2016"
}

@article{Wan:2026pjq,
    author = "Wan, Shi-Lin and Zhou, Shuang-Yong",
    title = "{Analytic Bootstrap of the Veneziano Amplitude}",
    eprint = "2605.11084",
    archivePrefix = "arXiv",
    primaryClass = "hep-th",
    month = "5",
    year = "2026"
}

@article{Lipstein:2012kd,
    author = "Lipstein, Arthur E. and Mason, Lionel",
    title = "{Amplitudes of 3d Yang Mills Theory}",
    eprint = "1207.6176",
    archivePrefix = "arXiv",
    primaryClass = "hep-th",
    doi = "10.1007/JHEP01(2013)009",
    journal = "JHEP",
    volume = "01",
    pages = "009",
    year = "2013"
}

@article{Leigh:2003gk,
    author = "Leigh, Robert G. and Petkou, Anastasios C.",
    title = "{Holography of the N=1 higher spin theory on AdS(4)}",
    eprint = "hep-th/0304217",
    archivePrefix = "arXiv",
    reportNumber = "CERN-TH-2003-095",
    doi = "10.1088/1126-6708/2003/06/011",
    journal = "JHEP",
    volume = "06",
    pages = "011",
    year = "2003"
}

@article{Aharony:2011jz,
    author = "Aharony, Ofer and Gur-Ari, Guy and Yacoby, Ran",
    title = "{d=3 Bosonic Vector Models Coupled to Chern-Simons Gauge Theories}",
    eprint = "1110.4382",
    archivePrefix = "arXiv",
    primaryClass = "hep-th",
    doi = "10.1007/JHEP03(2012)037",
    journal = "JHEP",
    volume = "03",
    pages = "037",
    year = "2012"
}

@article{Chang:2012kt,
    author = "Chang, Chi-Ming and Minwalla, Shiraz and Sharma, Tarun and Yin, Xi",
    title = "{ABJ Triality: from Higher Spin Fields to Strings}",
    eprint = "1207.4485",
    archivePrefix = "arXiv",
    primaryClass = "hep-th",
    reportNumber = "TIFR-TH-12-29",
    doi = "10.1088/1751-8113/46/21/214009",
    journal = "J. Phys. A",
    volume = "46",
    pages = "214009",
    year = "2013"
}

@book{Arkani-Hamed:2012zlh,
    author = "Arkani-Hamed, Nima and Bourjaily, Jacob L. and Cachazo, Freddy and Goncharov, Alexander B. and Postnikov, Alexander and Trnka, Jaroslav",
    title = "{Grassmannian Geometry of Scattering Amplitudes}",
    eprint = "1212.5605",
    archivePrefix = "arXiv",
    primaryClass = "hep-th",
    reportNumber = "PUPT-2435",
    doi = "10.1017/CBO9781316091548",
    isbn = "978-1-107-08658-6, 978-1-316-57296-2",
    publisher = "Cambridge University Press",
    month = "4",
    year = "2016"
}

@article{Bala:2026bdx,
    author = "Bala, Aswini and Jain, Sachin and S., Dhruva K. and Rao, Adithya A.",
    title = "{Super-Grassmannians for $\mathcal{N}=2$ to $4$ SCFT$_3$: From AdS$_4$ Correlators to $\mathcal{N}=4$ SYM scattering Amplitudes}",
    eprint = "2604.07503",
    archivePrefix = "arXiv",
    primaryClass = "hep-th",
    month = "4",
    year = "2026"
}

@article{Bala:2026hdm,
    author = "Bala, Aswini and Jain, Sachin and S., Dhruva K. and Rao, Adithya A.",
    title = "{The $\mathcal{N}=1$ Super-Grassmannian for CFT$_3$ and a Foray on AdS and Cosmological Correlators}",
    eprint = "2604.07446",
    archivePrefix = "arXiv",
    primaryClass = "hep-th",
    month = "4",
    year = "2026"
}

@article{Arundine:2026myr,
    author = "Arundine, Mattia and Pimentel, Guilherme L.",
    title = "{Cosmological Collider in the Grassmannian}",
    eprint = "2605.21581",
    archivePrefix = "arXiv",
    primaryClass = "hep-th",
    month = "5",
    year = "2026"
}

@article{Huang:2026tsh,
    author = "Huang, Yu-tin and Kuo, Chia-Kai and Liu, Yohan and Mei, Jiajie",
    title = "{Beyond Discontinuities: Cosmological WFCs from the Supersymmetric Orthogonal Grassmannian}",
    eprint = "2604.08512",
    archivePrefix = "arXiv",
    primaryClass = "hep-th",
    month = "4",
    year = "2026"
}

@article{Jain:2020rmw,
    author = "Jain, Sachin and John, Renjan Rajan and Malvimat, Vinay",
    title = "{Momentum space spinning correlators and higher spin equations in three dimensions}",
    eprint = "2005.07212",
    archivePrefix = "arXiv",
    primaryClass = "hep-th",
    doi = "10.1007/JHEP11(2020)049",
    journal = "JHEP",
    volume = "11",
    pages = "049",
    year = "2020"
}

@article{Anninos:2026hia,
    author = {Anninos, Dionysios and Baracco, Chiara and Letsios, Vasileios A. and M{\"u}hlmann, Beatrix},
    title = "{dS$^4$ Metamorphosis}",
    eprint = "2602.19812",
    archivePrefix = "arXiv",
    primaryClass = "hep-th",
    month = "2",
    year = "2026"
}

@article{Anninos:2020hfj,
    author = "Anninos, Dionysios and Denef, Frederik and Law, Y. T. Albert and Sun, Zimo",
    title = "{Quantum de Sitter horizon entropy from quasicanonical bulk, edge, sphere and topological string partition functions}",
    eprint = "2009.12464",
    archivePrefix = "arXiv",
    primaryClass = "hep-th",
    doi = "10.1007/JHEP01(2022)088",
    journal = "JHEP",
    volume = "01",
    pages = "088",
    year = "2022"
}

@article{Giombi:2013fka,
    author = "Giombi, Simone and Klebanov, Igor R.",
    title = "{One Loop Tests of Higher Spin AdS/CFT}",
    eprint = "1308.2337",
    archivePrefix = "arXiv",
    primaryClass = "hep-th",
    reportNumber = "PUTP-2451",
    doi = "10.1007/JHEP12(2013)068",
    journal = "JHEP",
    volume = "12",
    pages = "068",
    year = "2013"
}

@article{Chang:2013afa,
    author = "Chang, Chi-Ming and Pathak, Abhishek and Strominger, Andrew",
    title = "{Non-Minimal Higher-Spin DS4/CFT3}",
    eprint = "1309.7413",
    archivePrefix = "arXiv",
    primaryClass = "hep-th",
    month = "9",
    year = "2013"
}

@article{Anninos:2014hia,
    author = "Anninos, Dionysios and Mahajan, Raghu and Radicevic, Djordje and Shaghoulian, Edgar",
    title = "{Chern-Simons-Ghost Theories and de Sitter Space}",
    eprint = "1405.1424",
    archivePrefix = "arXiv",
    primaryClass = "hep-th",
    reportNumber = "SU-ITP-14-11",
    doi = "10.1007/JHEP01(2015)074",
    journal = "JHEP",
    volume = "01",
    pages = "074",
    year = "2015"
}

@article{Hertog:2017ymy,
    author = "Hertog, Thomas and Tartaglino-Mazzucchelli, Gabriele and Van Riet, Thomas and Venken, Victoria",
    title = "{Supersymmetric dS/CFT}",
    eprint = "1709.06024",
    archivePrefix = "arXiv",
    primaryClass = "hep-th",
    doi = "10.1007/JHEP02(2018)024",
    journal = "JHEP",
    volume = "02",
    pages = "024",
    year = "2018"
}

@article{Didenko:2014dwa,
    author = "Didenko, V. E. and Skvortsov, E. D.",
    title = "{Elements of Vasiliev Theory}",
    eprint = "1401.2975",
    archivePrefix = "arXiv",
    primaryClass = "hep-th",
    doi = "10.1007/978-3-031-59656-8_3",
    journal = "Lect. Notes Phys.",
    volume = "1028",
    pages = "269--456",
    year = "2024"
}

@article{Diaz:2006nm,
    author = "Diaz, Danilo E. and Dorn, Harald",
    title = "{On the AdS higher spin / O(N) vector model correspondence: Degeneracy of the holographic image}",
    eprint = "hep-th/0603084",
    archivePrefix = "arXiv",
    reportNumber = "HU-EP-06-10",
    doi = "10.1088/1126-6708/2006/07/022",
    journal = "JHEP",
    volume = "07",
    pages = "022",
    year = "2006"
}

@article{Maldacena:2015iua,
    author = "Maldacena, Juan and Simmons-Duffin, David and Zhiboedov, Alexander",
    title = "{Looking for a bulk point}",
    eprint = "1509.03612",
    archivePrefix = "arXiv",
    primaryClass = "hep-th",
    doi = "10.1007/JHEP01(2017)013",
    journal = "JHEP",
    volume = "01",
    pages = "013",
    year = "2017"
}

@article{Sleight:2017pcz,
    author = "Sleight, Charlotte and Taronna, Massimo",
    title = "{Higher-Spin Gauge Theories and Bulk Locality}",
    eprint = "1704.07859",
    archivePrefix = "arXiv",
    primaryClass = "hep-th",
    doi = "10.1103/PhysRevLett.121.171604",
    journal = "Phys. Rev. Lett.",
    volume = "121",
    number = "17",
    pages = "171604",
    year = "2018"
}

@article{Ponomarev:2017qab,
    author = "Ponomarev, Dmitry",
    title = "{A Note on (Non)-Locality in Holographic Higher Spin Theories}",
    eprint = "1710.00403",
    archivePrefix = "arXiv",
    primaryClass = "hep-th",
    reportNumber = "IMPERIAL-TP-DP-2017-02",
    doi = "10.3390/universe4010002",
    journal = "Universe",
    volume = "4",
    number = "1",
    pages = "2",
    year = "2018"
}

@article{Ponomarev:2019ltz,
    author = "Ponomarev, Dmitry and Sezgin, Ergin and Skvortsov, Evgeny",
    title = "{On one loop corrections in higher spin gravity}",
    eprint = "1904.01042",
    archivePrefix = "arXiv",
    primaryClass = "hep-th",
    doi = "10.1007/JHEP11(2019)138",
    journal = "JHEP",
    volume = "11",
    pages = "138",
    year = "2019"
}

\end{document}